\documentstyle[emulateapj]{article}

\lefthead{Evrard \etal\ (Virgo Consortium)} 
\righthead{Clusters in Hubble Volume Sky Surveys}
\submitted{}

\def\apj{ApJ~}

\def\apjs{ApJS}

\def\aa{A\&A~}
\def\mnras{MNRAS~}

\def\spose#1{\hbox to 0pt{#1\hss}}
\def\lta{\mathrel{\spose{\lower 3pt\hbox{$\mathchar"218$}}
     \raise 2.0pt\hbox{$\mathchar"13C$}}}
\def\gta{\mathrel{\spose{\lower 3pt\hbox{$\mathchar"218$}}
     \raise 2.0pt\hbox{$\mathchar"13E$}}}

\def\\{\hfil\break}

\def \mpc       {{\rm\ Mpc}}
\def \gpc       {{\rm\ Gpc}}
\def \kpc       {{\rm\ kpc}}

\def \kms       {\hbox{ km s$^{-1}$}}

\def \eg        {\hbox{\it e.g.},}

\def \etal      {{\it et al.}}
\def \dln       {\hbox{\rm d$\,$ln }}

\def \Ho        {{\rm H_{0}}}

\def \kev       {{\rm\ keV}}
\def \msol      {{\rm M}_\odot}
\def \hinv      {\hbox{$\, h^{-1}$}}
\def \ergs      {\hbox{$\,$ erg s$^{-1}$}}
\def \rms       {\hbox{\sl rms}}
\def \cgsflux   {{\rm\ erg\ s^{-1}\ cm^{-2}}}

\def \se        {\!=\!}

\def \ssim      {\! \sim \!}
\def \sims      { \sim \!}
\def \ssimeq    {\! \simeq \!}
\def \sequiv    {\! \equiv \!}
\def \spropto   {\! \propto \!}

\def \xray {\hbox{X--ray}}

\def \rtwoh {\hbox{$r_{200}$}}

\def \Mtwoh {\hbox{$M_{200}$}}
\def \mtwoh {\hbox{$M_{200}$}}
\def \Deltac {\hbox{$\Delta$}}
\def \rdeltac {\hbox{$r_{\Delta}$}}
\def \Mdeltac {\hbox{$M_{\Delta}$}}

\def \rhocrit {\hbox{$\rho_c$}}
\def \rhobar {\hbox{$\bar{\rho}_m$}}

\def \tcdm {$\tau$CDM}
\def \lcdm {$\Lambda$CDM}

\def \sigmagal {\hbox{$\sigma_{\rm gal}$}}
\def \sigmadm {\hbox{$\sigma_v$}}

\def \Tx {\hbox{$T$}}

\def \lxt {\hbox{$L_X$--$T$}}

\def \betat {\hbox{$\beta_{\tau}$}}
\def \betal {\hbox{$\beta_{\Lambda}$}}

\def \lnsiginv {\hbox{$\ln\sigma^{-1}$}}
\def \dlnM {\hbox{$\dln \! M$}}
\def \lnM {\hbox{$\ln M$}}
\def \dlnsiginv {\hbox{$\dln\sigma^{-1}$}}

\def \Mdndz {\hbox{$M_{N^\prime}$}}
\def \Tdndz {\hbox{$T_{N^\prime}$}}

\def \dnovrn {\hbox{$\langle (\delta n / n)^2 \rangle^{1/2}$}}
\def \nJMF {\hbox{$n_{\rm JMF}$}}
\def \zmed {\hbox{$z_{\rm med}$}}

\begin{document}

\title{GALAXY CLUSTERS IN HUBBLE VOLUME SIMULATIONS: \\
COSMOLOGICAL CONSTRAINTS FROM SKY SURVEY POPULATIONS}

\author{A.E. Evrard$^{1}$, T.J. MacFarland$^{2,3}$,
H.M.P. Couchman$^{4}$, J.M. Colberg$^{5,6}$,
N. Yoshida$^{5}$, S.D.M. White$^{5}$, 
A. Jenkins$^{7}$, C.S. Frenk$^{7}$, 
F.R. Pearce$^{7,8}$,  
J.A. Peacock$^{9}$, P.A. Thomas$^{10}$ 
({\em The Virgo Consortium\/}) }

\affil{$^1$Departments of Physics and Astronomy, University of
Michigan, Ann Arbor, MI 48109-1120 USA } 
\affil{$^2$Rechenzentrum Garching, Boltmannstr. 2, D-85740
Garching, Germany} 
\affil{$^4$Department of Physics and Astronomy, McMaster
University, Hamilton, Ontario, L8S 4M1, Canada}
\affil{$^5$Max-Planck-Institut f\"ur Astrophysik,
Karl-Schwarzschild-Str. 1, D-85740 Garching, Germany} 
\affil{$^7$Department of Physics, Durham University, South Road,
Durham DH1 3LE, UK} 
\affil{$^8$Department of Physics and Astronomy, University of
Nottingham, Nottingham, NG7 2RD, UK} 
\affil{$^9$Institute for Astronomy, University of Edinburgh,
Royal Observatory, Edinburgh EH9 3HJ, UK}
\affil{$^{10}$Astronomy Centre, CEPS, Universty of Sussex, BrightonBN1
9QJ, UK}   

\altaffiltext{3}{1Click Brands LLC, 704 Broadway, New York, NY 10003, USA}
\altaffiltext{6}{Econovo Software, 1 State Street, Boston MA 02109 USA} 

\authoremail{evrard@umich.edu}

\slugcomment{ApJ, accepted 7 Mar 2002}

\begin{abstract}

We use giga-particle N-body simulations to study galaxy cluster
populations in Hubble volumes of \lcdm\ ($\Omega_m=0.3$,
$\Omega_\Lambda=0.7$) and \tcdm\ ($\Omega_m=1$) world
models.  Mapping past light-cones of locations in the computational
space, we create mock sky surveys of dark matter structure to 
$z \ssimeq 1.4$ over $10,000$ sq deg and to $z \ssimeq 0.5$ over two full
spheres.  Calibrating the Jenkins mass function at $z \se 0$ with
samples of $\sims 1.5$ million clusters, we show that the fit
describes the sky survey counts to $\lta 20\%$ acccuracy 
over all redshifts for systems more massive than
poor galaxy groups ($5 \times 10^{13} \hinv \msol$).  

Fitting the observed local temperature function determines the
ratio $\beta$ of specific thermal energies in dark matter and
intracluster gas.  We derive a scaling with power spectrum
normalization $\beta \spropto \sigma_8^{5/3}$, and find that the
\lcdm\ model requires $\sigma_8 \se 1.04$ to match $\beta \se 1.17$
derived from gas dynamic cluster simulations.  
We estimate a $10\%$ overall systematic uncertainty in $\sigma_8$,
$4\%$ arising from cosmic variance in the local sample and the bulk
from uncertainty in the absolute mass scale of clusters. 

Considering distant clusters, 
the \lcdm\ model matches EMSS and RDCS \xray-selected survey
observations under economical assumptions for intracluster gas 
evolution.  
Using transformations of mass-limited cluster samples that mimic
$\sigma_8$ variation, we explore Sunyaev--Zel'dovich (SZ) search
expectations for a 10 sq deg survey complete above $10^{14}
\hinv\msol$.  Cluster counts are shown to be extremely sensitive to
$\sigma_8$ uncertainty while redshift 
statistics, such as the sample median, are much more stable.  
Redshift information is crucial to extract the full cosmological
diagnostic power of SZ cluster surveys. 

For \lcdm, the characteristic temperature at
fixed sky surface density is a weak function of redshift, implying an
abundance of hot clusters at $z >1$.  Assuming constant $\beta$, four 
$kT>8 \kev$ clusters lie at $z>2$ and 40 $kT>5 \kev$ clusters lie at
$z>3$ on the whole sky.  Too many such clusters
can falsify the model; detection of clusters more massive than Coma 
at $z>1$ violates \lcdm\ at $95\%$ confidence if their surface
density exceeds $0.003$ per sq deg, or 120 on the whole sky.  

\end{abstract}


\bigskip
\setcounter{footnote}{0}

\section{Introduction}

Studies of galaxy clusters provide a critical interface between
cosmological structure formation and the astrophysics of galaxy
formation.  Spatial statistics of the cluster population provide
valuable constraints on cosmological parameters while multi-wavelength
studies of cluster content offer insights into the cosmic mix of
clustered matter components and into the interactions between galaxies
and their local environments.

In the near future, the size and quality of observed cluster samples
will grow dramatically as surveys in optical, X--ray and sub-mm
wavelengths are realized.  In the optical, the ongoing wide-field 2dF
(Colless \etal\ 2001) and SDSS (Kepner \etal\ 1999; Nichol \etal\
2001; Annis \etal\ 2001) 
surveys will map the galaxy and cluster distributions over large
fractions of the sky to moderate depth ($z \ssim 0.3$), while deeper
surveys are probing of order tens of degrees of sky to $z \ssim 1$
(Postman \etal\ 1996; Dalton \etal\ 1997; Zaritsky \etal\ 1997;
Ostrander \etal\ 1998; Scoddeggio \etal\ 1999; Gal \etal\ 2000;
Gladders \& Yee 2000; Willick \etal\ 2001; Gonzalez \etal\ 2001).  
In the \xray, ROSAT archival surveys (Scharf \etal\
1997; Rosati \etal\ 1998; Ebeling \etal\ 1998; Viklinin \etal\ 1998; 
deGrandi \etal\ 1999;
Bohringer \etal\ 2001; Ebeling, Edge \& Henry 2001; Gioia \etal\ 2001)
have generated redshift samples of many hundreds of clusters.  Similar
surveys to come from developing Chandra and XMM archives (\eg\ Romer \etal\
2001) will lead to order of magnitude improvements in sample size and
limiting sensitivity.  Finally, the detection of clusters via their
spectral imprint on the microwave background (Sunyaev \& Zel'dovich
1972; Birkinshaw 1999) offers a new mode of efficiently surveying for
very distant ($ z > 1$) clusters with hot, intracluster plasma
(Barbosa \etal\ 1996; Holder \etal\ 2000; Kneissl \etal\ 2001).

Deciphering the cosmological and astrophysical information in the
coming era of large survey data sets requires the ability to
accurately compute expectations for observables within a given cosmology.  
Given some survey observation $R$ at redshift $z$, a
likelihood analysis 
requires the probability $p(R, z \,|\, {\cal C},{\cal A})$
that such data would arise within a model described by sets 
of cosmological ${\cal C} \sequiv \{ C_i \}$ and astrophysical
${\cal A} \sequiv \{ A_j \}$ parameters.  Considering clusters 
as nearly a one-parameter family ordered by total mass $M$, 
the likelihood of the observable $R$ can be written
\begin{equation} \label{eq:pprod}
p(R, z \,|\, {\cal C},{\cal A}) \ = \frac{\int dM \ p(M, z \,|\, {\cal C})
\ p(R \,|\, M, z, {\cal A})}{ \int dM \  p(M, z \,|\, {\cal C})} .
\end{equation}
where $p(M, z \,|\, {\cal C})$ is the likelihood that a cluster of mass
$M$ exists at redshift $z$ in cosmology ${\cal C}$ 
within the survey of interest, and $p(R \,|\, M,z,{\cal A})$ 
is the likelihood that observable $R$ is associated with such a cluster
given the astrophysical model ${\cal A}$.

Separating the problem in this way assumes its pieces to be
independent.  
The space density $n(M,z\,| \, {\cal C})$, or {\em mass function\/}, 
describes the probability of finding a cluster at redshift $z$ with
total mass $M$ within comoving volume element $dV$
\begin{equation} \label{eq:pofm} 
p(M,z\,|\, {\cal C}) \ \propto \ n(M,z \,| \, {\cal C}) \, dV .
\end{equation} 
The absence of explicit astrophysical dependence in the mass function 
is based on the assumption that weakly interacting dark matter
dominates the matter energy density.  If a cluster's total mass $M$ is
relatively immune to astrophysical processes, then the mass function
is well determined by the gravitational clustering of dark matter.  
On the other hand, the likelihood of a particular observable feature 
$p(R \,|\, M,z,{\cal A})$, is dependent, often critically, on the 
astrophysical model.  For optical and
\xray\ observations, it encapsulates the answer to the question ``how
do dark matter potential wells lights up?''

For Gaussian initial density fluctuation spectra, Press \&
Schechter (1974; PS) used a spherical collapse argument and $N$-body
simulation to show that the space density of the rarest clusters is
exponentially 
sensitive to the amplitude of density perturbations on $\sim 10 \hinv
\mpc$ scales.  The analytic form of PS was put on a more rigorous
footing by Bond \etal\ (1991), but recent extensions to ellipsoidal
collapse (Sheth \& Tormen 1999; Lee \& Shandarin 1999) revise
the original functional form.  Calibration by N-body
simulations has led to a functional shape for the mass function 
that retains the essential character of the original PS derivation
(Jenkins \etal\ 2001, hereafter J01, and references therein).  
For cluster masses defined using threshold algorithms tied to the
cosmic mean mass density $\rhobar(z)$, J01 show that the mass fraction
in collapsed objects is well described by a single function that
depends only on the shape of the filtered power spectrum of initial
fluctuations $\sigma^2(M)$. 

Complications arise in determining the mass function
$n(M,z)$ from both simulations and observations.  The first is 
semantic.  Clusters formed from hierarchical clustering 
do not possess unique, or even distinct, physical
boundaries, so it is not obvious what mass to assign to a particular
cluster.  This issue is solvable by convention, 
and we choose here a commonly employed measure $\Mdeltac$, defined as 
the mass interior to a sphere within which the mean interior 
density is a fixed multiple $\Delta$ of the critical density
$\rhocrit(z)$ at the epoch of identification $z$.  Acknowledging the
non-unique choice of threshold, we develop in an appendix 
a model, based on the mean density profile of
clusters derived from simulations (Navarro, Frenk \& White 1996;
1997), that transforms the mass function fit parameters to threshold
values different from that used here.

Attempts to empirically constrain the mass function are complicated 
by the inability to directly observe the theoretically defined
mass.  Instead, a surrogate estimator $\hat{M}$ must be employed
that is, in general, a biased and noisy representation of $\Mdeltac$.
For example, 
estimates derived from the weak gravitational lensing distortions
induced on background galaxies tend to overestimate $\Mdeltac$
by $\ssim 20\%$, with a dispersion of order unity (Metzler,
White \& Loken 2001; White, van Waerbeke \& Mackey 2001)). 

The temperature $T$ of
the intracluster medium (ICM) derived from \xray\ spectroscopy is an
observationally accessible mass estimator.  Gas
dynamic simulations predict that the ICM rarely strays far from virial
equilibrium (Evrard 1990; Evrard, Metzler \& Navarro 1996; Bryan
\& Norman 1998; Yoshikawa, Jing \& Suto 2000; Mathiesen \& Evrard
2001), so that $p(M \,|\, T)$ is well described by a mean power--law
relation with narrow ($\lta 15\%$ in mass) intrinsic scatter.
Observations are generally supportive of this picture (Hjorth, Oukbir
\& van Kampen 1998;  Mohr, Mathiesen \& Evrard 1999; Horner \etal\
1999; Nevalainen \etal\ 2000), but the detailed form of $p(M \,|\, T)$
remains uncertain.  The overall normalization is a particular concern;
we cannot prove that we know the median mass of, say, a $6 \kev$
cluster to better than $25\%$ accuracy.

Even with this degree of uncertainty, the space density of clusters as
a function of $T$ (the {\em temperature function\/}) has been used to place
tight constraints on $\sigma_8$, the present, linear-evolved amplitude
of density fluctuations averaged within spheres of radius $8 \hinv
\mpc$.  Henry \& Arnaud (1991) derived $\sigma_8 \se 0.59 \pm 0.02$
from temperatures of 25 clusters in a bright, \xray\ flux limited
sample, assuming $\Omega_m \se 1$.  Subsequent analysis of this sample
(White, Efstathiou \& Frenk 1993; Eke, Cole \& Frenk 1996; Vianna
\& Liddle 1996; Fan, Bahcall \& Cen 1997; Kitayama \& Suto 1997;
Pen 1998) and revised samples (Markevitch 1998; Blanchard \etal\ 2000)
generated largely consistent results and extended constraints to
arbitrary $\Omega_m$.  For example, Pierpaoli, Scott \& White (2001),
reanalyzing the Markevitch sample using revised temperatures of White
(2000), find
\begin{equation} \label{eq:PSWsig8}
\sigma_8 \ = \ 0.495\, ^{+0.034}_{-0.037} \ \ \Omega_m^{-0.60}.
\end{equation}

Accurate determination of $\sigma_8$ is a prerequisite for deriving
constraints on the clustered mass density $\Omega_m$ from a
differential measurement of the local and high redshift cluster
spatial abundances.  Most studies have excluded the possibility that $\Omega_m
\se 1$ from current data (Luppino \& Kaiser 1997; Bahcall, Fan \&
Cen 1997; Carlberg, Yee \& Ellingson 1997; Donahue \etal\ 1998; Eke
\etal\ 1998; 
Bahcall \& Fan 1998) but others disagree (Sadat, Blanchard \& Oukbir
1998; Blanchard \& Barlett 1998; Vianna \& Liddle 1999).
Uncertainty in $\sigma_8$ plays a role in this ambiguity, as recently
illustrated by Borgani \etal\ (1999a).  In their analysis of 16 CNOC
clusters at redshifts $0.17 \le z \le 0.55$, the estimated value of
$\Omega_m$ shifts by a factor 3, from $0.35$ to $1.05$, as $\sigma_8$
is varied from $0.5$ to $0.6$.

Motivated by the need to study systematic effects in both local and
distant cluster samples, we investigate the spatial distribution of
clusters in real and redshift space samples derived from $N$-body
simulations of cosmic volumes comparable in scale to the Hubble Volume
$(c/\Ho)^3$.  A pair of $10^9$ particle realizations of flat cold,
dark matter (CDM) cosmologies are evolved with particle mass
equivalent to that associated with the extended halos of bright
galaxies.  The simulations are designed to discover the rarest and
most massive clusters (by maximizing volume) while retaining force and mass
resolution sufficient to determine global quantities (mass, shape, 
low-order kinematics) for objects more massive than poor groups of
galaxies ($\sims 5 \times 10^{13} \hinv\msol$).   
To facilitate comparison to observations, we
generate output that traces the dark matter structure along the past
light-cone of two observing locations within the computational volume.
These virtual {\em sky surveys\/}, along with usual fixed proper time
{\em snapshots\/}, provide samples of millions of clusters
that enable detailed statistical studies.  We publish the cluster
catalogs here as electronic tables.  

In this paper, we extend the detailed cluster mass function analysis
of J01 to the sky survey output, updating results using a cluster
finding algorithm with improved completeness properties for poorly
resolved groups.  We match the observed local \xray\ temperature
function by tuning the proportionality factor $\beta$ between the 
specific energies of dark matter and intracluster gas.  The required
value of $\beta$ depends on the assumed $\sigma_8$, and we derive a 
scaling $\beta \ssim \sigma_8^{5/3}$ based on virial equilibrium and
the Jenkins' mass function form.  From $z \se 0$ subvolumes sized to local 
temperature samples, we show that sample variance of temperature-limited
samples contributes $\sims 4\%$ uncertainty to determinations of
$\sigma_8$.  Uncertainty in converting temperatures to masses remains
the dominant source of systematic error in $\sigma_8$, and we
investigate the influence of a $25$ per cent uncertainty in mass scale
on expectations for Sunyaev-Zel'dovich searches.  

In \S\ref{sec:HVSims}, we describe the simulations, including the
process of generating sky survey output, and the model used to convert
dark matter properties to \xray\ observables.  
The cluster mass function is examined in \S\ref{sec:ClusPops}.
Million cluster samples at $z \se 0$ are used to determine 
the best fit parameters of the Jenkins mass function, and we show that
this function reproduces well the sky survey populations extending to
$z \!>\! 1$.  The interplay 
between the fit parameters, $\sigma_8$ and the normalization of cluster
masses is explored, and this motivates a procedure for
transforming the discrete cluster sets to mimic variation in 
$\sigma_8$.  

In \S\ref{sec:sigmavar}, we use observations analyzed by
Pierpaoli \etal\ to calibrate the specific energy factor $\beta$ 
for each model.  We explore properties of the
high redshift cluster population in \S\ref{sec:highz}, 
emphasizing uncertainties from $\sigma_8$ error, 
intracluster gas evolution and possible \xray\ selection biases 
under low signal-to-noise conditions.  The $\sigma_8$ 
transformations developed
in \S\ref{sec:ClusPops} are used to explore cluster yields anticipated
from upcoming SZ surveys, and the median redshift in mass-limited
samples is identified as a robust cosmological discriminant.
Characteristic properties of the \lcdm\ cluster
population are summarized in \S\ref{sec:lcdm}, and we review our 
conclusions in \S\ref{sec:concl}.


\section{Hubble Volume Simulations}
\label{sec:HVSims}

\begin{deluxetable}{lcccccc}
\tablewidth{0pt} \tablecaption{Model Parameters. \label{tab:models}}
\tablehead{ \colhead{Model} & \colhead{$\Omega_m$} &
\colhead{$\Omega_\Lambda$} & \colhead{$\sigma_8$} &
\colhead{$z_{init}$} & \colhead{$L$\tablenotemark{a}} &
\colhead{$m$\tablenotemark{b}} } 
\startdata 
\lcdm & 0.3 & 0.7 & 0.9 & 35 & 3000 & $2.25$\\ 
\tcdm & 1.0 & 0 & 0.6 & 29 & 2000 & $2.22$\\
\enddata
\vspace* {-0.3truecm} 
\tablenotetext{a}{Cube side length in $\hinv$ Mpc.}  
\tablenotetext{b}{Particle mass in $10^{12}\hinv\msol$.}
\end{deluxetable}

After an upgrade in 1997 of the Cray T3E at the Rechenzentrum
Garching\footnote{The Max-Planck Society Computing Center at
Garching.}  to 512 processors and 64Gb of memory, we carried out a
pair of one billion ($10^9$) particle simulations over the period Oct
1997 to Feb 1999.  A memory-efficient version of Couchman, Pearce and 
Thomas' Hydra N-body code (Pearce \& Couchman 1997) parallelized using
{\em shmem} message-passing utilities was used to perform the
computations.  MacFarland \etal\ (1998) provide a description and
tests of the parallel code.

We explore two cosmologies with a flat spatial metric, a \lcdm\ model
dominated by vacuum energy density (a non-zero cosmological
constant) and a \tcdm\ model dominated by non-relativistic, cold dark
matter.  The \tcdm\ model completed May 1998 while the \lcdm\
model finished Feb 1999.  Published work from these simulations 
includes an extensive analysis of counts-in-cells
statistics (to $>10^{\rm th}$ order) by Colombi \etal\ (2000) and
Szapudi \etal\ (2000), investigation of the clustering behavior of
clusters (Colberg \etal\ 2000; Padilla \& Baugh 2001), 
analysis of two-point function
estimators (Kerscher, Szapudi \& Szalay 2000), a description of the
mass function of dark matter halos (J01), a study of confusion on the
\xray\ sky due to galaxy clusters (Voit, Evrard \& Bryan 2001), and 
statistics of pencil--beam surveys (Yoshida \etal\ 2000).  Kay, Liddle
\& Thomas (2001) use the sky survey catalogs to predict Sunyaev--Zel'dovich
(SZ) signatures for the planned {\sl Planck Surveyor\/} mission while 
Outram \etal\ (2001) use the deep mock \lcdm\ surveys to test 
analysis procedures for the 2dF QSO Redshift Survey.

\subsection{Simulation description}

Table~\ref{tab:models} summarizes parameter values for each model,
including the final epoch matter density $\Omega_m$, vacuum energy
density $\Omega_\Lambda$, power spectrum normalization $\sigma_8$,
starting redshift $z_{init}$, simulation side length $L$ and particle
mass $m$. 

Values of $\sigma_8$ were chosen to agree approximately with both the
amplitude of temperature anisotropies in the cosmic microwave background
as measured by COBE and with the nearby space density of rich \xray\
clusters.  The degree of uncertainty in these constraints allows the 
final space density of clusters as a function of mass to differ
between the two simulations.  However, as we discuss below, it is
possible to ensure that the observed space density of clusters as a
function of \xray\ temperature is matched in both models by adjusting a 
free factor $\beta$ used to link X-ray temperature to dark matter
velocity dispersion.  

To initiate the numerical experiments, particle positions and momenta
at $z_{init}$ are generated by perturbing a replicated `glass' of
one million particles with a set of discrete waves randomly drawn from
power spectra computed for each cosmology.  Initial Fourier modes of
the applied perturbations have amplitudes drawn from a Gaussian
distribution with variance given by the power spectrum $P(k) \se T(k)
P_{\rm prim}(k)$.  A Harrison--Zel'dovich 
primordial spectrum $P_{\rm prim}(k) \spropto k$ is assumed for both
models.  For the \lcdm\ model, the 
transfer function $T(k)$ is computed using CMBFAST (Seljak \& Zaldarriaga
1996) assuming $h=0.7$ and baryon density $\Omega_b h^2 =0.0196$
(Burles \& Tytler 1998).  The \tcdm\ model uses  
transfer function $T(k) = (1+\big[aq + (bq)^{3/2}
+ (cq)^2\big]^{\nu})^{-1/\nu}$, where $q = k/\Gamma$, $\Gamma \se
\Omega_m h$, $a = 6.4 \hinv
\mpc$, $b = 3 \hinv \mpc$, $c = 1.7\hinv \mpc$ and $\nu=1.13$ (Bond
\& Efstathiou 1984).  

The simulations are designed to resolve the collapse of a Coma-sized
cluster with 500 particles.  Although this resolution is sufficient to
capture only the later stages of the hierarchical build-up of
clusters, convergence tests (Moore \etal\ 1998; Frenk \etal\ 1999) show
that structural properties on scales larger than a few times the
gravitational softening length are essentially converged.  From tests
presented in J01 and in an Appendix to this work, cluster
identification is robust down to a level of about $20$ particles.
Using $M_{\rm Coma} \se 1.1 \times 10^{15} \hinv\msol$ (White \etal\
1993), leads to particle mass $2.2\times 10^{12} \hinv\msol$ in both
models, comparable to the total mass within $\sims 300 \hinv \kpc$ of
bright galaxies (Fischer \etal\ 2000; Smith \etal\ 2001; Wilson \etal\
2001).  The mass associated with one billion particles at the mean
mass density sets the length $L$ of the periodic cube used for the
computations, resulting in a Hubble Length $L \se c/\Ho \se 3
\hinv\gpc$ for the \lcdm\ model and $L=2 \hinv\gpc$ for \tcdm.

A Newtonian description of gravity is assumed, appropriate for
weak-field structures.  A non-retarded
gravitational potential is employed because the peculiar acceleration
converges on scales well below the Hubble length. The good agreement
between the higher-order clustering statistics of the simulations and
expectations derived from an extended perturbation theory treatment of
mildly non-linear density fluctuations provides indirect evidence that
this treatment of gravity is accurate (Colombi \etal\ 2000; Szapudi
\etal\ 2000).

Gravitational forces on each particle are calculated as a sum of a
long-range component, determined on a uniform spatial grid of $1024^3$
elements using Fast Fourier Transforms, and a short-range component
found by direct summation.  The latter force is softened with a
spline-smoothing roughly equivalent to a Plummer law gravitational
potential $\phi(r) \spropto (r^2+\varepsilon^2)^{-1/2}$ with smoothing
scale $\varepsilon \se 0.1 \hinv \mpc$.  A leapfrog time integration
scheme is employed with 500 equal time steps for each calculation.

\begin{deluxetable}{lcccc}
\tablewidth{0pt} \tablecaption{Light-cone Sky Surveys.\label{tab:surveys}} 
\tablehead{ \colhead{Name} & \colhead{center} & \colhead{solid angle}
& \colhead{$z_{\rm max} $  ($\Lambda$)} & \colhead{$z_{\rm max} $ ($\tau$)} }
\startdata 
MS & $(L/2,L/2,L/2)$ & $4\pi$ & 0.57 & 0.42 \\ 
VS & $(0,0,0)$ & $4\pi$ & 0.57 & 0.42 \\ 
PO & $(0,0,0)$\tablenotemark{a} & $\pi/2$ & 1.46 & 1.25 \\ 
NO\tablenotemark{b} & $(L,L,L)$\tablenotemark{a} & $\pi/2$ & 1.46 & 1.25 \\ 
DW & $(0,0,0)$\tablenotemark{a} & $10^\circ \times 10^\circ$ & 4.4 & 4.6 \\ 
XW & $(0,0,0)$ & $16^\circ \times 76^\circ$ & $6.8$ & $-$ \\ 
\enddata
\vspace* {-0.3truecm} 
\tablenotetext{a}{Orientation centered on cube diagonal.}  
\tablenotetext{b}{NO/PO have opposite orientations about a common center.}
\end{deluxetable}

Processor time for these computations was minimized by employing a
parallel algorithm well matched to the machine architecture
(MacFarland \etal\ 1998) and by simulating large volumes that entail a
minimum of message passing overhead.  The Cray-T3E offers high
interprocessor communication bandwidth along with a native
message-passing library ({\em shmem}) to control data flow.  A
two-dimensional, block-cyclic domain decomposition scheme allocates
particles to processors.  Each processor advances particles lying
within a disjoint set of rectangular regions of dimension $L \times
(L/32) \times (L/16) $ that subdivide the computational space.  Each
calculation required approximately 35,000 processor hours, or three
days of the 512-processor machine.  This corresponds to advancing
roughly 4000 particles per second on an average step.  The
computations were essentially limited by I/O bandwidth rather than cpu
speed.  Execution was performed in roughly twenty stages spanning a
calendar time of three to four months, with data archived to a mass
storage system between stages.  Approximately 500 Gb of raw data were
generated by the pair of simulations.

\subsection{Sky survey output}

In addition to the traditional simulation output of snapshots of the
particle kinematic state at fixed proper time, we introduce here sky
survey output that mimics the action of collecting data along the past
light-cone of hypothetical observers located within the simulation
volume.  The method extends to wide-angle surveys an approach
pioneered by Park \& Gott (1991) in simulating deep, pencil-beam
observations.  Since there is no preferred location in the volume, we
chose two survey origins, located at the vertex and center of the
periodic cube for convenience.

In a homogeneous world model, a fixed observer at the present epoch
$t_0$ receives photons emitted at $t < t_0$ that have traversed
comoving distances $r(t) \ = c \int_{t}^{t_0} dt \, a^{-1}(t)$ where
$a(t)$ is the scale factor of the metric ($a(t_0) \se 1$) and $c$ the
speed of light.  The set of events lying along the continuum of
concentric spheres $\{t,\vec{r}(t)\}$ for $t<t_0$ defines the past
light-cone of that observer.  In the discrete environment of the
numerical simulation, we construct the light-cone survey by choosing
spherical shells of finite thickness such that each particle's state
is saved at a pair of consecutive timesteps that bound the exact time
of intersection of the light-cone with that particle's trajectory.
Defining $t_i$ as the proper time at step $i$ of the computation, we
choose inner and outer radii $(1-\eta)r(t_{i+1})$ and
$(1+\eta)r(t_{i-1})$.  Here $\eta \ssimeq 0.02$ is a small parameter
that safeguards against a particle appearing only once in the output
record due to peculiar motion across the discrete shells during a
step.  The inner radius is set to zero on the final two steps of the
computation.

With successive states for particles in the output record, a linear
interpolation is performed to recover the original second-order time
accuracy of the leap-frog integrator.  Given a particle's position
relative to the survey origin $\vec{x}_i$ and, at the subsequent step,
$\vec{x}_{i+1} \sequiv \vec{x}_i + \Delta\vec{x}$, we solve for
interpolation parameter $\alpha$ defining position $\vec{x} \se
\vec{x}_i + \alpha \Delta\vec{x}_i$ such that $|\vec{x}| = r(t_i
+\alpha \Delta t)$, with $\Delta t$ the timestep.  For a spherical
survey, this implies
\begin{equation} \label{eq:alpha}
\alpha \ = \ \frac{ r^2(t_i) - x_i^2 }{ 2 ( \vec{x_i}^. \Delta \vec{x}
+ r(t_i) \Delta r ) }
\end{equation}
with $\Delta r \se r(t_i) - r(t_{i+1}) > 0$.  After solving for
$\alpha$, the particle's position and velocity are interpolated and
the result stored to create the processed sky survey data sets.
Comoving coordinates and physical velocities are stored as two-byte
integers, sufficient to provide $\sims \varepsilon/10$ positional
accuracy and km/s accuracy in velocity.  Data are stored in binary
form in multiple files, each covering a spatial subcube of side $L/16$ of
the entire computational volume.  

\begin{figure*}
\epsfxsize=20.0cm 
\hbox{\hskip -0.5truecm \epsfbox{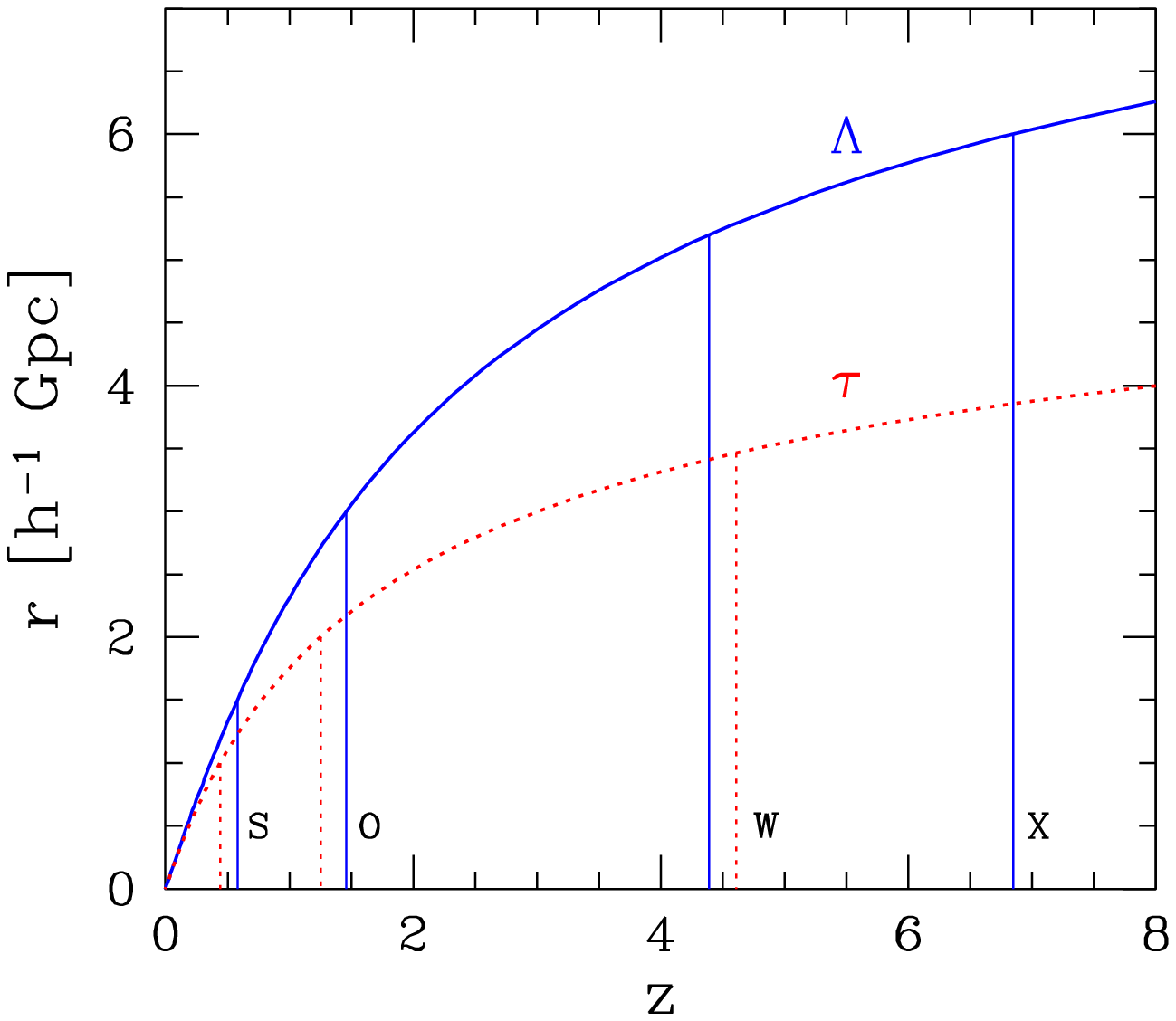}}
\vskip -1.0truecm
\caption{
Comoving look--back distance as a function of redshift 
for \lcdm\ (solid) and \tcdm\ (dashed).  Vertical lines
indicate redshift limits of the spherical (S), octant (O), deep 
wedge (W) and extended wedge (X) surveys (Table~\ref{tab:surveys}).  
\label{fig:lc} 
}
\end{figure*}

The scale $L$, along with the comoving distance--redshift relations 
shown in Figure~\ref{fig:lc}, determine the redshift extents of the
surveys listed in Table~\ref{tab:surveys}.  Two principal survey types
--- spheres and octants --- extend to distances $L/2$ and $L$,
respectively.  From the cube center, the MS full-sphere surveys extend
to redshifts $z_{\rm max}  \se 0.57$ (\lcdm) and $0.42$ (\tcdm).  From the
origin and its diagonally opposite image, octant surveys (PO and NO,
respectively) extend to redshifts $1.46$ (\lcdm) and $1.25$ (\tcdm).  The
surveys have opposite orientation; both view the interior
region of the computational space.  The VS sphere centered on the
origin is created using translational symmetries of separate octant
surveys conducted from the eight vertices of the fundamental cube.
The interior portions of the PO and NO surveys are thus subsets
(opposite caps) of the VS survey.  The combined volumes of the spheres
and octants sample the computational volume roughly
once for each type.  In terms of cosmic time, the octants
extend over the last 74\% and 71\% of the age of the universe (\lcdm\
and \tcdm, respectively), equivalent to roughly a 10 Gyr look-back
time.

In addition to these surveys, smaller solid angle wedge surveys reach
to greater depth.  A $10 \times 10$ sq deg deep wedge (DW) extends
along the cube diagonal to the opposite corner and reaches redshifts
$4.4$ (\lcdm) and $4.6$ (\tcdm).  For the \lcdm\ model only, a $16
\times 76$ sq deg extended wedge (XW) uses periodic images of the
fundamental cube to reach $z_{\rm max}  \se 6.8$.  This wedge 
is a partial extension of the PO survey.

\subsection{Connecting to \xray\ observations}

Connecting to observations of clusters requires a model that relates
luminous properties to the underlying dark matter.  We focus here on
the ICM temperature \Tx\ under the assumption that both \Tx\ and the
dark matter velocity dispersion \sigmadm\ are related to the
underlying dark matter gravitational potential through the virial
theorem (Cavaliere \& Fusco-Femiano 1976).  Empirical support for
this assumption comes from the observation that $\Tx \ssim
\sigmagal^2$ (e.g., Wu, Xue \& Fang 1999), the scaling expected if
both galaxies and the ICM are thermally supported within a common
potential.  High resolution simulations of galaxy formation within a
cluster indicate that $\sigmagal$ should accurately reflect the dark
matter $\sigmadm$ except for the brightest, early-type galaxies which
display a mild bias toward lower velocity dispersion
(Springel \etal\ 2000).

Rather than map $T$ to $M$ directly, we prefer to use a
one-to-one mapping between $T$ and dark matter velocity dispersion
$\sigmadm$.  This approach has the advantage of naturally building in
scatter between $T$ and $M$ at a fractional amplitude $\sims 10\%$
that is consistent with expectations from direct, gas dynamic modeling
(\eg\ Mathiesen \& Evrard 2001; Thomas \etal\ 2001).  
Define a one-dimensional velocity dispersion $\sigmadm$ by 
\begin{equation} \label{eq:sigmadm}
\sigmadm^2  \ = \ \frac{1}{3N_p} \ \sum_{i=1}^{N_p} \sum_{j=1}^3\ 
| v_{i,j} - \bar{v_j} |^2 
\end{equation}
where the index $i$ ranges over the $N_p$ cluster members identified
within $\rdeltac$, $j$ sums over principal directions, and $\bar{v}$
is the mean cluster velocity defined by the same $N_p$ members.  
We assume that $\sigmadm^2$ maps directly to \xray\
temperature and introduce as an adjustable parameter the ratio of
specific energies $\beta 
\sequiv \sigmadm^2/(k \Tx/\mu m_p)$, where $k$ is Boltzmann's
constant, $\mu$ the mean molecular weight of the plasma (taken to be 
$0.59$) and $m_p$ the proton mass.  We fit $\beta$ by requiring
that the models match the local temperature function.  

Varying the power spectrum normalization $\sigma_8$ leads to 
shifts in the space density of clusters as a function of
mass and velocity dispersion.  Values of $\beta$ required to fit
\xray\ observations thus depend on $\sigma_8$, and we derive 
an approximate form for this dependence in \S\ref{subsec:degen}
below.


\section{Cluster Populations}
\label{sec:ClusPops}

We begin this section by visualizing the evolution of clustering 
in the octant surveys.  
Details of the cluster finding algorithm are then presented, and fits
to the $z \se 0$ mass functions performed for $\Delta \se 200$.  A
simple model for evolving the fit parameters with redshift 
in the \lcdm\ case is given, and predictions based on the $z \se 0$
fits compared to the sky survey output in three broad redshift
intervals.  

Additional details are provided in two appendices.  Appendix~A 
compares output of the SO algorithm employed here to that used by J01.
Appendix~B presents a model for extending the mass function fits 
under variation of the density threshold $\Delta$. 
More generally, it provides a means to transform
the discrete cluster sample under variation of $\sigma_8$.  
Appendix~C presents the sky survey
and $z \se 0$ cluster catalogs as electronic tables. 
Truncated versions in the print edition list the ten most massive
clusters in each survey.  Electronic versions list all clusters above
a mass limit of $5 \times 10^{13} \hinv \msol$ (22 particles).  

\subsection{Evolution of the matter distribution }\label{subsec:evol}

\begin{figure*}
\epsfxsize 16.0truecm 
\hbox {\hskip 0.5truecm \epsfbox{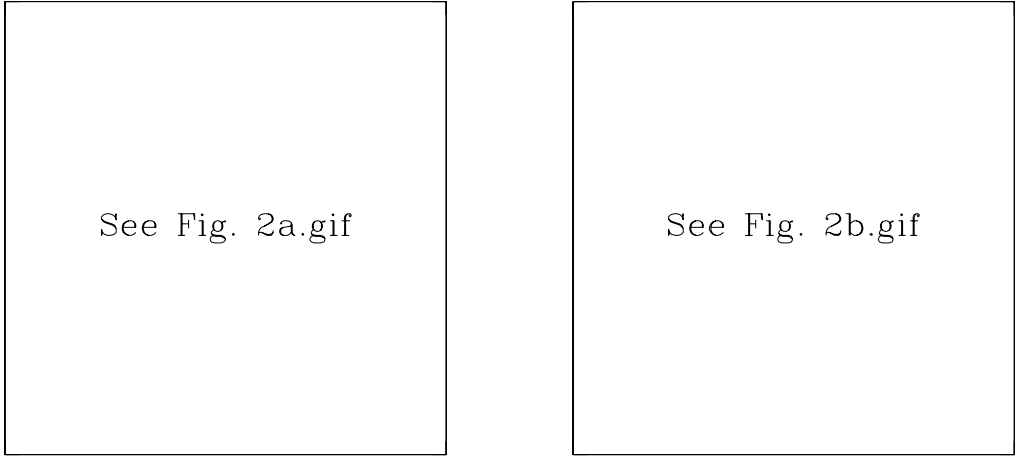} }
\vskip 0.0truecm
\caption{ Maps of the dark matter in slices through the deep octant
surveys in (a) \tcdm\ and (b) \lcdm\ world models.   Survey
origins are at the vertices (O), and color represents
mass density relative to the mean ranging from 
$0.05$ (black) to 50 (white) on a logarithmic scale.  Density is mapped
onto a two-dimensional grid using an adaptive (Lagragian) smoothing kernel
with scale $2 \times 10^{13} \hinv \msol$.  For each model, 
two representations of a $45^\circ$ slice extending to $z \se 1.25$
are shown.  Horizontal maps display structure in
the comoving metric while the vertical maps display the same comoving
region, reflected about the diagonal, in redshift space.  Positions of
clusters at the  intersection of filaments are evident in redshift
space through the radial distortions arising from their internal 
velocity dispersions (so--called `fingers of God').   The inset of
the \tcdm\ image shows the relation between comoving distance $r$ and
redshift $z$ over the range mapped by the images.  
The inset of the \lcdm\ image shows
a close-up of the particle distribution around the largest cluster of
the \lcdm\ octant surveys, located at $z \se 1.04$.  
Particles colored white lie within a sphere of physical radius $1.5
\hinv \mpc$ that encompasses a mean density 200 times the critical
value. 
\label{fig:RZimage}
}
\end{figure*}

In Figure~\ref{fig:RZimage}, we present maps of the Lagrangian smoothed 
mass density in slices through the octant surveys that extend to $z
\se 1.25$.  Horizontal and vertical maps show comoving and redshift space
representations, respectively.  Since the Hubble Length far exceeds
the characteristic clustering length of the mass, the feature most
immediately apparent in the density maps is their overall homogeneity.
Gravitational enhancement of the clustering amplitude over time is
evident from the fact that the density fluctuations are more
pronounced near the survey origin (vertex of each triangular slice)
compared to the edge.  The effect is subtle in this image because the
dynamic range in density, from black to white, spans three orders of
magnitude, much larger than the linear growth factors of $1.8$ (\lcdm)
and $2.2$ (\tcdm) for large-scale perturbations in the interval
shown.  

\begin{figure*}
\epsfxsize=18.0cm 
\hbox{\hskip -0.0truecm  \epsfbox{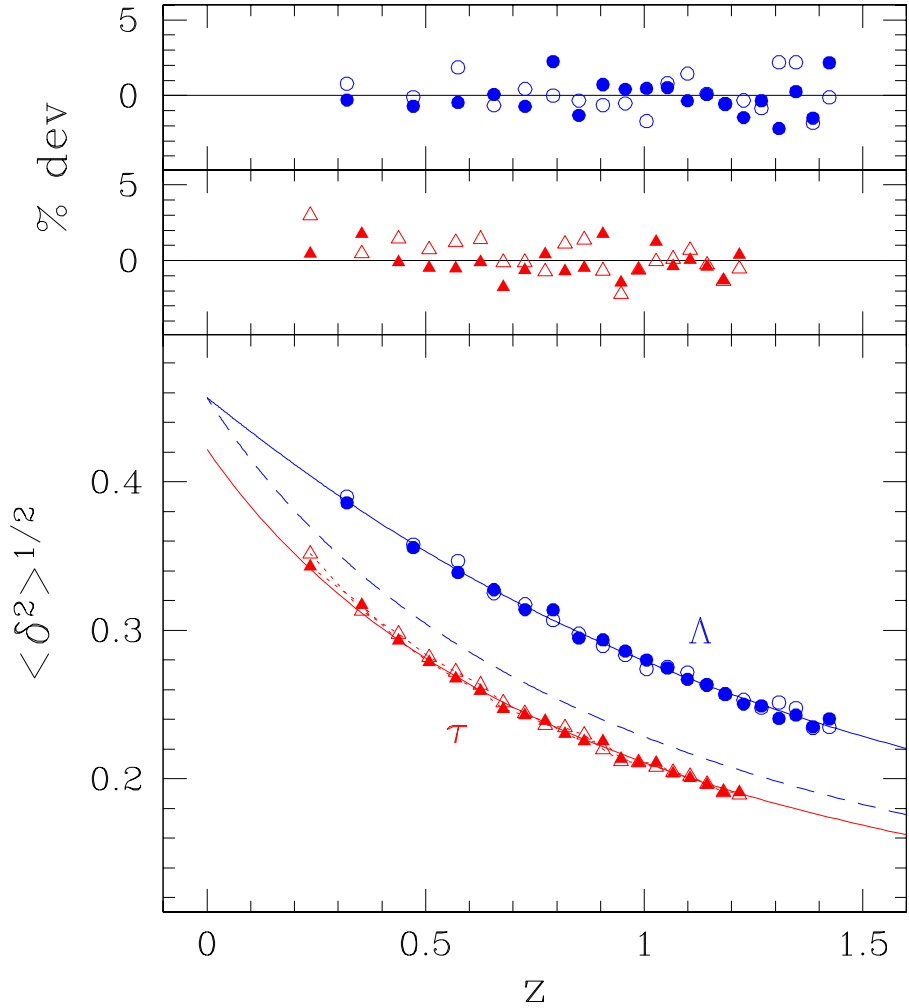} }
\vskip -1.5truecm
\caption{
Redshift evolution of the \rms\/ amplitude of density fluctuations
$ \langle \delta^2 \rangle^{1/2}$ in top-hat spheres containing, on 
average, a mass of $2.2 \times 10^{15} \hinv \msol$ 
(1000 particles).  Points
are octant survey measurements (filled, PO; open, NO) from the
\lcdm\ (circles) and \tcdm\ (triangles) simulations, obtained
by randomly sampling twenty 
radial shells of equal comoving volume and plotted at the
volume-weighted redshift of each shell.  Solid lines are predictions from
linear theory based on the input fluctuation spectra.  The upper
panel demonstrates agreement with linear theory at the $\ssim 1\%$
level, except for the non-linear departure of \tcdm\ fluctuations at
late times.  
The dashed line in the lower panel 
shows the evolution that \lcdm\ fluctuations would have if they 
followed the \tcdm\ linear growth evolution.   
\label{fig:delta_z}
}
\end{figure*}

To verify the accuracy of the clustering evolution in the octant
surveys, we show in Figure~\ref{fig:delta_z} the behavior of the \rms\
amplitude of density fluctuations $\langle \delta^2 \rangle^{1/2}$,
where $\delta \sequiv \rho/\rhobar(z)-1$, within spheres encompassing,
on average, a mass of $2.2 \times 10^{15} \hinv \msol$ (1000
particles).  Points in the figure show $\langle \delta^2
\rangle^{1/2}$ determined by randomly sampling locations within 
twenty radial shells of equal volume in the octant surveys.  Values are
plotted at the volume-weighted redshift of each shell.  Solid lines
are not fits, but 
show the expectations for $\langle \delta^2 \rangle^ {1/2}$ based on
linear evolution of the input power spectrum.  Deviations between the
measured values and linear theory, shown in the upper panels of
Figure~\ref{fig:delta_z}, are at the $1\%$ level.  Although we do not
attempt here to model these deviations explicitly, Szapudi \etal\
(2000) show that the higher-order
clustering properties of these simulations at the final epoch are 
well described by an extended perturbation theory treatment of
fluctuation evolution.  

\begin{deluxetable}{lccc}
\tablewidth{0pt} 
\tablecaption{Fit parameters for $\lnsiginv(M)$.\tablenotemark{a}
\label{tab:lnsiginv}} 
\tablehead {\colhead{Model} & $\sigma_{15}$ & $a$ & $b$ } 
\startdata
\lcdm & 0.578 & 0.281 & 0.0123 \\ 
\tcdm & 0.527 & 0.267 & 0.0122 \\
\enddata
\vspace* {-0.3truecm}
\tablenotetext{a}{equation~(\ref{eq:lnsiginvfit}).}
\end{deluxetable}

\begin{figure*}
\epsfxsize=17.0cm 
\hbox{\hskip 0.5truecm  \epsfbox{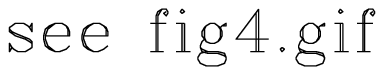} }
\vskip 0.5truecm
\caption{
Redshift-space structure in $75,000\kms$ wide regions centered at
$z \se 1.1$ in \tcdm\ (left) and \lcdm\ (right).  The
grey-scale shows only overdense material $\delta>0$.  The \lcdm\ image 
includes the most massive cluster in the octant surveys, visible as
the long streak at the lower right.  
The regions shown lie just 
interior to the vertical edges of the redshift-space maps of 
Figure~\ref{fig:RZimage}.  
\label{fig:z1_closeup}
}
\end{figure*}

The orientation of the slice shown in Figure~\ref{fig:RZimage} is
chosen to include the most massive cluster in the \lcdm\ octant
surveys.  It lies at the surprisingly high redshift $z \se 1.04$.  
The inset of Figure~\ref{fig:RZimage}b shows that this cluster is
actively forming from mergers fed by surrounding filaments.  
In Figure~\ref{fig:z1_closeup}, we show a
close-up of the redshift space structure in $75,000\kms$ wide regions
centered at $z \se 1.1$ and lying just interior to the vertical edges
of the redshift-space views of Figure~\ref{fig:RZimage}.  
The grey-scale shows only overdense material $\delta>0$ and 
the region includes the most massive \lcdm\ cluster.  
With rest-frame line-of-sight velocity dispersion $1964 \kms$,it 
produces a $\sim 16,000 \kms$ sized `finger-of-God' feature 
in the lower right image.

Along with this extreme object, close inspection of
Figure~\ref{fig:z1_closeup} reveals many more smaller fingers
representing less massive clusters in the \lcdm\ image.  
The \tcdm\ panel contains many fewer such clusters.  
It is this difference that motivates high redshift cluster
counts as a sensitive measure of the matter density parameter
$\Omega_m$.  To perform quantitative analysis of the cluster
population, we first describe the method used to identify
clusters in the particle data sets.

\subsection{Cluster finding algorithm}

A number of methods have been developed for identifying clusters
within the particle data sets of cosmological simulations.   
We refer the reader to J01, White (2000) and
Lacey \& Cole (1994) for discussion and intercomparisons of common 
approaches.  Two algorithms are employed by J01.  One is a percolation
method known as ``friends-of-friends'' (FOF) that
identifies a group of particles whose members have at least one other
group member lying closer than some threshold separation.  The
threshold separation, typically expressed as a fraction $\eta$ of the
mean interparticle spacing, is a parameter whose variation leads to
families of groups, referred to as FOF($\eta$), with favorable nesting
properties (Davis \etal\ 1985). 

The other algorithm of J01 is a spherical overdensity (SO) method that
identifies particles within spheres, centered on local
density maxima, having radii defined by a mean enclosed iso-density
condition.  We use here an SO algorithm that differs slightly from
that of J01.  The iso-density condition requires the mean mass density
within radius $r_\Delta$ to be a factor $\Deltac$ times the critical
density $\rhocrit(z)$ at redshift $z$.  J01 define spherical regions
that are overdense with respect to the mean background, rather than
critical, mass density.  For clarity, we refer to the approaches
as `mean SO' and `critical SO' algorithms.  If not
stated explicitly, reference to SO($\Deltac$) should be read as the
critical case evaluated at contrast $\Deltac$.  By definition, a
critical SO($\Deltac$) population is identical to a mean SO($\Deltac /
\Omega_m$) population. 

Our method employs a code independent of that used by J01.  The codes
produce matching output for well-resolved groups, but differ 
at low particle number.  Appendix~A provides a discussion of
completeness based on direct comparison of group catalogs from the 
algorithms. 

\subsection{Mass function fits at $z \se 0$}\label{subsec:massftn}

By coadding 22 snapshots of 11 Virgo Consortium simulations
ranging in scale from $84.5 \hinv \mpc$ to $3000 \hinv \mpc$, J01
showed that the space density of clusters defined by either FOF(0.2) 
or SO(180) algorithms are well described by a single functional
form when expressed in terms of $\lnsiginv(M)$, where $\sigma^2(M)$ is the
variance of the density field smoothed with a spherical top-hat filter
enclosing mass $M$ at the mean density.  Define the mass
fraction $f(\lnsiginv)$ by
\begin{equation}\label{eq:fsiginvdef}
 f(\lnsiginv) \ \equiv \ \frac{M}{\rhobar(z)} {{\rm d}n(<M,z) \over
\dlnsiginv}
\end{equation}
with $\rhobar(z)$ the background matter density at the epoch of
interest and $n(<M,z)$ the cumulative number density of clusters
of mass $M$ or smaller.  The general form found by J01 for the mass
function is
\begin{equation} \label{eq:HVMF}
 f(\lnsiginv) \ = \ A \, \exp\big[-|\ln\sigma^{-1}+ B
 \,|^{\epsilon}\big]
\end{equation}
where $A$, $B$ and $\epsilon$ are fit parameters.  The amplitude
$A$ sets the overall mass fraction in collapsed objects, $e^B$ plays
the role of a (linearly evolved) collapse perturbation threshold,
similar to the parameter $\delta_c$ in the Press-Schechter model or
its variants,
and $\epsilon$ is a stretch parameter that provides the correct shape
of the mass function at the very dilute limit. 

Values of these
parameters depend on the particular cluster finding scheme
implemented, but J01 show them to be independent of cosmology and redshift
when cluster masses are based on algorithms tied to the mean mass
density.   For the FOF(0.2) group
catalogs, J01 find that $A \se 0.315$, $B \se 0.61$ and $\epsilon \se 3.8$
provide a fit that describes all of the numerical data to $\lta 20\%$
precision over eight orders of magnitude in number density.  

We fit here the SO mass function by employing a quadratic relation
describing the filtered power spectrum shape
\begin{equation} \label{eq:lnsiginvfit}
  \lnsiginv(M) \ = \ - \ln \sigma_{15} + a \, \lnM + b \, (\lnM)^2
\end{equation}
where $M$ is mass in units of $10^{15} \hinv \msol$ and the \rms\
fluctuation amplitude $\sigma_{15}$ at that mass scale is simply
related to the fiducial power spectrum normalization by $\ln
\sigma_{15} \se \ln(\sigma_8) + const$.  Table~\ref{tab:lnsiginv}
lists parameters of the fit to equation~(\ref{eq:lnsiginvfit}).  The
maximum error in the fit is $2\%$ in \lnsiginv\ for masses above
$10^{13} \hinv\msol$.  For both models, the effective logarithmic
slope, 
\begin{equation} \label{eq:alphaeff}
  \alpha_{\rm eff}(M) \ \equiv \ \dlnsiginv(M)/\dlnM \ = \ a + 2b \,
  \lnM ,
\end{equation}
slowly varies between $0.2$ and $0.3$ for masses in the range 
$10^{13}$ to $10^{15.5}
\hinv\msol$.  The Jenkins mass function (JMF) expression for the
differential number density as a function of
mass and redshift
\begin{equation}\label{eq:JMF}
\nJMF(M,z) = \frac{A\rhobar(z)}{M} \alpha_{\rm eff}(M) \exp
\big[-|\lnsiginv(M) + B|^{\epsilon}\big]
\end{equation}
is the form that we fit to the simulated cluster catalogs.

\begin{deluxetable}{lcccc}
\tablewidth{0pt} 
\tablecaption{SO(200) mass function parameters.\tablenotemark{a}
\label{tab:HVMF}}  
\tablehead {\colhead{Model} & $A$ & $B$ & $\epsilon$ & $\dnovrn$ } 
\startdata \lcdm & 0.22 & 0.73 & 3.86 & 0.026 \\ 
\tcdm & 0.27 & 0.65 & 3.77 & 0.028 \\
\enddata
\vspace* {-0.3truecm} \tablenotetext{a}{equation~(\ref{eq:JMF}).}
\end{deluxetable}

\begin{figure*}
\vskip -2.0truecm
\epsfxsize=20.0cm 
\hbox{\hskip -1.0truecm  \epsfbox{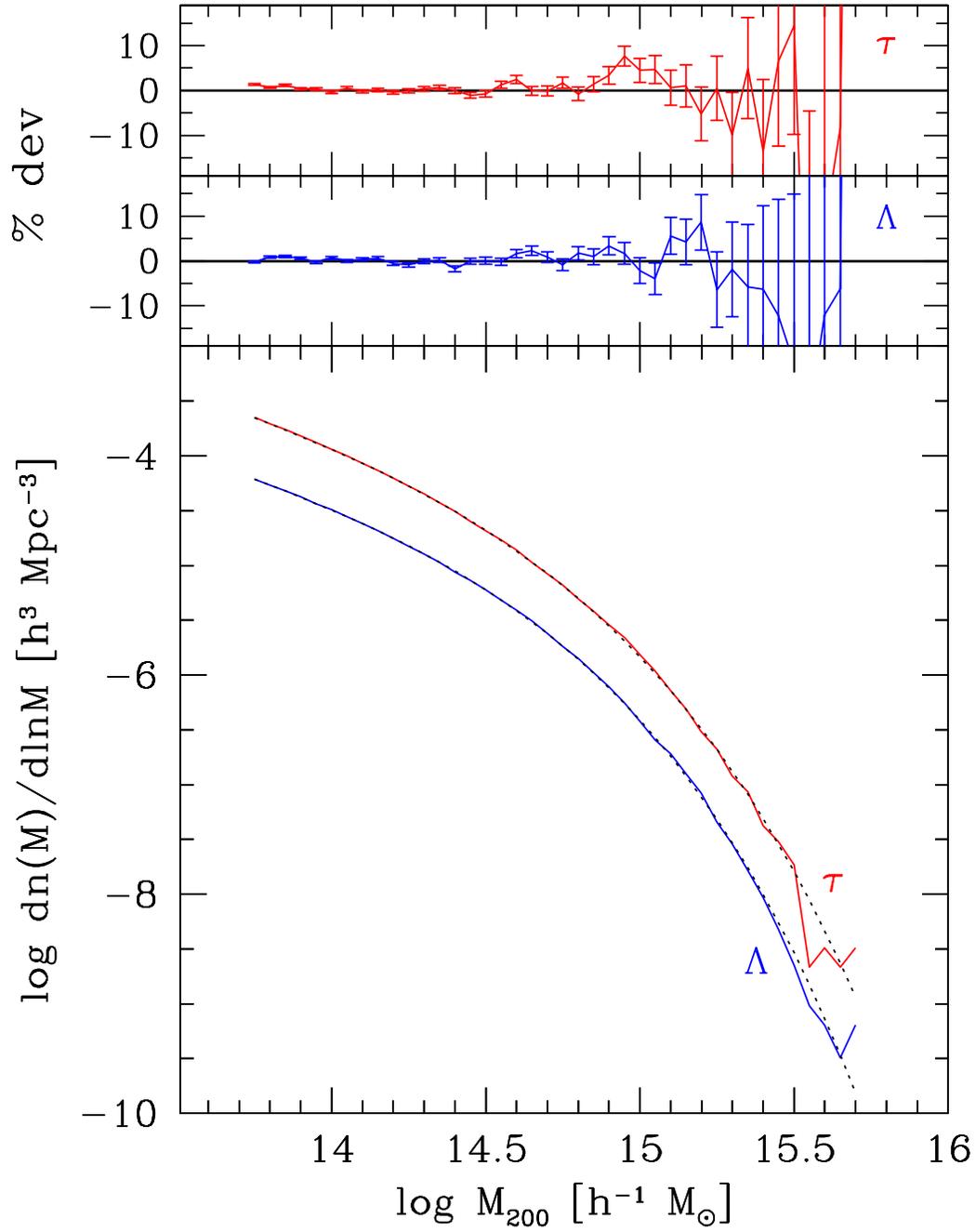} }
\vskip 0.0truecm
\caption{
The lower panel shows critical SO(200) mass functions derived from the
$z \se 0$ HV cluster catalogs (solid lines) along with fits to the
Jenkins' mass function, equation~(\ref{eq:JMF}), using
parameters listed in Table~\ref{tab:HVMF} (dotted lines).  
Upper panels show the percent deviation in number density between the
HV data and the fits.  Error bars are based on Poisson statistics in
each mass bin.
\label{fig:ndiff_M}
}
\end{figure*}

The critical SO(200) mass functions at $z \se 0$ for both models are
shown in Figure~\ref{fig:ndiff_M}, derived from samples of $1.39$
million (\lcdm) and $1.48$ million (\tcdm) clusters above $5 \times
10^{13} \hinv\msol$.  Fits to equation~(\ref{eq:JMF}) are shown as
dotted lines, with fit parameters listed in Table~\ref{tab:HVMF}.  The
upper panels of Figure~\ref{fig:ndiff_M} show the fractional
deviations $\delta n/n \se n/n_{JMF}-1$ in bins of width $0.12$ in
$\lnM$.  Error bars assume Poisson statistics.  For bins with 100 or
more clusters ($\Mtwoh \lta 2 \times 10^{15} \hinv \msol$), the \rms\
deviations $\dnovrn$ are $\lta 3\%$ (Table~\ref{tab:HVMF}).  

The high statistical precision of these fits is a lower bound on the
absolute accuracy of the mass function calibration.  Based on the fits
performed by J01 to a large ensemble of simulations covering a wider
dynamic range in scale than the HV models alone, we estimate that the
normalization $A$ may be systematically low by $\sims 10\%$ (see
Appendix~A).  Considering that this degree of uncertainty in $A$
corresponds to an uncertainty in mass of only $2-3\%$, this level of 
accuracy is sufficient for the practical purpose of comparing to
current and near future observations, for which the level of systematic
uncertainty in mass is an order of magnitude larger.

Another estimate of systematic error is provided by
comparing our results to the recent large simulations of Bode \etal\
(2001).  Their $1 \hinv$ Gpc \lcdm\ simulation assumes identical
cosmological parameters to our model, but uses a particle mass
and gravitational softening smaller by factors of 30 and 7,
respectively.  Bode \etal\ employ a measure of mass within a comoving
radius $1.5 \hinv \mpc$; this scale encompasses a critical density
contrast of 200 at the present epoch for mass $\Mtwoh \se 7.94 \times
10^{14} \hinv \msol$.  From their Figure~6, the space density of
clusters above that mass scale is $10^{-6.6} \ssim 2.5 \times 10^{-7}
h^3 \mpc^{-3}$.  In our $27 h^{-3} \gpc^3$ volume, we find 6499
clusters above this mass limit, implying a space density $2.4 \times
10^{-7} h^3 \mpc^{-3}$.   

Figure~\ref{fig:ndiff_M} shows that the two models do not produce
identical mass functions at the present epoch; the \lcdm\ space density is
lower by a factor $\ssim 4$ than that of \tcdm.  Two factors combine
to make this difference.  The first is that our chosen values of
$\sigma_8$ straddle the constraint derived from fitting the local
\xray\ cluster space density, such as that quoted in
equation~(\ref{eq:PSWsig8}).  The sense of the differences --- the
\lcdm\ model has lower amplitude and \tcdm\ higher, both by about
$10\%$ -- pushes the models in opposite directions.  
The second factor is that our choice of fixed critical threshold
$\Delta \se 200$ leads to smaller masses for \lcdm\ clusters.   
Previous work has typically employed the lower
$\Omega_m$--dependent thresholds derived by Eke \etal\ (1996) from
the spherical collapse solutions of Lahav \etal\ (1991) and Lilje
(1992).  For $\Omega_m \se 0.3$, Eke \etal\ calculate critical
threshold $\Delta \se 97.2$, leading to masses are larger by a factor
$\sims 1.25$ compared to $\Delta \se 200$.

\begin{deluxetable}{lrrrrrrrr}
\tablewidth{0pt} 
\tablecaption{counts of clusters with $hM_{200}/\msol$ above mass
limits.\label{tab:counts}} 
\tablehead{ & & & \colhead{\lcdm} & & & & \colhead{\tcdm} & \\ 
\colhead{Survey} & & $5 \times 10^{13}$ & $10^{14}$ & $10^{15}$ & & $5
\times 10^{13}$ & $10^{14}$ & $10^{15}$} 
\startdata 
MS && 564,875 & 178,223 & 322 && 377,043 & 102,742 & 120 \\ 
VS && 565,886 & 178,483 & 285 && 378,548 & 103,157 & 111 \\ 
PO && 255,083 & 64,608 & 45 && 107,900 & 22,853 & 10 \\ 
NO && 259,279 & 64,930 & 42 && 108,807 & 23,216 & 13 \\ 
DW && 5,238 & 1,316 & 1 && 1,833 & 411 & 0 \\
Total && 1,504,620 & 441,833 & 623 && 878,356 & 226,602 & 231 
\enddata
\vspace* {-0.3truecm}
\end{deluxetable}

\subsection{Sky survey cluster populations}\label{subsec:skysurv}

We define cluster catalogs in the sky survey output using the 
same SO algorithm applied to the snapshots at fixed proper time.  
A minor modification for the \lcdm\ model must be 
made in order to define the threshold $\Delta$ with respect to the
epoch-dependent critical mass density
$\rhocrit(z) = \rhocrit(0) [\Omega_m(1+z)^3 + \Omega_\Lambda]$.

For the choice $\Deltac \se 200$, Table~\ref{tab:counts} lists counts
of clusters identified in the sky survey catalogs above mass limits $5
\times 10^{13}$, $10^{14}$ and $10^{15} \hinv
\msol$.  The lower mass limit corresponds to 22 particles and the
maximum redshifts of the catalogs are given in
Table~\ref{tab:surveys}.  

Total counts of $1.5$ and $0.9$ million clusters offer large
statistical samples.  On the other hand, the small numbers of objects 
at the most massive end of the spectrum put the
finite size of the visible universe into context and provide
additional motivation for near-future surveys to define an absolutely
complete sample of the largest clusters in the universe.  Only a few
hundred Coma-like or larger clusters are expected on the sky at all
redshifts in either cosmology.

Figure~\ref{fig:wedge_Mcut} shows redshift-space maps of clusters with
$\Mtwoh > 10^{14} \hinv \msol$ and $z \le 1.25$ in $3 \times 90$ 
deg strips taken from the PO octant surveys of the \tcdm\ (left) and
\lcdm\ (right) models.  The 
surveys display markedly different evolution at $z > 0.5$; 
distant clusters are more abundant in the low mass density cosmology 
(Efstathiou, Frenk \& White 1993; Eke, Cole \& Frenk 1996; Bahcall Fan
\& Cen 1997).  Within the 3
deg slice --- a width equivalent to two Sloan Digital Sky
Survey scans --- the \lcdm\ model contains 3084 clusters above the
$10^{14} \hinv \msol$, half lying beyond $z \se 0.70$.  The
\tcdm\ model contains 1122 clusters, with median $z \se 0.39$.

The sky surface density of clusters within three broad redshift
intervals are shown as a cumulative function of mass in
Figure~\ref{fig:Ncum_M}.  The ranges in redshift are chosen to
represent three classes of observation: local, $ z <0.2$;
intermediate, $0.2 \le z < 0.5$, and high, $0.5 \le z < 1.2$.  
Counts measured within the octant surveys are shown as 
points while solid lines show the number expected from
integrating the Jenkins mass function 
$N(>\!M) \se \int_{z_{\rm min}}^{z_{\rm max}} \, dz \,
(dV/dz) \, \int_{{\rm ln}M}^\infty \, \dlnM^\prime \,
\nJMF(M^\prime,z)$.  

For \tcdm, the integral is performed using the fit parameters
determined at $z \se 0$ (Table~\ref{tab:HVMF}).  For \lcdm, we must
recognize the fact that, because $\Omega_m(z)$ varies along the
light-cone, the fit parameters will evolve with redshift.  As
$\Omega_m$ tends to unity at high redshifts, we expect the
parameters to converge to the \tcdm\ values.  Since differences in
$A$, $B$ and $\epsilon$ between the two models are small at $z \se 0$,
we take a simple approach and vary the parameters linearly in
$\Omega_m$.  For example, we assume for $A$ that
\begin{equation} 
\label{eq:fit_omegaz}
A(\Omega_m) \ = \ (1-x) A(1) \, + \, x A(0.3) ,
\end{equation} 
where $x \sequiv (1-\Omega_m)/0.7$ and $A(1)$ and $A(0.3)$ are the
$z \se 0$ fit parameters for \tcdm\ and \lcdm\ from
Table~\ref{tab:HVMF}.  Similar interpolations are assumed for $B$ and
$\epsilon$.

The predictions of this model agree very well with the measured counts
in the octant surveys.  The model is accurate to $\lta 10\%$ in number
for \lcdm\ at all masses and redshifts shown.  Similar accuracy is
displayed for the \tcdm\ model at low and intermediate redshifts, but
the model systematically underestimates counts in the high redshift
interval by $\sims 25\%$.  

Dashed lines in Figure~\ref{fig:Ncum_M} show numbers expected by the
Press-Schechter model in its simplest form (see J01 for details). For
the \tcdm\ model, the PS curve tends to underestimate the space
density at high masses.  For the \lcdm\ model, the use of mass
measured within a critical, rather than mean, mass density threshold
leads to an offset in mass between the measured counts and PS curves
at low redshifts.  The offset declines as
$\Omega_m$ approaches unity, resulting in a relatively good 
match to the simulated counts in the high redshift interval.

In the \lcdm\ panels, we plot the \tcdm\ JMF curves as dotted lines
for comparison.  At low redshifts, the two models exhibit an offset in
the direction of \tcdm\ being overabundant relative to \lcdm, a
difference already discussed in \S\ref{subsec:massftn} for the $z \se 0$
population.  In the intermediate redshift interval, this offset is
reversed at nearly all masses above the $5 \times 10^{13} \hinv\msol$
limit.  At high redshifts, the \lcdm\ counts are typically an order of
magnitude higher than those of \tcdm.

\subsection{Sky survey completeness}\label{subsec:completeness}

Comparison of the octant counts with the JMF expectations provides a
measure of the incompleteness of the HV sky catalogs that arises from their
finite redshift extent.  
Figure~\ref{fig:Nsky_z} plots the cumulative sky surface
density of clusters above mass limits $5 \times 10^{13}$, $3 \times
10^{14}$, and $10^{15} \hinv\msol$ as a function of redshift.  Points
show the sky densities of clusters lying at redshift $z$ or higher
with masses above the stated limits (top to bottom, respectively), 
determined by combining the octant surveys of each model.  Lines in
the figure show the JMF expectations from integrating
equation~(\ref{eq:JMF}) using the linear evolution of the fit
parameters, equation~(\ref{eq:fit_omegaz}).  

At Coma mass scales ($> \! 10^{15} \hinv\msol$), the
catalogs are essentially complete, as fewer than one such object is 
expected over $\pi$ steradian beyond the survey redshift limit.  At $3
\times 10^{14} \hinv\msol$, the \lcdm\ model octants are missing
$\sims 100$ clusters expected above $z \se 1.5$, implying $\sims 98\%$
completeness.  The \tcdm\ model at this mass
limit is essentially complete; the small discrepancy between the
measured and JMF counts at redshifts $z \gta 0.5$ reflects the
systematic trend exhibited in the high redshift panel of 
Figure~\ref{fig:Ncum_M}.  At the mass
scale of groups, $5 \times 10^{13} \hinv\msol$, 
the incompleteness becomes more significant.  In the \lcdm\ model, for
example, $\sim 15\%$ of the group population should lie at $z > 1.5$.

The \lcdm\ model possesses a healthy population of very high redshift
clusters.  Across the whole sky, a cluster as massive as Coma is
expected at redshifts as high as $1.3$.  At $z > 2.5$, one $3 \times
10^{14} \hinv\msol$ cluster should lie somewhere on the sky, joined by
$\sims 20,000$ others above $5 \times 10^{13} \hinv\msol$, nearly one
per square degree.  Before getting carried away by such seemingly firm
predictions, we must investigate the effect of varying a 
degree of freedom that has so-far been kept fixed in the models: 
the amplitude of the fluctuation power spectrum $\sigma_8$.



\section{The Temperature Function, Absolute Mass Scale and Power
Spectrum Normalization Uncertainty}\label{sec:sigmavar}

In this section, we use the freedom in the mass-temperature relation
to tune $\beta$, 
the ratio of specific energies in dark matter and ICM gas, 
so that both models match the observed local temperature
function.  We show how $\beta$ should scale with $\sigma_8$ so as to
maintain consistency with observations.   The final snapshots 
are used to calibrate the level of uncertainty in $\sigma_8$ arising
from sample variance in local volumes.  We discuss the overall
systematic error in $\sigma_8$, then examine in \S\ref{sec:highz} how
this uncertainty affects predictions for the high redshift cluster
population.

\subsection{Fitting local temperature observations}\label{subsec:Tftn}

Pierpaoli, Scott \& White (2001, hereafter PSW) provide the most
recent study of the local temperature function and its constraints on
$\sigma_8$.  The sample of 38 clusters used in their analysis is
adapted from the \xray\ flux-limited sample of Markevitch (1998) and
is designed as an essentially volume-limited sample within redshifts
$0.03 < z < 0.09$ and galactic latitude $|b| > 20^\circ$ for clusters
with $k\Tx \gta 6 \kev$.  PSW update temperatures for 23 clusters in
the Markevitch (1998) sample with values given in White (2000).

PSW note that the White temperature values, derived from ASCA
observations using a multi-phase model of cluster cooling flow
emission, tend to be hotter (by 14\% on average), than those
Markevitch obtained 
through a single-temperature fit after exclusion of a core
emission.  Based on recent high-resolution studies of cooling flows
(David \etal\ 2001; McNamara 2001) that do not appear to support
the underlying cooling flow emission model used by White (2000), there
is cause for concern that the increased temperatures may be
artificial.  We therefore revert to the original values of Markevitch
(1998), and note that the resulting effect on the derived values of
$\beta$ is about $10\%$.

We use the data from Tables 3 and 4 of PSW and perform a maximum likelihood
fit to determine values of $\beta$ for each model.  Our procedure is
similar to that used by PSW\footnote{We note a typographical error in
their equation~(18), which should read $\ln {\cal L} \se \sum_i
[(\eta_i-1)\mu_i + \eta_i \ln (1- {\rm exp}(-\mu_i))]$.}, but rather
than use an analytic model as a reference, we use the binned $z \se 0$
differential velocity distribution $n(\sigmadm)$ converted to a
set of temperature functions $n(T)$ for $\beta$ in the range
$0.2$ to 2.  For a given $\beta$, 300 Monte Carlo realizations of the
observational sample are generated, assuming Gaussian statistics and
temperature errors distributed evenly in number (half positive, half
negative) about the 
best fit value.  To consider those clusters for which the selection
volume is best defined and for which cluster ICM physics is better
understood, a lower limit of $6 \kev$ is applied to each random
realization.  

Figure~\ref{fig:Tftn_local} shows cumulative number of clusters
as a function of temperature.  The bold line in each panel gives
the observations while the thin solid line shows the $z \se 0$
snapshot number density obtained using best-fit values $\betal \se
0.92$ and $\betat \se 1.20$.  The dotted line shows sky survey results
for clusters lying within the range observed, $0.03 < z < 0.09$), using
combined MS and VS samples and values $\betal \se 0.92$ and $\betat
\se 1.10$.  

The fact that a single value of $\betal$ leads to acceptable fits for
both the snapshot and sky survey samples indicates that number
evolution in the sky survey sample is 
small for \lcdm.  The evolution in the \tcdm\ case is sufficient to
warrant a slightly lower value of $\betat$ for the sky survey data.
The likelihood analysis described above produces $1\sigma$ error
estimates $\betal \se 0.92 \pm 0.06$ and $\betat \se 1.10 \pm 0.06$.
Current samples constrain the temperature scale of the
cluster population to an accuracy of about $6\%$.  

\subsection{Cosmic variance uncertainty in $\sigma_8$ }\label{subsec:sigmacv}

The locally observed cluster sample is one realization of a cosmic
ensemble that varies due to shot noise and spatial clustering on
survey scales.  Except for extending the angular coverage of the
observations into the plane of the galaxy, there is no possibility of
observing another cluster sample in the same redshift range.  Cosmic
variance in the local sample can be investigated using the 
large sampling volumes of the simulations.

An impression of the magnitude of the sample variation is given in
Figure~\ref{fig:Mftn_var}.  Differential mass functions for 16
independent $5000$ sq deg survey volumes, extending to $z \se 0.15$
and extracted from the MS and VS samples, are shown for the \lcdm\
model.  A mass limit of $5 \times 10^{14} \hinv \msol$ leads to an
average sample size of 30 clusters.  
Dotted lines in each panel show the range in number density
of the \lcdm\ Jenkins mass function as $\sigma_8$ is raised and
lowered by 5 per cent about its default.  The sky survey sample
variations are largely confined within the $\pm 5\%$ range of
$\sigma_8$ shown.  

The full $z \se 0$ snapshot samples allow a more precise estimate of
the cosmic variance contribution to $\sigma_8$ error.  
We divide the full computational volumes into
cubic cells of size $375 \hinv \mpc$ (\lcdm) and $400 \hinv \mpc$
(\tcdm).  Offsetting the grid of cells by half a cell width along the
principal axes and resampling generates totals of $4096$ and $1000$
samples of clusters in cells of volume comparable to the $5 \times
10^7 (\hinv \mpc)^3$ sampled by
local temperature observations.  Within each cell, we determine
the most likely value of $\sigma_8$ using the JMF 
space density fit to mass-limited samples.  Mass limits of $6
\times 10^{14}$ (\lcdm) and $10^{15} \hinv \msol$ (\tcdm) produce
average counts of 30 clusters within each cell.  

The distributions of $\sigma_8$ resulting from this
exercise are nearly log--normal; we find 
$\sigma_8 \se 0.911 e^{\pm 0.030}$ and
$0.591 e^{\pm 0.023}$ for \lcdm\ and \tcdm, respectively.  
The error in $\sigma_8$ from the likelihood analysis is well 
approximated by 
\begin{equation} 
\label{eq:sigma8err}
\frac{\Delta \sigma_8}{\sigma_8} \ = \ 
\frac{\langle (N-\bar{N})^2 \rangle^{1/2}}{\bar{N}} \  
\biggl| \frac{\sigma_8}{n(M)} \, \frac{\partial n(M)}{\partial
\sigma_8} \biggr|_{M_{\rm lim}}^{-1}  
\end{equation} 
where ${\langle (N-\bar{N})^2 \rangle^{1/2}}$ is the \rms\ deviation
of counts-in-cells above the mass limit, $\bar{N}$ the mean count, and
the Jacobian is evaluated 
at the survey mass limit $M_{\rm lim}$.  The latter is a steep
function of mass, taking on values $5.8$ and $8.3$ for \lcdm\ and
\tcdm. 

Because of scatter in the temperature--mass relation, the variance of 
counts-in-cells for temperature-limited samples is slightly larger
than that of mass-limited samples.  
Performing a similar analysis based on counts for temperature limited
samples results in $\Delta \sigma_8 /\sigma_8 \se 0.039$ for \lcdm\
and $0.025$ for \tcdm.
Since observations are temperature limited, these values apply to
analysis of current temperature data.

\subsection{$M-T$ calibration and overall $\sigma_8$
uncertainty}\label{subsec:sig8err}

As emphasized by previous studies, uncertainty in the calibration of
$p(M\,|\,T)$ is the largest source of error in $\sigma_8$.  
The error in $\sigma_8$ associated with uncertainty in the 
absolute mass scale can be derived by solving for the zero in the
total derivative of the mass function, equation~(\ref{eq:JMF}).
Ignoring the weak mass dependence of $\alpha_{\rm eff}(M)$, 
the result is 
\begin{equation} \label{eq:dlnsig8dlnM}
\frac{\Delta \sigma_8}{\sigma_8} \ = 
\alpha^\prime(M)  \ \frac{\Delta M}{M}  
\end{equation}
where, at large masses ($\sigma(M) < e^B$),  
\begin{equation} \label{eq:alphaprime}
\alpha^\prime(M) \ = \ \biggl[ \alpha_{\rm eff}(M) +
  \frac{1}{\epsilon \, (\lnsiginv(M) + B)^{\epsilon-1}} \biggr] .
\end{equation}
The first term can be connected to a shift in the
characteristic collapsed mass (fixed $e^B$) 
while the second term, which arises from the $1/M$ factor in
equation~(\ref{eq:JMF}), is required to maintain constant mass
fraction in objects at fixed $\lnsiginv(M)$.  
The sensitivity $\alpha^\prime(M)$, plotted in
Figure~\ref{fig:alphaprime}, asymptotes to a value $0.4$ 
above $\sims 5 \times 10^{14} \hinv \msol$ in both cosmologies.
Below this mass, $\alpha^\prime(M)$ increases considerably, reaching
unity at $10^{14} \hinv \msol$.  The rarest, most massive clusters
place the most sensitive limits on $\sigma_8$.  

Attempts at calibrating the mass--temperature relation have been made
using numerical simulations and observations.  Simulation results by
different groups compiled by Henry (2000) and PSW display an overall
range of $\sims 50\%$ in temperature at fixed mass, equivalent to a
$75\%$ range in mass if one assumes $M \spropto T^{3/2}$.  
A complicating factor is 
that normalizations are typically quoted using a mass--weighted
temperature, and this measure can differ systematically at the $\sims
20\%$ level from the spectral temperatures derived from plasma
emission modeling of the simulated ICM (Mathiesen \& Evrard 2001).
Observational attempts at calibrating the relation (Horner, Mushotzky
\& Scharf 1999; Nevalainen, Markevitch \& Forman 2000; Finoguenov,
Reiprich, \& B\"ohringer 2001) display discrepancies of similar
magnitude to the simulations.   Part of this variation is due to 
the fact that these analyses are comparing $T$ to
estimators $\hat{M}$ that differ in their degree of bias and noise
with respect to the theoretical mass $\Mdeltac$.

With relatively little in the way of firm justification, 
we conservatively estimate
the $1\sigma$ uncertainty in the zero point of the mass--temperature 
relation to be $\Delta M / M \se 0.25$.  Assuming log-normal errors,
this assumption allows the absolute mass scale to lie within a factor
$2.3$ range at $90\%$ confidence.  We note that PSW assume a $15\%$
uncertainty in mass, somewhat smaller than the value used here.

From equation~(\ref{eq:dlnsig8dlnM}) in the high mass limit, the
uncertainty in power spectrum normalization is 
\begin{equation} \label{eq:sig8toterr} 
|\, \Delta \sigma_8 /\sigma_8 \,|\,_{\rm sys} \ = \ 0.10 ~~~~~~ (1\sigma) 
\end{equation}
or a $16\%$ uncertainty at $90\%$ confidence. 
We employ this level of error when 
exploring statistics of high redshift clusters in \S\ref{sec:highz}.

\subsection{Degeneracy in $\beta$ and $\sigma_8$}\label{subsec:degen}

The calibration uncertainty discussed above in terms of mass can be
rephrased in terms of temperature or, equivalently for this study, 
the parameter $\beta$ used to connect temperature to dark matter
velocity dispersion.  An advantage of $\beta$ is that it can be 
determined independently from gas dynamic simulations that model the  
gravitationally coupled evolution of the ICM and dark matter.  
In a comparison study of twelve, largely independent
simulation codes applied to the formation of a single cluster, Frenk
\etal\ (1999) find good agreement among the computed specific energy
ratios within $\Delta \se 200$, with mean and standard deviation 
$\beta_{\rm sim} \se 1.17 \pm 0.05$.  

At first glance, this determination agrees well with the \tcdm\ 
value of $\beta$ but is in mild ($2.7 \sigma$) disagreement with the 
\lcdm\ value derived from the local temperature sample in
\S\ref{subsec:Tftn}.   However, the uncertainties quoted previously
for $\beta$ are derived at the fixed values of $\sigma_8$ used in the
N--body simulations.  To incorporate the additional sources of error 
in $\sigma_8$ discussed above, 
we use the mass sensitivity, equation~(\ref{eq:dlnsig8dlnM}), 
and the virial scaling $T \spropto \beta^{-1} M^p$ with 
$p \simeq 2/3$ exhibited by gas dynamic simulations of clusters 
to derive the scaling 
\begin{equation} \label{eq:betasigreln}
  \beta \ \propto \ \sigma_8^{p/{\alpha^\prime}(M)} \ \sim \
 \sigma_8^{5/3} .
\end{equation}
An increase in $\betal$ sufficient to match the Frenk \etal\ (1999) 
simulation ensemble value requires $\sigma_8 \se 1.04$.  This value
is marginally within the range allowed by COBE microwave background
anisotropy constraints for Hubble parameter $h \ssim 0.7$ (Eke, Cole
\& Frenk 1996).  Tighter constraints on $\sigma_8$ could serve to
increase the tension between the two independent determinations of the 
specific energy ratio.


\section{Clusters at high redshift}\label{sec:highz}

We are now in a position to revisit the expected numbers of high
redshift clusters, incorporating into the analysis 
the systematic uncertainty in power
spectrum normalization.  We begin by noting the advantage of
predicting cluster counts as a function of \xray\ temperature rather
than mass, and compare the model predictions to the sky surface
density of high redshift clusters from the EMSS catalog (Henry \etal\
1992; Gioia \& Luppino 1994).   Redshift information 
from the RDCS catalog (Rosati \etal\ 1998; Borgani \etal\ 1999b)
supports the \lcdm\ model under conservative assumptions, but the
model predictions are sensitive to selection effects related to 
core luminosity evolution.  

We then return to mass selected samples and explore the sensitivity of
Sunyaev--Zel'dovich searches for distant clusters to $\sigma_8$
variation.  Finally, the redshift evolution of characteristic mass
and temperature scales at fixed sky surface density is used to compare
\lcdm\ and \tcdm\ expectations against redshift and temperature
extremes of the observed cluster population.

\subsection{\xray\ cluster counts}\label{subsec:Xray}  

Because models are constrained by observations of the local
temperature function, predictions of counts as a function of
temperature can be made with smaller uncertainty than predictions of
counts as a function of mass.  The mass function requires separate
knowledge of $\beta$ and $\sigma_8$ whereas the temperature function
requires only a unique combination of the pair.  This advantage
breaks down if $\beta$ (or an equivalent parameter linking $T$ to $M$)
evolves with redshift.  Current observations
support no evolution (Tran \etal\ 1999; Wu, Xue \& Fang 1999), at
least for the connection between galaxy velocity dispersion
and ICM temperature.  We therefore assume a non-evolving $\beta$
in order to examine the space density of clusters as a function of
temperature at arbitrary redshift.

Figure~\ref{fig:Ncum_T} shows the range of cumulative sky counts
expected as a function of temperature within the same three broad redshift
intervals used in Figure~\ref{fig:Ncum_M}.  The range in counts shown
within each panel corresponds to varying $\ln \beta$ within its $90\%$
confidence region with $\sigma_8$ held fixed for each model: 
$0.83 \le \betal \le 1.01$ and $0.91 \le \betat \le 1.20$.  
The constraint to match local 
observations produces nearly complete overlap in the temperature
functions of the two cosmologies at $z < 0.2$.  At intermediate
redshifts, the \lcdm\ counts are boosted by nearly an order of
magnitude compared to the low redshift range, while the \tcdm\ counts
grow by a factor $\sims 3$.  The $90\%$ confidence regions for the
models become disjoint in this redshift interval.  

In the high redshift region, the models separate further, with the
characteristic temperature at fixed sky surface density a factor 
$\sims 1.5$ times larger in \lcdm\ than \tcdm.  The steep nature of
the space density translates this moderate difference in $T$ into a
large factor difference in counts; at 8~keV, the counts differ by
a factor of about $20$.  
An estimate of the observed sky density in this redshift range, based
on the EMSS survey data, is shown as the square 
in the upper panel of Figure~\ref{fig:Ncum_T}.  
This point is based on three hot ($k\Tx > 8 \kev$) and distant
($z > 0.5$) clusters covering a search area of $278$ sq deg (Henry
2000), leading to a sky surface density $0.011$ per sq deg at 
$z > 0.5$.  The data are consistent with the \lcdm\ expectations and
rule out \tcdm\ at $>99\%$ confidence.  

A modest degree of evolution in $\betat$ with redshift could 
reconcile \tcdm\ with the EMSS data.
Additional information, such as the redshift
distribution of \xray\ flux-limited samples, provides independent
constraints capable of eliminating such a possibility (Oukbir \&
Blanchard 1992).  The RDCS survey (Rosati \etal\ 1998) is currently
the \xray-selected survey with the most extensive redshift data
available for distant clusters.  The survey, as analyzed by Borgani
\etal\ (1999b), is complete within 33 sq deg to limiting $0.5-2 \kev$
\xray\ flux $5 \times 10^{-14} \cgsflux$ and contains 70 clusters with
measured redshifts extending to near one.

To explore the compatibility of the octant survey populations with the
RDCS sample requires a model for the \xray\ luminosity $L_x$
anticipated from the simulated clusters.  As a base model, we assume a
mean bolometric \lxt\ relation $L_x \se 2.9 \times 10^{44} (T/6
\kev)^{2.88} h^{-2} \ergs$ (Arnaud \& Evrard 1999) that is assumed
not to evolve with redshift (Mushotzky \& Scharf 1997; Henry 2000;
Fairley \etal\ 2000).  
To account for the fact that the \lxt\ mapping is not
one-to-one, we add a uniformly distributed scatter of $\pm 0.4$ in
log$_{10}(L_x)$.  Fluxes in an observed $0.5-2 \kev$ \xray\ band are
derived from a {\tt mekal} spectral synthesis code assuming 0.3 solar
metallicity.  Applying a $5 \times 10^{-14} \cgsflux$ flux cut,
excluding $z < 0.05$ clusters, and scaling the PO and NO simulated
cluster surveys to $33$ sq deg area leads to predictions shown as the
solid lines in Figure~\ref{fig:RDCS}.  Under these economical
assumptions of ICM evolution, the \lcdm\ model provides an acceptable
fit to the observations.

Since the most distant cluster sources are typically detected at modest 
signal-to-noise, it is worth investigating the influence that 
additional sources of \xray\ emission would have on survey selection.
The steep nature of the mass function offers the opportunity
for clusters lying just below the survey flux limit to be pushed above
it, given some mechanism to enhance its \xray\ luminosity.  

For the purpose of illustration, we consider adding to the base model
described above random
additional sources of \xray\ luminosity whose influence increases
with redshift.  These sources may be thought of as
arising from cooling flows, active galaxies embedded within or
near the cluster, or mergers, any or all of which may be more likely
at higher redshift.  The specific model assumes that half of the
population has luminosities boosted by an amount drawn from a uniform
distribution of amplitude $\alpha(z) L_x$, with $\alpha(z)=2z$ and
$L_x$ the base luminosity.  
Although arguably extreme, this model raises the zero-point of the
\lxt\ relation by only $50\%$ at $z \se 0.5$.  
Expectations for RDCS based on this alternative, 
shown as dashed lines in Figure~\ref{fig:RDCS}, differ significantly
from the base model predictions at high redshift.  The \tcdm\ still
fails to match the observations at redshifts between $0.1$ and $0.3$,
but its high redshift behavior is much improved.  
The \lcdm\ model consistently overpredicts the counts beyond $z \se
0.4$.   

Deep \xray\ imaging with {\em Chandra} and XMM will help settle the issue
of whether this toy model is too extreme.  For now, we note that the
good agreement between the RDCS and the economical \lcdm\ model
predictions may signal that the ICM undergoes relatively simple
evolution dominated by gravitational shock heating after an initial,
early epoch of preheating (Evrard \& Henry 1991; Kaiser 1991; Bower
1997; Cavaliere, Menzi \& Tozzi 1999; Balogh, Babul \& Patton 1999; 
Llyod-Davies, Ponman \& Cannon 2000; Bower \etal\ 2001; 
Bialek, Evrard \& Mohr 2001; Tozzi \& Norman
2001).  The preheated cluster simulations of Bialek \etal\ (2001)
produce low redshift scaling relations for \xray\ luminosity, isophotal
size and ICM mass versus temperature that 
simultaneously match local observations and 
exhibit little evolution in the \lxt\ relation to $z \ssim 1$.




\subsection{Mass-selected samples}\label{subsec:MassSam}

Interferometric SZ surveys have been proposed that would survey $\sims
10$ sq deg of sky per year with sufficient sensitivity to detect all
clusters above a total mass limit $\sims 10^{14} \hinv \msol$, nearly
independent of redshift (Holder \etal\ 1999; Kneissl \etal\ 2001).
The mass limit assumes that the ICM mass fraction does not depend
strongly on cluster mass or redshift, an assumption supported by 
simulations.  Bialek \etal\ (2001) find that the  
ICM gas fraction within $\Delta \se 200$ remains a fair representation
of the baryon--to--total cosmic ratio: 
$f_{\rm ICM} \se (0.92 \pm 0.04) \, \Omega_b / \Omega_m $ 
above rest frame temperature $kT \se 4 \kev$.  We investigate
expectations for SZ surveys assuming that they will be sensitive to a
limiting total mass that is independent of redshift.

\begin{figure*}
\epsfxsize=20.0cm \epsfysize=10.0cm
\hbox{\hskip -1.0truecm  \epsfbox{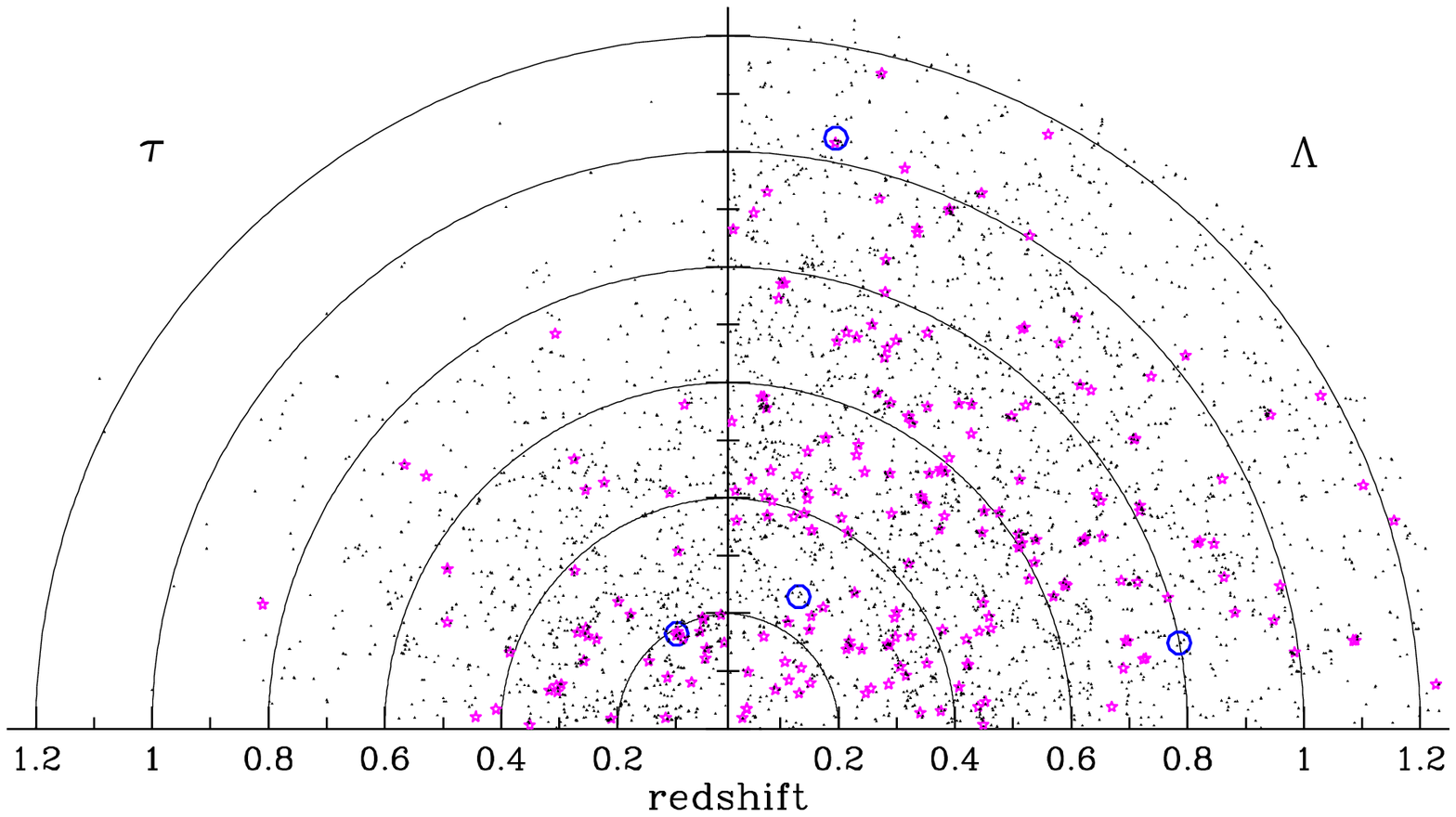} }
\vskip -4.0truecm
\caption{
Maps of clusters in $90^\circ \times 3^\circ$ slices extending to $z \se
1.25$, derived from the octant sky surveys of the \tcdm\ (left) and
\lcdm\ (right) 
models.  Symbols shows clusters of different masses: $h M_{200}/\msol
> 10^{15}$ (open circles); $\in 10^{14.5} - 10^{15}$ (stars) and $\in
10^{14} - 10^{14.5}$ (dots).  Numbers of clusters in these mass ranges
are 1, 50, 1071 (\tcdm) and 3, 185, 2896 (\lcdm).  
\label{fig:wedge_Mcut}
}
\end{figure*}

Maps of mass-limited cluster samples in SDSS--like survey slices 
were presented in Figure~\ref{fig:wedge_Mcut} for the default values of 
$\sigma_8$.  To illustrate the effect of $\sigma_8$ variation, we plot
clusters in 
the same spatial regions again in Figure~\ref{fig:wedge_bias}, after
applying an effective fractional variation in $\sigma_8$ of $+10\%$
(\tcdm) and $-10\%$ (\lcdm).  Although
equation~(\ref{eq:dlnsig8dlnM}) suggests a simple shift in 
mass threshold to mimic a change in $\sigma_8$, the mass dependence of
$\alpha^\prime(M)$ (Figure~\ref{fig:alphaprime}) introduces cumbersome
non-linearity into the shift.  We adopt instead an
equivalent procedure that adjusts both masses $M$ and 
number densities $n(M)$ in the HV cluster catalogs by amounts 
\begin{eqnarray}\label{eq:Mbiastrans}
& M^\prime = \ e^\mu M , \nonumber \\ & n(M^\prime) \, \dln \! M^\prime
= \ e^{-\mu} \, n(M) \, \dln \! M
\end{eqnarray} 
with 
\begin{equation} \label{eq:Mbias}
 \mu \ = \ \frac{\ln \, (1 + \Delta \sigma_8 / \sigma_8)} 
{\langle \alpha_{\rm eff} \rangle} 
\end{equation}
and $\langle \alpha_{\rm eff} \rangle \se 0.25$.  
Tests of these transformations 
using the Jenkins mass function verify their  
accuracy to better than $10\%$ in number for masses $10^{13.7} - 10^{15.3}
\hinv\msol$ and variations of power spectrum normalization within the
$90\%$ confidence region $|\Delta \sigma_8 /\sigma_8| \le 0.16$.  The
practical value of these simple transformations is in allowing the
discrete simulation output to represent a family of models covering a
range of normalizations $\sigma_8$.  

When compared to Figure~\ref{fig:wedge_Mcut}, the intermediate redshift 
cluster populations of the two cosmologies shown in
Figure~\ref{fig:wedge_bias} appear much more similar.  Unlike
Figure~\ref{fig:wedge_Mcut}, the overall counts above $10^{14} \hinv
\msol$ in the 3 degree slice are now nearly identical --- 1696 for
\tcdm\ compared to 1843 for \lcdm.  However, their redshift
distributions remain different; the \tcdm\ clusters stay 
concentrated at lower redshifts while the \lcdm\ clusters are more
broadly distributed (Oukbir \& Blanchard 1992).

Figures~\ref{fig:wedge_Mcut} and \ref{fig:wedge_bias} imply that a
redshift statistic, such as the sample median, will be superior to
counts as a means to constrain cosmology.  Motivated by the
aforementioned planned SZ surveys, we perform a specific investigation
of expectations for a random 10 sq deg survey complete above a
mass-limit $\Mtwoh \se 10^{14} \hinv \msol$.  We sample clusters in
3000 randomly located, square fields of 10 sq deg area, 
divided equally between the PO and NO surveys and chosen to avoid survey
boundaries.  We use the transformations in
equation~(\ref{eq:Mbiastrans}) to define the cluster population at
values of $\sigma_8$ different from the default.  To drive the models
in directions that minimize their differences, we increase $\sigma_8$
in the \tcdm\ model and decrease it in the \lcdm\ case.

The distributions of counts at all redshifts ($z\!<\!1.25$), counts at
high redshift ($0.8\!<\! z\!<\!1.25$) and the median redshift for
clusters above the survey mass limit derived from the random 
samples are presented in Figure~\ref{fig:N_zmed_bias}.  At the
default values of $\sigma_8$ (left column), the distributions of
number expected either at all redshifts $z \!<\! 1.25$ (bottom row) or
at high redshift (middle row) would allow unambiguous discrimination
between the models using a single 10 sq deg field.  At high redshift,
the \lcdm\ model predicts, on average, a factor 15 more clusters than
\tcdm.  Overall, the mean counts in 10 sq deg are 117 and 45,
respectively.  Biasing $\sigma_8$ by $10\%$ in the chosen directions
(middle column) produces essentially identical expectations for the
overall cluster yield, with both models expecting $72 \pm 12$ clusters
per field.  At high redshift, the ability to discriminate is weakened.
For a $16\%$ bias (right column), the sense of the overall counts are
reversed, with the \tcdm\ model having a $60\%$ larger yield, on
average, than \lcdm.  The high redshift count distributions of the
models possess considerable overlap.

In contrast to the count behavior, the distributions of sample median
redshift $\zmed$ are extremely stable to variations in $\sigma_8$.
The $95$-th percentile value of $\zmed$ for \tcdm\ moves from $0.498$
to $0.528$ to $0.538$ at $0$, $10\%$ and $16\%$ bias.  As a
frequentist measure of
discrimination we quote the power (Sachs 1982), defined as the
probability of rejecting \tcdm\ at the chosen level ($95\%$) of
significance given \lcdm\ as the true model.  Measuring the power by
integrating the \lcdm\ distributions of $\zmed$ above the $95$-th
percentile \tcdm\ value, results in power of $99.9\%$, $98.8\%$ and
$94.8\%$.  These power measures, and others calculated in a similar
manner for the counts, are listed in corresponding panels of
Figure~\ref{fig:N_zmed_bias}.  High redshift counts lose power to
discriminate between the models as the applied bias on $\sigma_8$ is
increased.

The large shift in the expected counts as $\sigma_8$ is
varied provides an appropriate lever arm to use for placing firmer
constraints on this parameter with SZ surveys.  Holder, Haiman and
Mohr (2001) estimate that a 10 sq deg survey as assumed here could,
with complete redshift information and assuming complete knowledge of
the relation between SZ signal and cluster mass, constrain $\sigma_8$
at the $3-5\%$ level.

\subsection{Sky surface density of distant clusters}

{\em Chandra} \xray\ Observatory detections of extended \xray\
emission from three clusters at $z \! > \! 1$ have recently been
reported.  Stanford \etal\ (2001) report detection of hot ICM in a
pair of RDCS-selected clusters separated by only 4 arcmin on the sky
and 0.01 in redshift, RX~J0848+4453 at $z \se 1.27$ and RX~J0849+4452
at $z \se 1.26$ (Stanford \etal\ 1997; Rosati \etal\ 1999).
RX~J0848+4453 appears to have a complex morphology and a cool
temperature $kT \se 1.6^{+0.8}_{-0.6} \kev$ while RX~J0849+4452 
appears to be a relaxed system with higher temperature $kT \se
5.8^{+2.8}_{-1.7} \kev$.  In addition to these systems, Fabian \etal\
(2001) present {\em Chandra} evidence for extended ICM emission at
temperature $ kT \se 5.0^{+2.6}_{-1.5}\kev$ around the radio galaxy
3C294 at $z \se 1.786$.  Quoted errors in these temperature estimates
are $68\%$ confidence values.

From the temperature--mass relation calibrated by the local
temperature function sample in \S\ref{sec:sigmavar} and assuming a
non-evolving $\beta$, we can estimate the masses of these clusters.
Results for \lcdm\ (\tcdm) are $4.4 \ (3.9) \times 10^{13} \hinv
\msol$ and $\Mtwoh \se 3.1 \ (2.7) \times 10^{14} \hinv \msol$ for
RX~J0848+4453 and RX~J0849+4452, respectively, and $\Mtwoh \se 1.9 \
(1.6) \times 10^{14} \hinv \msol$ for 3C294.  To explore the
likelihood of finding such clusters, we employ a statistic that links
physical properties to measurable sky surface density.  

The statistics we consider are sky surface density characteristic mass
and temperature, defined as the mass $\Mdndz(z)$ and temperature
$\Tdndz(z)$ at which the differential sky surface density $N^\prime(z)
\sequiv dN/dz$ of inversely rank-ordered clusters at redshift $z$
takes on fixed values.  The mass 
scale $\Mdndz(z)$ is defined by the relation
\begin{equation} \label{eq:dNdz}
  N^\prime(z) \ = \ (1/\Omega_{\rm surv})
 \int_{M_{N^\prime}(z)}^\infty \dlnM \ n(M,z) \, dV/dz
\end{equation}
where $\Omega_{\rm surv}$ is the survey sky area.  The
characteristic temperature is defined in a similar manner. 
As a practical approximation to the redshift differential, we employ
counts in redshift bins of width $0.1$ to 
derive this statistic from the HV sky survey data.

Figure~\ref{fig:MTnofz_lcdm} shows the redshift behavior of the sky
surface density characteristic (SSDC) mass and temperature for the
\lcdm\ model.  Filled points are values based on the combined octant
survey populations.  Solid lines are predictions from the Jenkins mass
function, derived by computing equation~(\ref{eq:dNdz}) using
equation~(\ref{eq:JMF}) for $\nJMF(M,z)$ and integrating in bins of
width 0.1 in redshift.  Sky surface density thresholds $dN/dz$ vary by
factors of 10 from $0.001$ to 10 per sq deg per unit redshift, as
labeled.  Open circles show results for the SSDC at $0.01$ per sq deg
per unit redshift extending to $z \ssim 3$ using the $16^\circ \times
76^\circ$ extension to the PO survey.  Thick dashed lines in each
panel show the limiting resolved mass of $5 \times 10^{13} \hinv
\msol$ (22 particles) and the corresponding limiting resolved virial
temperature.  The good 
agreement between the Jenkins model and the discrete cluster sample
measurements is to be expected from the results of
Figure~\ref{fig:Ncum_M}; the POX extension data verify the utility of
the model to $z \ssim 3$.

The vertical bar in each panel of Figure~\ref{fig:MTnofz_lcdm} shows
the $90\%$ uncertainty range in the local calibration of each
quantity; $e^{\pm 0.11}$ in $kT$ and $e^{\pm 0.42}$ in
$M$. The HV simulation and Jenkins model results for the SSDC
measures can be varied vertically by these amounts in
Figure~\ref{fig:MTnofz_lcdm}.  The narrow spacings between $\Mdndz(z)$
and $k\Tdndz(z)$ contours reflect the steepness of the cumulative
counts at fixed redshift;  the terrain of the
counts is steep in the mass and temperature directions.  At a
particular redshift, the calibration uncertainties translate into a 
large range of allowed sky surface densities for a given mass, and a
smaller but still significant range for a given temperature.  

Although steep in the temperature direction, the contours in the lower
panel of Figure~\ref{fig:MTnofz_lcdm} are remarkably flat in the redshift
direction.   Over the entire redshift interval $0.1<z<1.5$, the JMF
expectations for the SSDC temperature at 0.01 per sq deg per unit redshift 
lie in a narrow range between 8 and 10 keV.  In the \lcdm\ model,
distant, hot clusters should be as abundant on the sky as those nearby.

Temperatures of the aforementioned observed distant clusters
are plotted in the lower panel of Figure~\ref{fig:MTnofz_lcdm} as open
triangles (the RX clusters) and open square (3C294).  Temperature
uncertainties at $90\%$ confidence are shown, assuming Gaussian
statistics to convert $1\sigma$ errors.  The central values of the
hotter pair are consistent with a sky surface density of 1 per 10 
sq deg per unit redshift, but within the temperature 
measurement uncertainties, these
objects could be up to a factor 100 more common or a factor $\ssim
1000$ more rare.  The lower temperature system at $z \se 1.27$ is
consistent with a surface density of several per sq deg per unit
redshift.

Figure~\ref{fig:MTnofz_tcdm} shows that the \tcdm\ model is less able
to accommodate the existence of these $z \! > \! 1$ clusters.  The
central temperatures correspond to surface densities of 1 per 1000 sq
deg per unit redshift, a factor 100 times more dilute than the \lcdm\
values.  Given that only 40 such clusters would be expected on the
whole sky between redshifts one and two, it would be remarkable that
two would already be identified by these observations.

At the most dilute sky surface density plotted in these figures, each
filled circle represents the hottest or most massive cluster within
its $0.1$-wide redshift bin.  Even at this highest rank, the
variance in the discrete sample SSDC values remains remarkably small.
An exception is the unusual \lcdm\ at $z \se 1.04$.  This monster
lies nearly a factor two above the Jenkins model expectations and its
deviation is extreme compared to that displayed by values at the same
source density and other redshifts.  We note that its expected
temperature of $21 \kev$ exceeds that of the hottest known cluster
1E~0657-56 at $z \se 0.296$, with $kT \se 17.4\pm 2.5 \kev$ 
(Tucker \etal\ 1998).   This cluster is the 
asterisk in Figures~\ref{fig:MTnofz_lcdm} and \ref{fig:MTnofz_tcdm}.

As the hottest known cluster, 
it is natural to expect 1E~0657-56 to lie at the extreme end of the
surface density distribution in the redshift range $0.2-0.3$.  That is
indeed the outcome of comparing its location to \lcdm\ expectations in
Figure~\ref{fig:MTnofz_lcdm}.  For the case of \tcdm, its existence is
more troublesome, but given the combination of $T$ calibration 
uncertainty and scatter demonstrated by the first-ranked values of the 
discrete sample, this system is consistent
at the $\sims 2\sigma$ level with the expectations of
Figure~\ref{fig:MTnofz_tcdm}.  A similar statement of significance 
can be made for the comparably hot and more distant cluster 
RX~J1347-1145, with $kT \se 14.48^{+1.76}_{-1.46} \kev$ (Ettori, Allen
\& Fabian 2001) at $z \se 0.451$.  This analysis does not support the
interpretation of Ettori \etal\ (2001) that the existence of
RX~J1347-1145 alone can be used to place an upper limit on the matter
density parameter $\Omega_m < 0.5$.

To summarize, interpretation of distant cluster counts is complicated by
uncertainty in $\sigma_8$, variation of which can lead to large factor
changes in yield, as well as uncertainty due to possible evolution in
$\beta$ and other aspects of astrophysical evolution.  If a constant
$\beta$ assumption is valid for \lcdm, then $\sims 8 \kev$ clusters at
$z \se 1.2-1.3$ should be as numerous on the sky as those lying at $z \se
0.1-0.2$.


\section{Summary of \lcdm\ expectations}\label{sec:lcdm}

Given the increasing likelihood that the \lcdm\ model is an accurate
representation of our universe (Pryke \etal\ 2001; Netterfield \etal\
2001), we provide here
a brief summary and discussion of the characteristics of its cluster
population.

{\sl Coma-mass systems}.  The population of clusters with $\Mtwoh$ in
excess of $10^{15} \hinv\msol$ is potentially numerous, but not
overwhelmingly so.  With $\sigma_8 \se 0.9$, 400 Coma's
are expected on the whole sky (Figure~\ref{fig:Nsky_z}), but that
number ranges between 40 and 2000 as $\sigma_8$ is varied within its
$90\%$ confidence limits.  The median
redshift of this sample is expected to be $\zmed \se 0.48$, nearly
independent of $\sigma_8$.  Detection of Coma equivalents at $z \! >
\! 1$ in excess of $.003$ per sq deg ($\sims 120$ across the sky)
would rule out \lcdm\ at $95\%$ confidence.  A complete sample of
these objects could be obtained with an all-sky \xray\ imaging survey
only moderately more sensitive than the ROSAT All-Sky Survey
(B\"ohringer \etal\ 2001).  Such a survey would be
unique in being the first to be {\em absolutely complete}, meaning
complete in identifying {\em all\/} members of a class of
astrophysical objects within the finite volume of our past light-cone.

{\sl Hot \xray\ clusters}.  A characteristic feature of the \lcdm\
model is that the hottest clusters populate the sky at nearly fixed
surface density over a broad redshift interval 
(Figure~\ref{fig:MTnofz_lcdm}).  This implies a testable prediction of
a nearly flat redshift distribution, within $z \ssimeq 0.2 -1$, 
for a temperature-limited sample identified in a fixed angular survey 
area.  Within the 10,000 sq deg SDSS area, one $8 \kev$
cluster is expected to lie at $z \ge 2$.

{\sl Clusters at $z \ssim 3$}.  Looking to higher redshifts, clusters
with $\mtwoh > 10^{14} \hinv \msol$ and rest frame $kT > 4 \kev$
(apparent $kT \gta 1 \kev$) should exist at the level of one cluster
per 100 sq deg per unit redshift under the default $\sigma_8$ and
$\beta$ normalizations (Figure~\ref{fig:MTnofz_lcdm}).  Of order one
hundred such clusters are to be expected within the SDSS survey area
in the redshift interval $2.5 - 3.5$.  Of order ten clusters will have
rest frame $kT > 5 \kev$ and $z > 3$.  The vicinity of bright quasars
may be a natural place to search for these systems.  Verification of a
hot ICM at these redshifts would benefit from the large collecting
area of the planned Constellation-X Observatory.

{\sl Clusters at $z \lta 0.5$}.  The SDSS and 2dF optical surveys will
provide large numbers of clusters selected in redshift-space and
extending to redshifts $z \ssim 0.5$.  Although these samples offer an 
opportunity to place more sensitive constraints on $\sigma_8$, 
a number of systematic effects, such as biases in the
selection process and the mapping between properties measured in
redshift space (optical
richness or velocity dispersion) and underlying cluster mass $M$, must 
first be carefully calibrated.  Such systematic effects can be profitably 
studied by combining semi-analytic models of galaxy formation
with N-body models of dark matter halo evolution (\eg\ Springel \etal\
2001).  An \xray\ imaging survey to bolometric flux $10^{-14}
\cgsflux$, capable of identifying all clusters with 
$\mtwoh \! > \! 10^{14} \hinv \msol$ within $z \se
0.4$ (assuming a non-evolving \lxt\ relation), would provide the
ability to separate truly deep potential wells from redshift space
superpositions of smaller systems (Frenk \etal\ 1990).  

{\sl ICM temperature evolution}.  In predicting that the redshift
distribution of hot clusters at fixed sky surface density is flat over
observationally accessible redshifts,
we have implicitly assumed that the \xray\ temperature and mass follow
the virial relation $T \spropto (H(z)\Mtwoh)^{2/3}$.  It is
important to pursue high resolution imaging and spectroscopy of known
high redshift clusters with {\em Chandra} and XMM in order to test
whether more complex heating and cooling processes may be occurring,
particularly at high redshift.  Such processes would 
affect attempts to determine the geometry of the universe through the
\xray\ size--temperature relation (Mohr \& Evrard 1997; Mohr \etal\
2000).  

{\sl Precise parameter estimation}.  The accuracy of constraints on
$\sigma_8$ from the cluster temperature function is fundamentally
limited by the error in normalization of the mass--temperature
relation of hot clusters, equation~(\ref{eq:dlnsig8dlnM}).  
One per cent errors on $\sigma_8$ will require knowing 
the absolute mass scale of clusters to better than $3$ per cent.  This 
challenging prospect is currently beyond the capabilities of direct 
computational modeling and traditional observational approaches, such
as mass estimates based on hydrostatic equilibrium.  Weak
gravitational lensing, especially in 
the form of ``field'' lensing (see Mellier \& Waerbeke 2001 for a
review), appears the most promising approach; for example, 
Hoekstra \etal (2002) 
find $\sigma_8 \se 0.81^{+0.14}_{-0.19}$ at 95\% confidence from
analysis of relatively bright (limiting magnitude $R_C=24$) galaxies
in CFHT and CTIO fields covering 24 sq deg.  Imposing such constraints 
as priors will focus future studies on breaking existing degeneracies
between dark matter/dark energy densities and astrophysical evolution.  


\section{Conclusions}\label{sec:concl}

We present analysis of a pair of giga-particle simulations designed to
explore the emergence of the galaxy cluster population in large cosmic
volumes of flat world models dominated by matter energy density
(\tcdm) and a cosmological constant (\lcdm).  Besides shear scale,
these Hubble Volume simulations are unique in their production of 
sky survey catalogs that map structure of the dominant dark
matter component over large solid angles and
to depths $z \ssimeq 1.5$ and beyond.  Application of a
spherical overdensity (SO) cluster finding algorithm to the sky survey
and fixed epoch simulation output results in discrete samples of
millions of clusters above the mass scale of galaxy groups ($5 \times
10^{13} \hinv \msol$).   These samples form the basis of a
number of studies;  we focus here on precise 
calibration of the mass function and on systematic
uncertainties in cosmological parameter determinations caused 
by imprecise determination of the absolute mass scale of clusters.  A
summary of our principal findings is as follows.  

\medskip

\begin{itemize}
\item{We calibrate the SO(200) mass function to the Jenkins form
with resulting statistical precision of better than $3\%$ in number for
masses between $10^{13.5}$ and $10^{15.3} \hinv \msol$.  A preliminary
estimate of the overall theoretical uncertainty in this calibration
is approximately $20\%$. 
}

\item{We fit the local temperature function under the assumption
that the disordered kinetic energy in dark matter predicts the ICM
thermal temperature, leading to specific energy ratios $\betal \se
(0.92 \pm 0.06) \,(\sigma_8/0.9)^{5/3}$ and  $\betat \se (1.10 \pm 0.06) \,  
(\sigma_8/0.6)^{5/3}$.  For the \lcdm\ model, $\sigma_8 \se
1.04$ is required to match the value $\beta_{\rm sim} \se 1.17$
preferred by gas dynamic simulations of ICM evolution.  
}

\item{Based on the Jenkins form for the mass function, we derive
transformations of the discrete cluster sample that mimic variation in
$\sigma_8$.  Using these transformations, we show that 
the redshift distribution of mass-limited samples is a more powerful
cosmological diagnostic 
than cluster counts;  the median redshift of clusters more massive
than $10^{14} \hinv \msol$ in a single 10 sq deg field of a \lcdm\
cosmology can rule out \tcdm\ at a minimum of $95\%$ confidence.
}

\item{The \lcdm\ model, under conservative assumptions for
intracluster gas evolution, is consistent with high redshift
cluster samples observed in the \xray-selected EMSS and RDCS surveys.
}

\item{The statistics of sky surface density characteristic (SSDC) mass
and temperature are introduced to more naturally 
account for observational and theoretical uncertainties in measured
physical scales.  The \lcdm\ model predicts flat redshift behavior in the
SSDC temperature; a randomly chosen 8~keV cluster on the
sky is nearly equally likely to lie at any redshift in the interval 
$0.2$ to $1.2$. 
}

\item{With $\sigma_8 \se 0.9$, the \lcdm\ model predicts roughly
400 Coma-mass ($10^{15} \hinv \msol$) clusters across the sky at all
redshifts, with the
most distant lying just beyond $z \se 1$.  Pushing $\sigma_8$ to its
95\% confidence upper limit, the \lcdm\
model could accommodate up to $120$ Coma equivalents on the sky
at $z \! > \! 1$.  
}

\end{itemize}

\medskip

Larger and deeper cluster samples with 
accurate determinations of temperature or mass will lead to improved
constraints on cosmological and astrophysical parameters.  The
developing 2dF and SDSS surveys will provide large numbers of
clusters with galaxy velocity dispersion $\sigmagal$ serving as a
temperature measure and optical richness $N_{\rm gal}$ serving as a
surrogate for mass.  Gravitational lensing mass estimates will also be
possible for co-added ensembles of clusters (Sheldon \etal\ 2001). 
Extracting cosmological information from these data will require
likelihoods such as $p(\sigmadm \,|\,\sigmagal)$ or $p(M\,|\,N_{\rm gal})$.  
The challenge to the theoretical community will be to model
these likelihoods at a level of precision warranted by the large data
sets.  Almost certainly, the theoretical uncertainty associated with
this aspect of the modeling will dominate statistical errors, since
samples of many thousands, perhaps tens of thousands, of groups and
clusters will be available in the complete 2dF and SDSS surveys.
By imposing external constraints on selected parameters and
requiring model consistency across independent observables (\eg,
sub-mm, optical and X-ray), constraints on cosmological and
astrophysical parameters can be derived from a number of alternate 
paths.  

Valuable complementary information is available at \xray\
wavelengths.  An \xray\ imaging survey reaching to limiting flux $\sim 3
\times 10^{-14} \cgsflux$ in the $0.5-6 \kev$ band would
be capable of detecting a cluster with 6 keV rest frame temperature
to $z \se 1$ in either of the cosmologies studied here, assuming a
non-evolving \lxt\ relation.  The redshift distribution of such a
sample would be a powerful cosmological diagnostic, as long as
astrophysical evolution of the ICM could be sufficiently well
constrained.  A program of deep pointed observations with {\em
Chandra} and XMM (to constrain the astrophysical evolution), coupled
with a deep \xray\ imaging survey covering a significant
portion of the SDSS area (to identify a large cluster population in
redshift space) would be a powerful combination.  SZ surveys over 
large solid angle with bolometer arrays could play a similar role to
an \xray\ imaging mission, and interferometric arrays will probe to
smaller masses and higher redshifts than can be achieved by any
current search techniques.  Ultimately, the combination of {\em all\/} these 
approaches, along with deep, optical imaging and spectroscopy, 
will allow determinations of cosmological parameters to be
made not only more precise, by shear statistical weight, but
also be made more accurate by improving our understanding of the
astrophysical processes that govern the evolution of the visible 
components of clusters. 


\bigskip
This work was funded by the PPARC in the UK, the Max-Planck Society in
Germany, NSERC in Canada, NASA and NSF in the US, and NATO in all the
countries involved. Some of this work was carried out as part of the
EU Network for Galaxy Formation and Evolution. CSF acknowledges a
Leverhulme Research Fellowship.  We thank the staff of the
Rechenzentrum Garching for outstanding computational support.  
AEE acknowledges support from NSF AST-9803199 and NASA NAG5-7108,
clarifying conversations with A. Blanchard and J. Bartlett, and the
benefits of the Scientific Visitor Program at Carnegie Observatories
in Pasadena.



\clearpage



\clearpage

\clearpage

\clearpage

\clearpage

\clearpage

\clearpage

\begin{figure*}[t]
\epsfxsize=20.0cm 
\hbox{\hskip -1.0truecm  \epsfbox{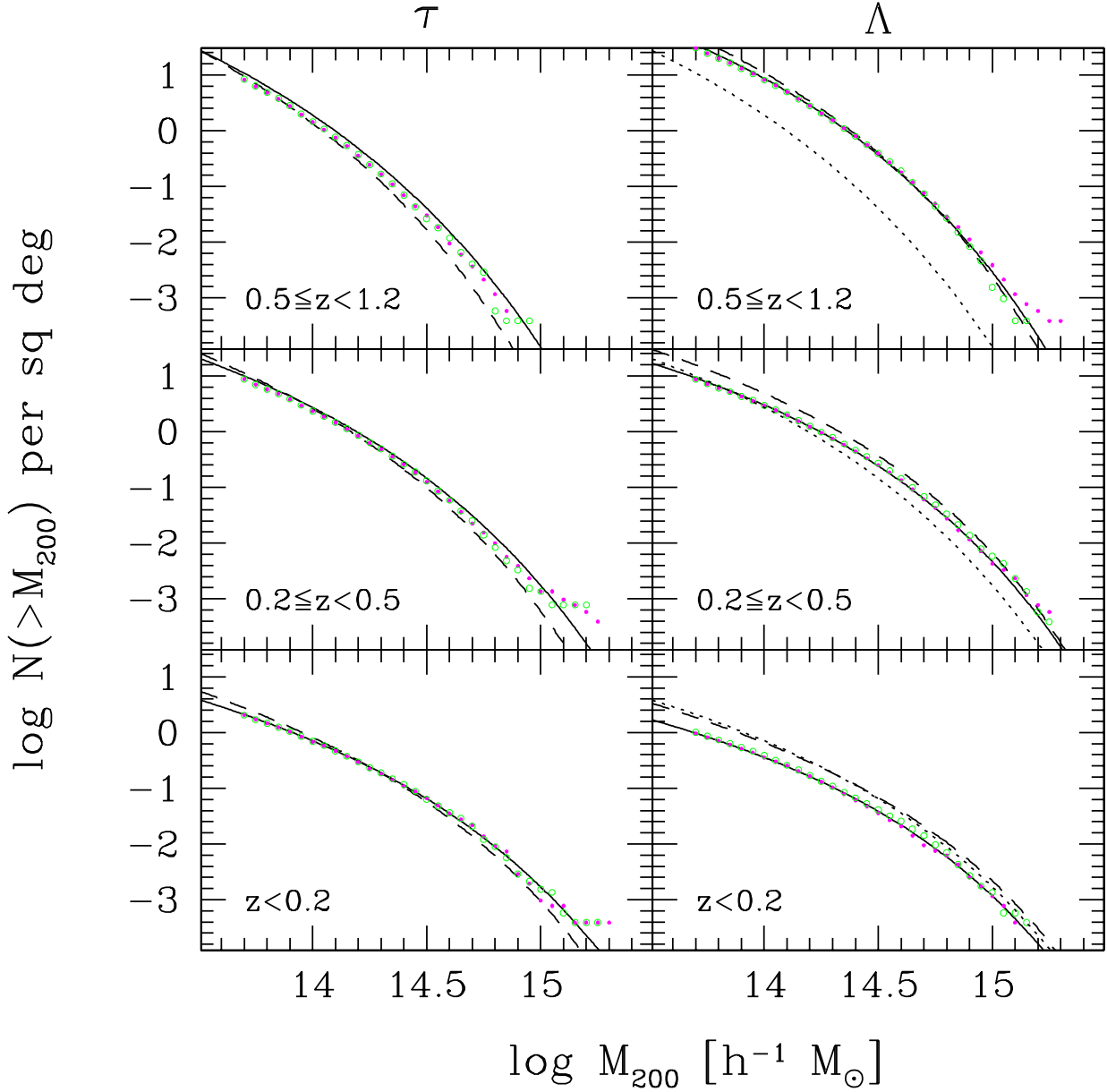} }
\vskip -1.0truecm
\caption{ 
Cumulative sky counts of clusters as a function of mass 
for low, intermediate and high redshift
intervals (bottom to top) for \tcdm\ (left) and \lcdm\ (right).  
Smooth solid lines in each panel 
give the expectations from integrating the Jenkins mass function,
equation~(\ref{eq:JMF}), over the appropriate volumes.  
Points show counts from the PO (filled circles) and NO
(open) octant surveys.  Dashed lines are standard Press--Schechter
estimates.  In the \lcdm\ panels, dotted lines display the
corresponding JMF expectations for the \tcdm\ cosmology.  
\label{fig:Ncum_M}
}
\end{figure*}

\clearpage

\begin{figure*}[t]
\epsfxsize=20.0cm 
\hbox{\hskip -1.0truecm  \epsfbox{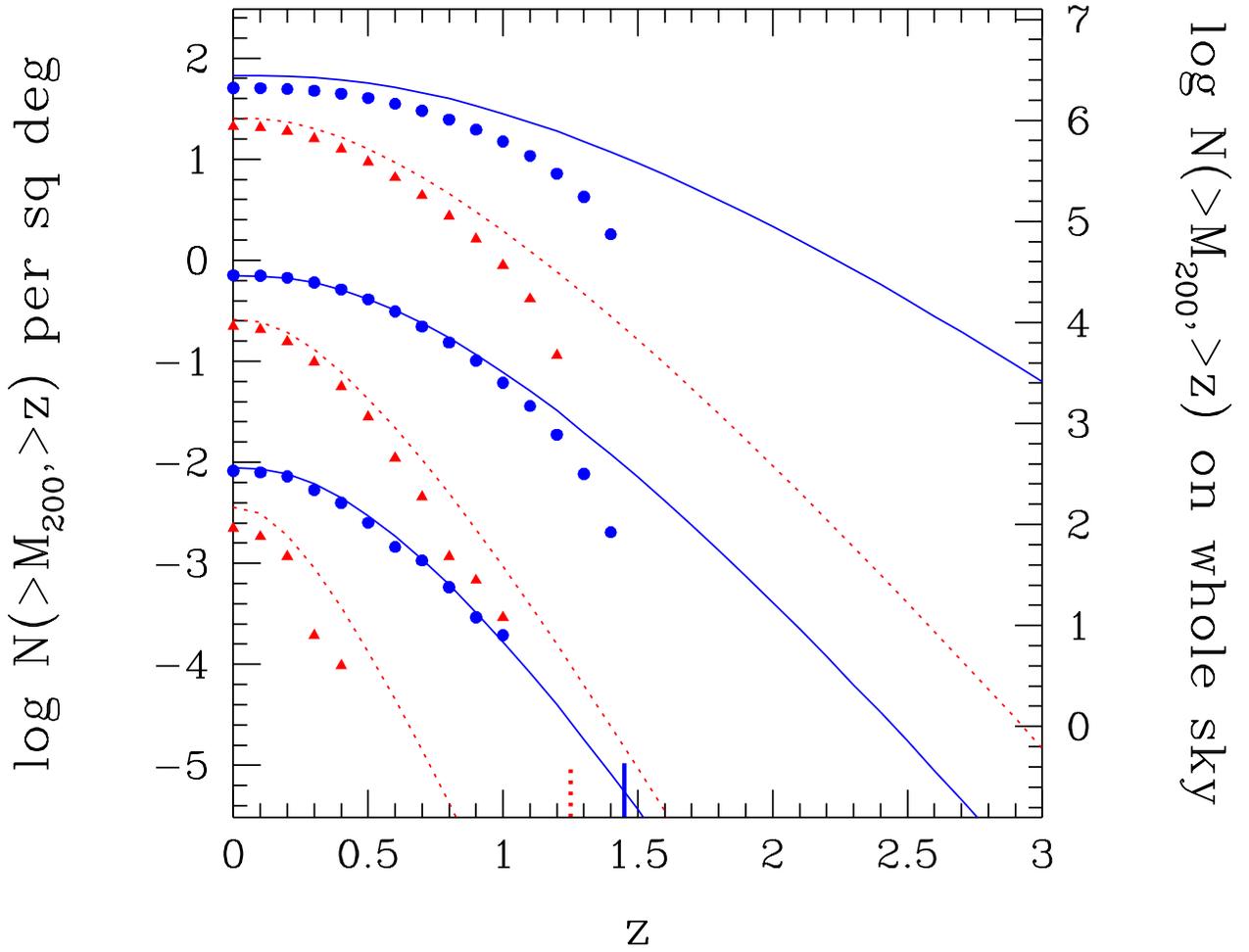} }
\vskip -1.0truecm
\caption{ 
Sky surface density of clusters lying at redshift $z$
or higher in \lcdm\ (filled circles, solid lines) and 
\tcdm\ (filled triangles, dotted lines).  Points give numbers derived
from the combined octant surveys with masses above 
$\Mtwoh \se 5 \times 10^{13}$ (top), $3 \times 10^{14}$ (middle) and
$10^{15} \hinv\msol$ (bottom).  Short vertical
lines mark the limiting redshifts of the octant surveys.  
Lines are expectations at each mass limit derived from
integrating the Jenkins mass function.  
\label{fig:Nsky_z}
}
\end{figure*}

\clearpage

\begin{figure*}
\epsfxsize=20.0cm 
\hbox{\hskip -1.0truecm  \epsfbox{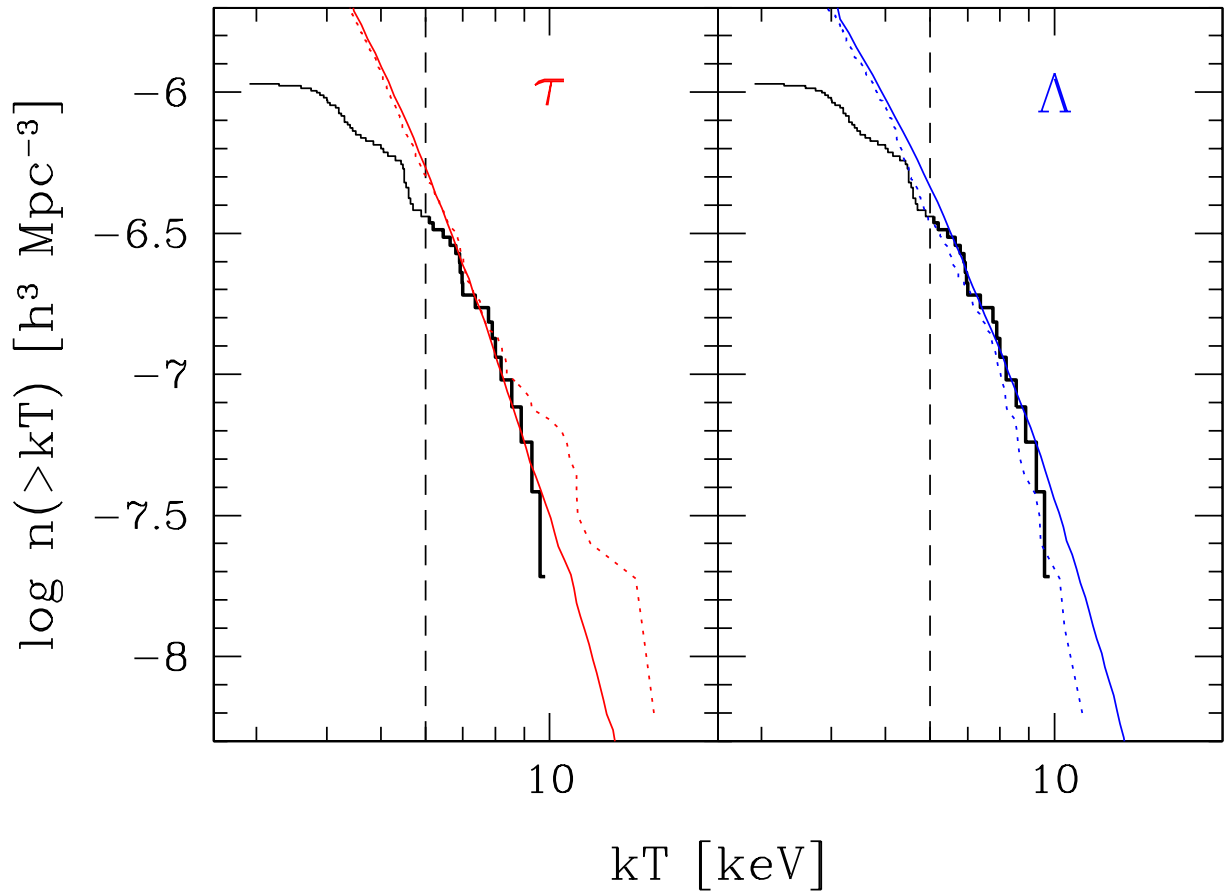} }
\vskip -1.0truecm
\caption{Thick lines show the local temperature function of Pierpaoli
\etal\ (2001) based largely on data of Markevitch (1998).  The HV simulation 
expectations, using best-fit values $\betal \se 0.92$ and
$\betat \se 1.20$, are shown from the light-cone (dotted) and $z \se
0$ snapshot (solid) outputs.  The former uses clusters 
within the combined MS and VS surveys lying in the redshift interval
$0.03 < z < 0.09$.  The latter uses the cluster population of the
entire computational volume.  The vertical dashed line in each panel 
shows the approximate completeness limit of the observations.  
\label{fig:Tftn_local}
}
\end{figure*}

\clearpage

\begin{figure*}
\epsfxsize=20.0cm 
\hbox{\hskip -1.0truecm  \epsfbox{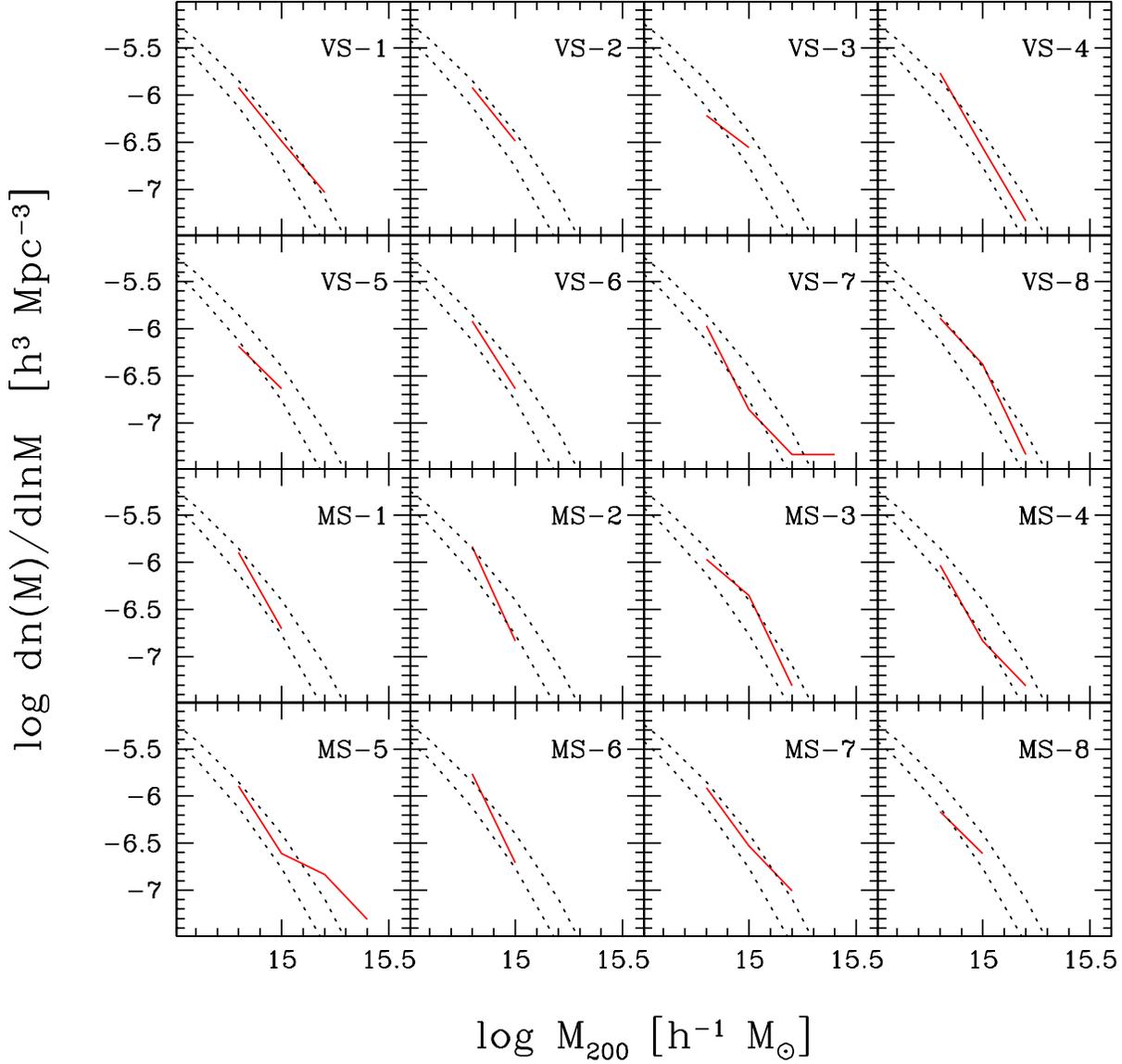} }
\vskip -1.0truecm
\caption{Differential mass functions within 16 independent 
$\pi/2$ steradian regions extending to $z \se 0.15$, derived from the
MS and VS surveys of the \lcdm\ model.  Dashed lines show JMF
expectations, equation~(\ref{eq:JMF}), for $\sigma_8$ values varied by
$\pm 5\%$ about the input value linearly evolved to $z \se 0.1$.  
The volume of the samples is comparable to that of the local observed
sample used to constrain $\sigma_8$.   
\label{fig:Mftn_var}
}
\end{figure*}

\clearpage

\begin{figure*}
\epsfxsize=20.0cm 
\hbox{\hskip -1.0truecm  \epsfbox{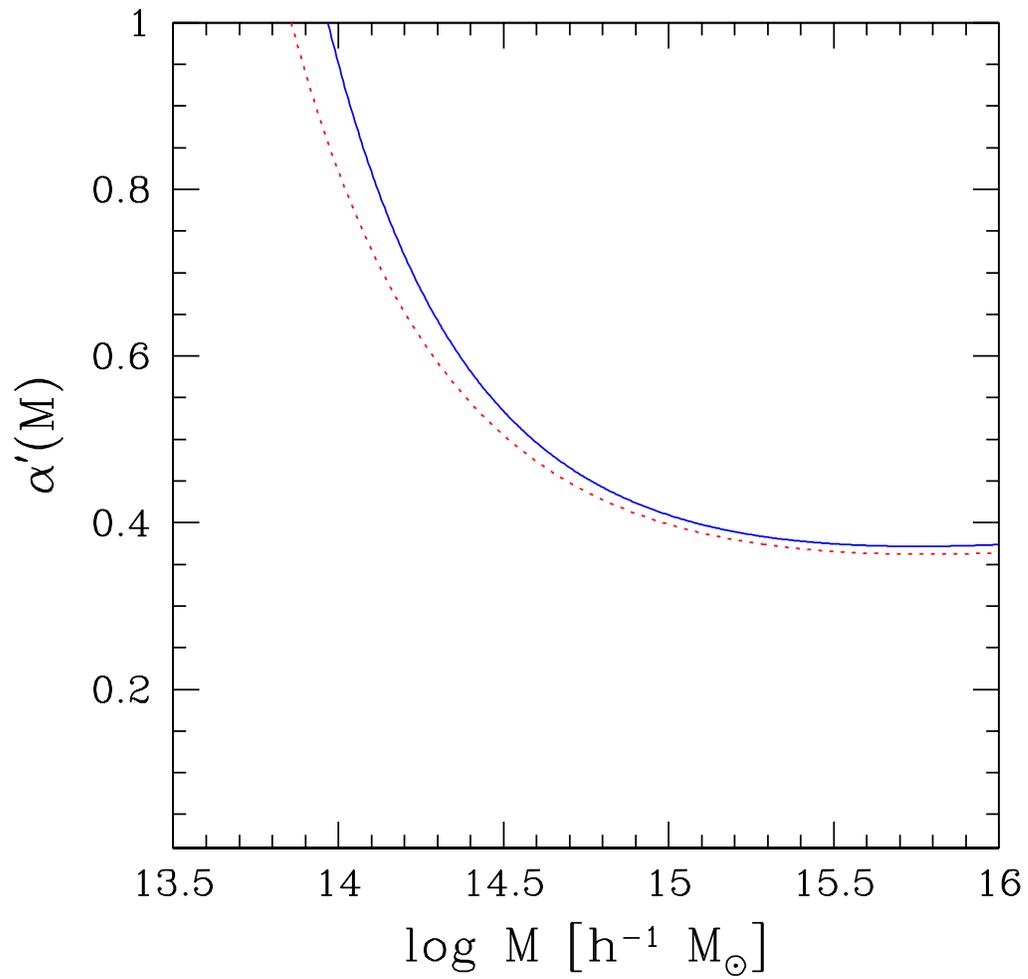} }
\vskip -1.0truecm
\caption{The sensitivity $\alpha^\prime(M)$,
equation~(\ref{eq:alphaprime}), for the \lcdm\ (solid) and \tcdm\
(dotted) models.
\label{fig:alphaprime}
}
\end{figure*}

\clearpage

\begin{figure*}
\epsfxsize=20.0cm 
\hbox{\hskip -1.0truecm  \epsfbox{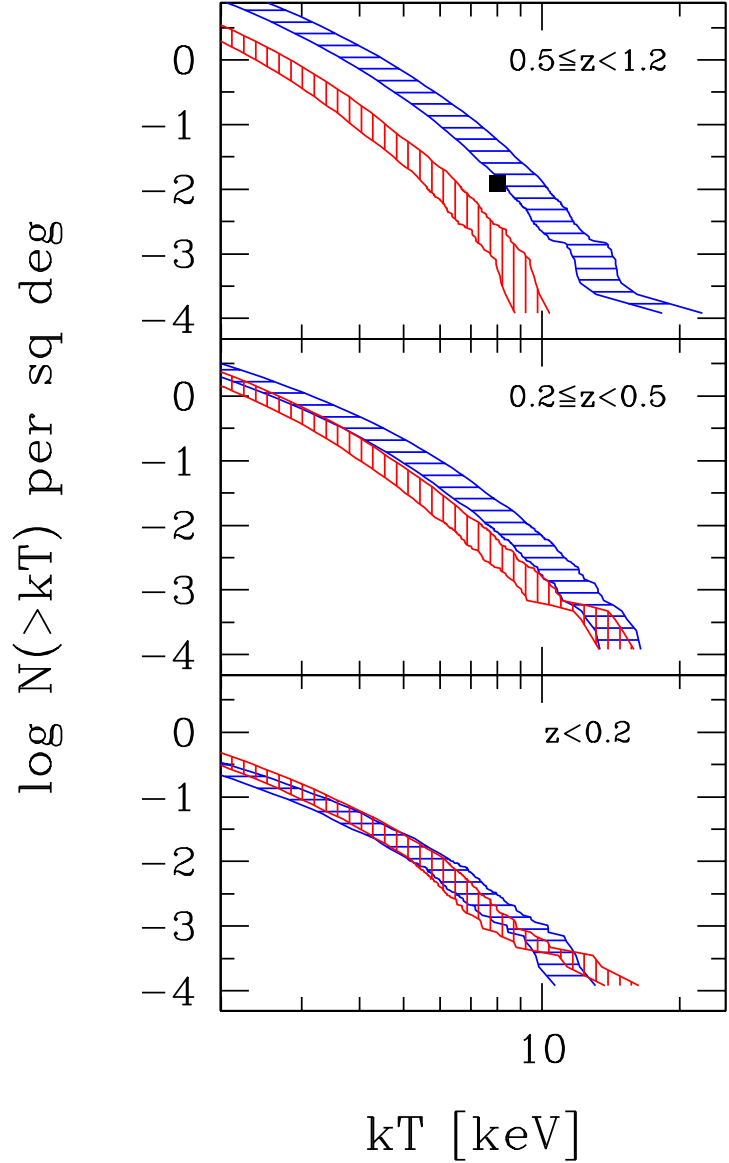} }
\vskip -1.0truecm
\caption{ 
The range of cumulative sky surface densities of clusters as a function of
temperature in three redshift intervals for the \lcdm\ (horizontal
hatched) and \tcdm\ (vertical) cosmologies.  
The range is determined from the combined octant survey counts by
varying $\beta$ within its overall $5$ to $95\%$ confidence range.  
The solid square
denotes the EMSS observational result for clusters hotter than 8 keV
at $z>0.5$. 
\label{fig:Ncum_T}
}
\end{figure*}

\clearpage

\begin{figure*}
\epsfxsize=20.0cm 
\hbox{\hskip -1.0truecm  \epsfbox{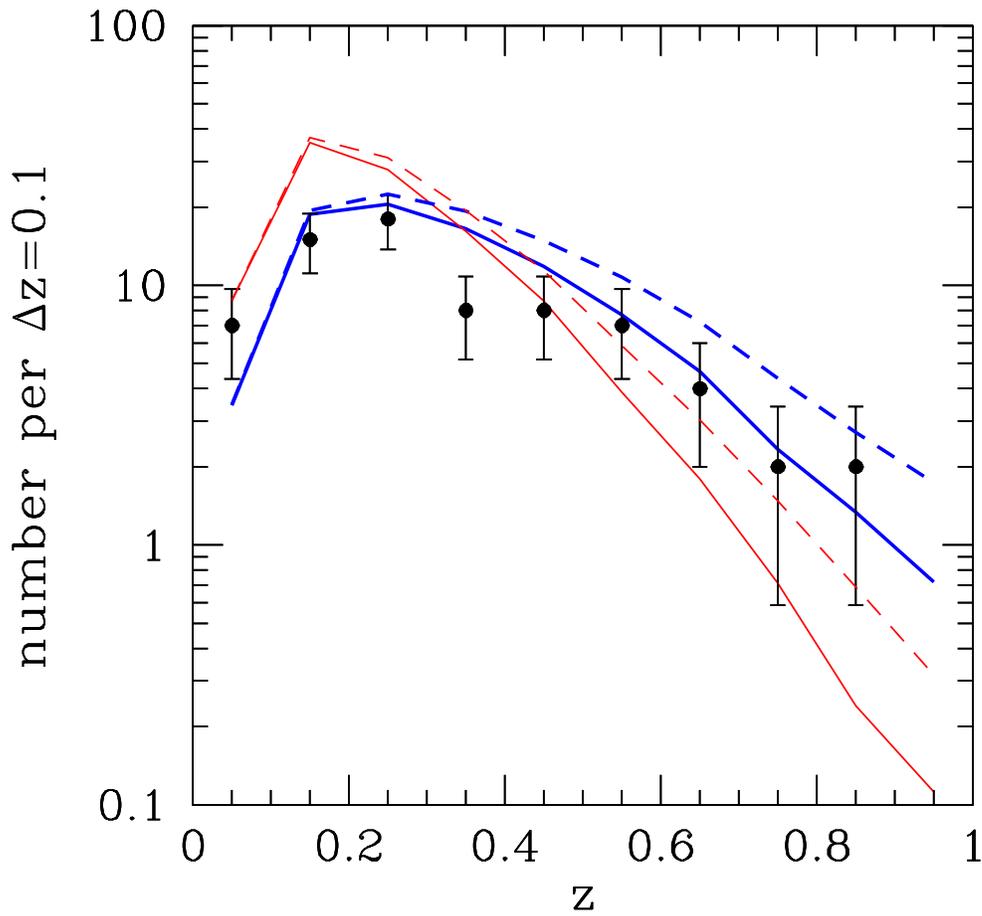} }
\vskip -1.0truecm
\caption{ 
Points show the redshift distribution of the \xray\
flux-limited RDCS survey (Rosati \etal\ 1998; Borgani \etal\ 1999b),
with Poissonian errors.  Solid lines show expectations for a $33$
sq deg survey derived from 
the combined octant surveys by assuming constant $\beta$ and a
non-evolving \lxt\ relation.  Dashed lines show plausible confusion 
effects of core luminosity contamination on the \xray-selection
(see text for details).  
\label{fig:RDCS}
}
\end{figure*}

\clearpage

\begin{figure*}
\epsfxsize=20.0cm \epsfysize=10.0cm
\hbox{\hskip -1.0truecm  \epsfbox{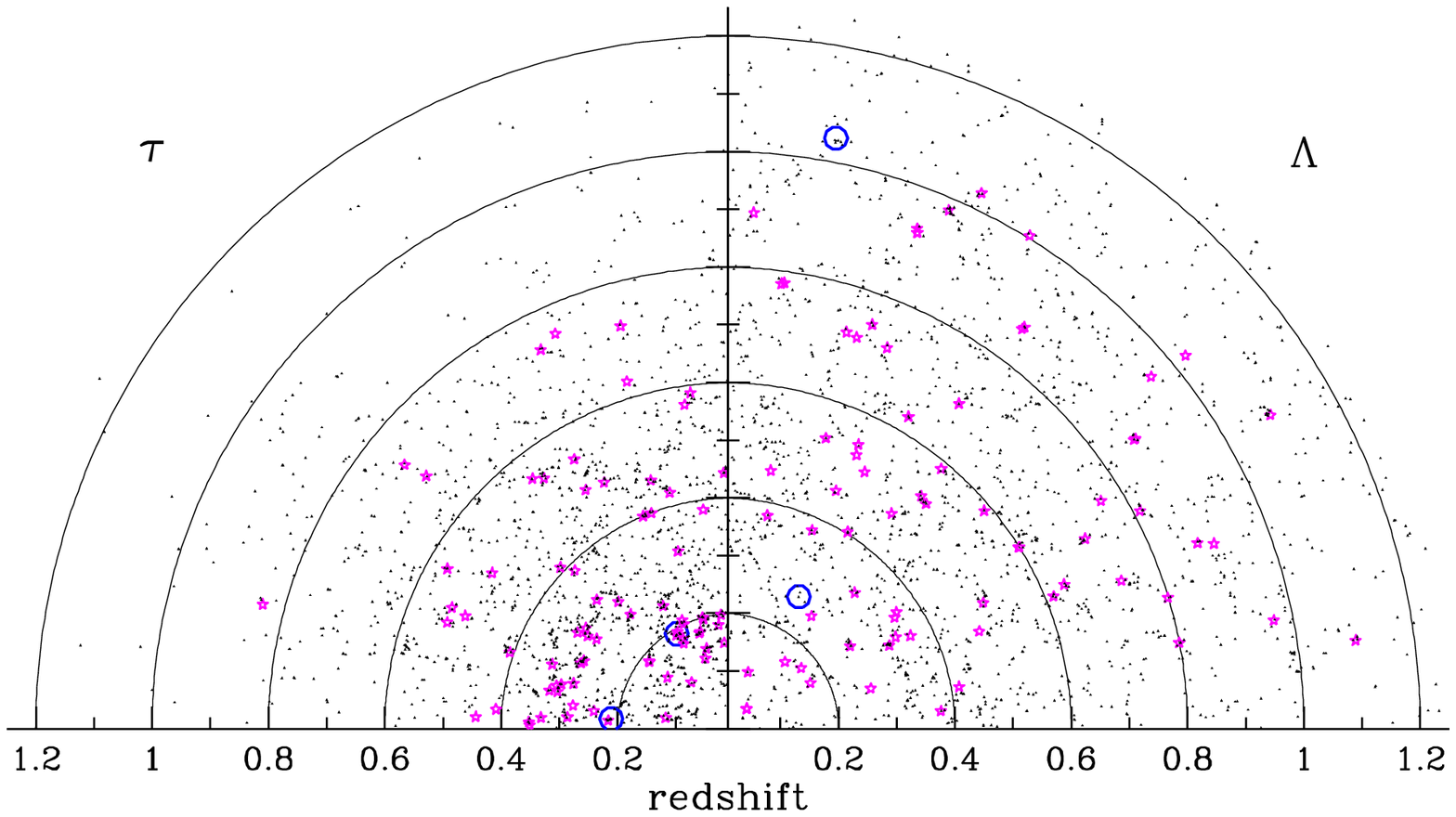} }
\vskip -4.0truecm
\caption{
Clusters expected in the same $90^\circ \times 3^\circ$ slices shown in
Figure~\ref{fig:wedge_Mcut}, but shown here after application of 
effective biases in $\sigma_8$ of $+10\%$ (\tcdm) and $-10\%$ (\lcdm).
\label{fig:wedge_bias}
}
\end{figure*}

\clearpage

\begin{figure*}
\vskip -2.0truecm
\epsfxsize=20.0cm 
\hbox{\hskip -1.0truecm  \epsfbox{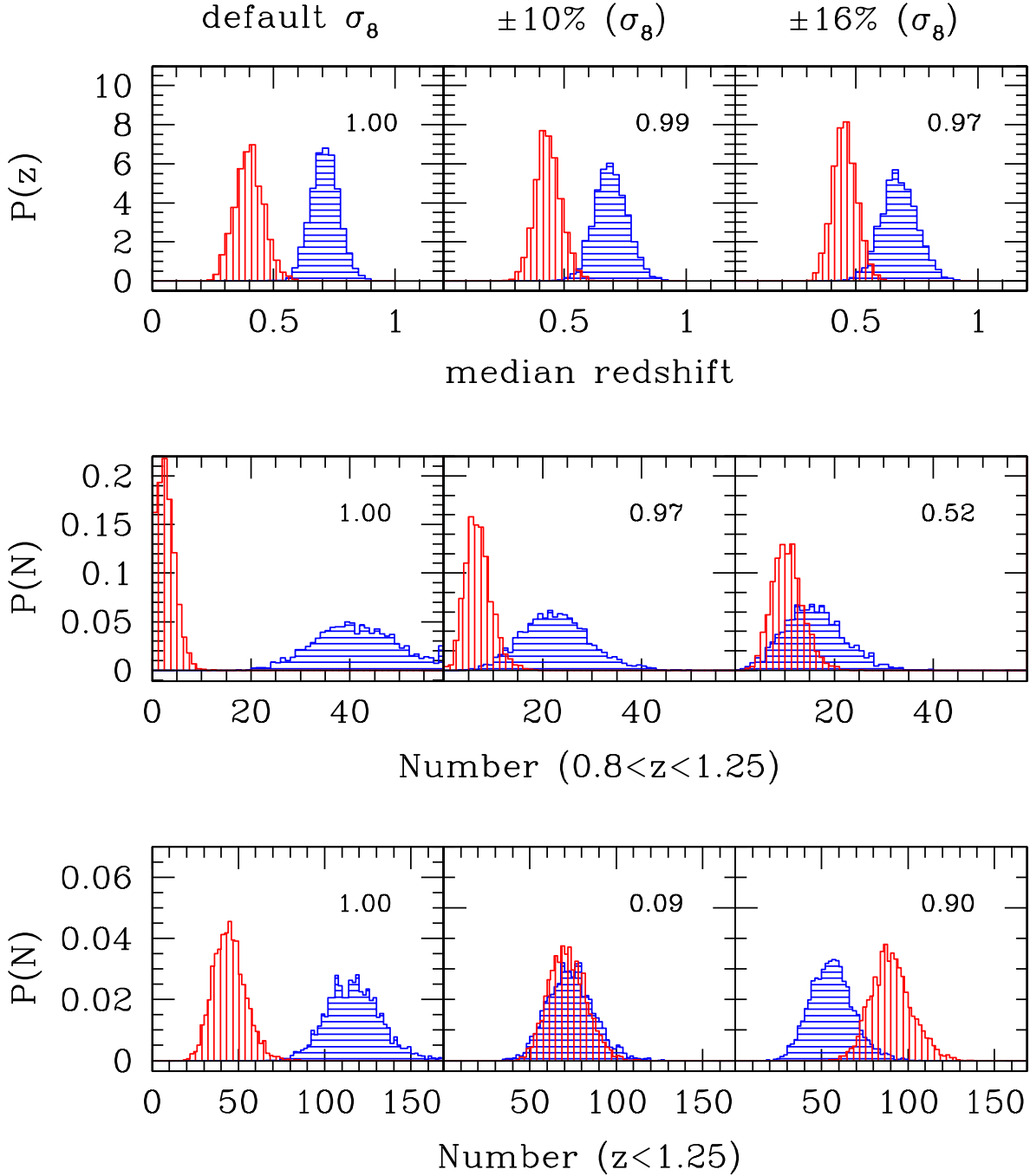} }
\caption{
The influence of varying $\sigma_8$ on the population of 
clusters more massive than $10^{14} \hinv\msol$ expected in 10 sq deg
survey fields.  Columns (left to right) show probability distributions 
at the default $\sigma_8$ values, $(+/-)10\%$ 
and $(+/-)16\%$ variation ($+$ for \tcdm/$-$ for \lcdm) for the
counts at redshifts $z<1.25$ (bottom row), counts in the 
high redshift interval $0.8<z<1.25$ (middle) and 
the median cluster redshift (top).  Vertical hatched distributions are
\tcdm, horizontal are \lcdm.  Numbers in each panel give the power
statistic described in the text.  The distributions 
are generated by sampling 10 sq deg fields
around 3000 randomly chosen pointings in the combined octant surveys
of each model.  Shifts in mass and number, equation~(\ref{eq:Mbiastrans}) are
used to effectively vary $\sigma_8$ in the cluster catalogs.
\label{fig:N_zmed_bias} 
}
\end{figure*}

\clearpage

\begin{figure*}
\epsfxsize=18.0cm 
\hbox{\hskip -1.0truecm  \epsfbox{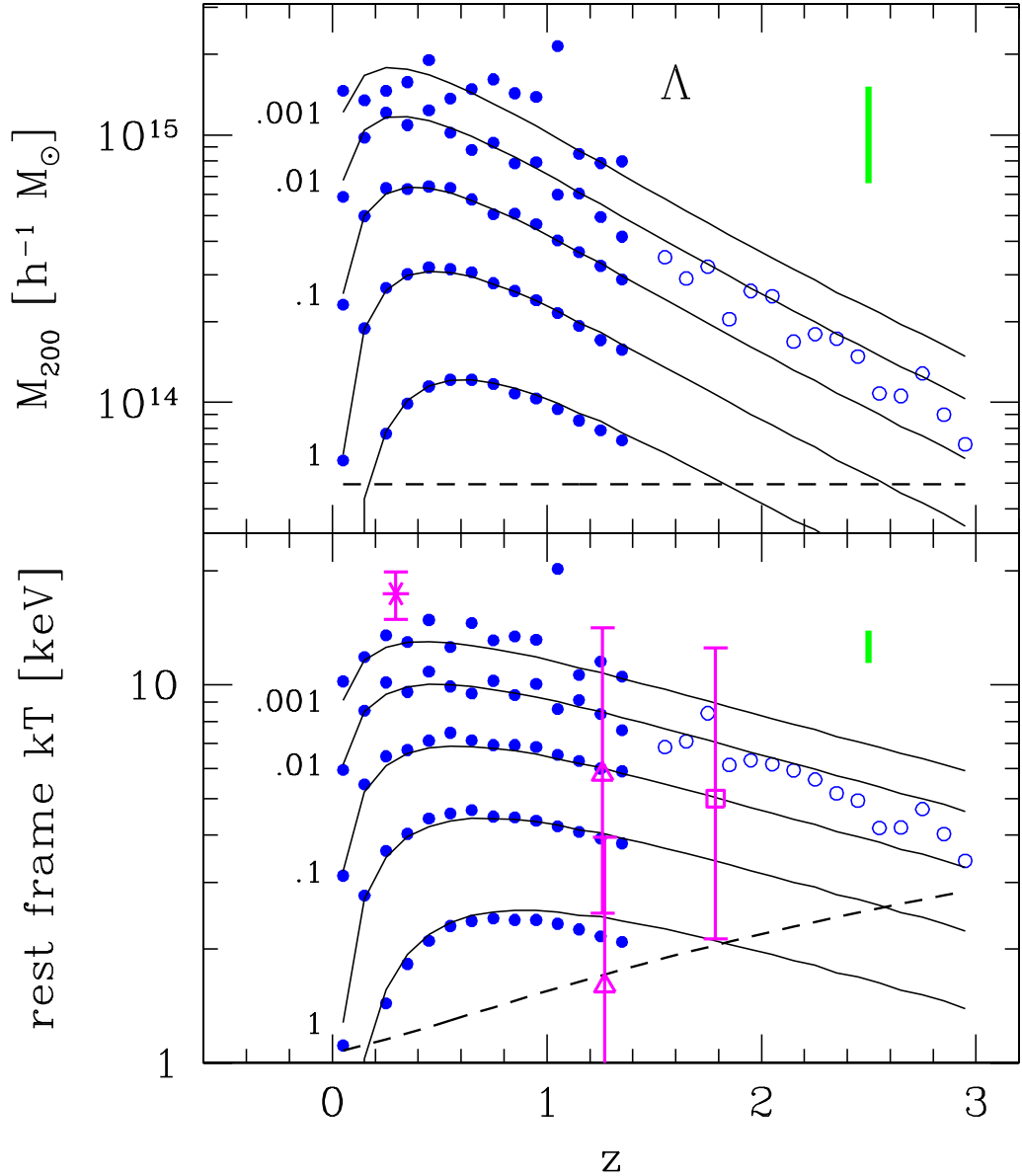} }
\vskip -0.5truecm
\caption{ Sky surface density characteristic mass (upper) and
temperature (lower) in the \lcdm\ model.  
Points from the 10,000 sq deg combined
octant surveys (filled circles) and the 1000 square degree extension
(open circles) show values 
above which the cluster sky surface density in the redshift interval 
$z-0.05$ to $z+0.05$ exceeds values $0.001$, $0.01$,
$0.1$, 1 and 10 per sq deg per unit redshift.   
Clusters at the lowest sky density
shown are the most massive or hottest in the particular 
redshift interval within the combined octants.  
Note the exceptional `monster' cluster at $z \se 1.04$.  
Solid lines are JMF expectations described in the text.  Vertical bars in
each panel denote the $90\%$ confidence uncertainty range in the 
absolute calibration of mass and temperature scales.  Open triangles
and squares plot extremes of the known \xray\
cluster population:  RX J0849+4452 at $ z \se 1.26$ and RX J0848+4453 at 
$ z \se 1.27$ (Stanford \etal\ 2000, triangles); 3C294 at $z \se 1.786$
(Fabian \etal\ 2001, square) and 1E~0657-56 at $ z \se 0.296$
(Tucker \etal\ 2001, asterisk).  
Dashed lines show a mass limit of 22 simulation particles 
(upper panel) and the virial temperature at that mass (lower).
\label{fig:MTnofz_lcdm}
}
\end{figure*}

\begin{figure*}
\epsfxsize=18.0cm 
\hbox{\hskip -1.0truecm  \epsfbox{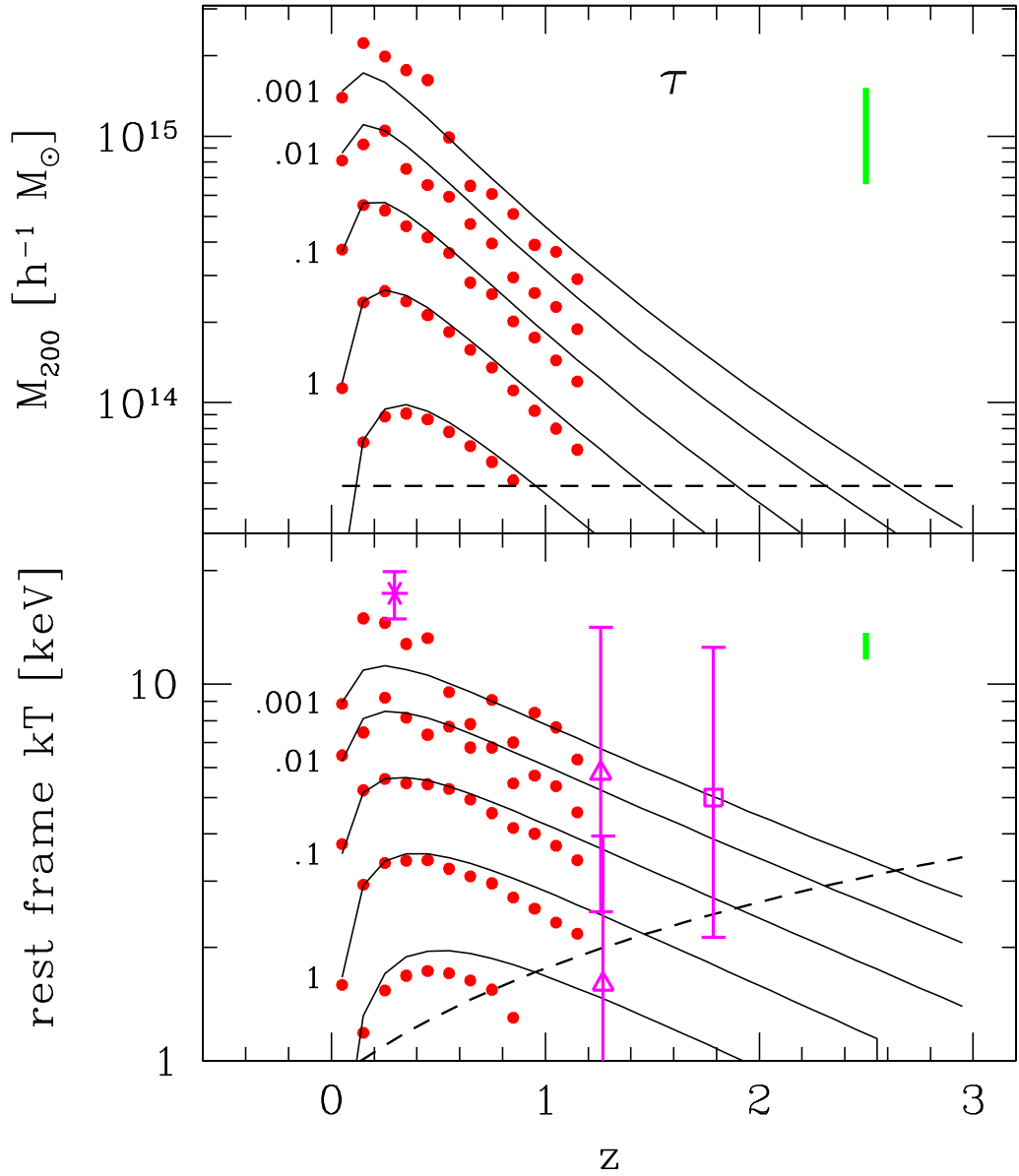} }
\vskip -0.5truecm
\caption{ Same as Figure~\ref{fig:MTnofz_lcdm}, but for the \tcdm\ model. 
\label{fig:MTnofz_tcdm} 
}
\end{figure*}

\clearpage


\appendix 
\section{A. Cluster Finding Details and Completeness Checks}

The cluster finding algorithm that produced the catalogs in this paper 
begins by generating a density estimate for each particle using the
distance to its eighth nearest neighbor (Casertano \& Hut 1985),
equivalent to Lagrangian filtering on a mass scale $2 \times 10^{13}
\hinv\msol$.  Sorting density values in decreasing order provides a
list of potential sites for cluster centers.  The list is pruned by
eliminating particles whose densities lie below the threshold $\Deltac
\rhocrit(z)$.  Beginning with the first member of the sorted list, a
sphere of radius $\rdeltac$ enclosing mass $\Mdeltac$ is defined about
that particle so that the enclosed density $\rho \sequiv 3 \Mdeltac /
4\pi {\rdeltac}^3 \se \Deltac \rho_c(z)$.  Particles lying within this
sphere are recorded as members of this group and are removed from the
list of potential cluster centers.  The process is repeated
sequentially, centering on the next available particle in the list ordered by
decreasing density, until the list is exhausted.  Particles may belong
to more than one group, but the center of a given group never 
lies within the spherical boundary of another group.

In analyzing SO(180) and mean SO(324) populations of the \tcdm\ and
\lcdm\ models, respectively, J01 noted a problem of incompleteness in
the SO cluster finding algorithm at particle counts $\lta 100$.
Resolution tests in J01 indicated that space densities of groups comprised
of 20 particles could be underestimated by $\ssim 30 \%$.  We
employ here an independent SO algorithm with improved  
completeness properties at small numbers of particles.  
Figure~\ref{fig:SO180_comp} compares the SO(180) and mean SO(324) 
abundance functions (for \tcdm\ and
\lcdm, respectively) at $z \se 0$ based on the new algorithm to the fits 
published in Appendix~B of J01. 

In the figure, the thick solid line show J01 functional fits while the
thin solid 
and dotted lines show discrete mass functions derived with the algorithm
employed here.  The dot-dashed line is the  discrete \tcdm\ mass
function derived by J01 using the previous SO algorithm.  The top
panel shows the percent deviation between the discrete sample
measurements and the fit expectations.   

For the \tcdm\ case, both the old and new algorithms compare well
against each other and against the fit above $\sims 10^{15} \hinv
\msol$.  At lower masses, the J01 algorithm displays an increasing
underestimate in number density with respect to the fit, approaching
a $30\%$ underestimate at the mass limit $5 \times 10^{13} \hinv
\msol$ used in this work.  The new SO algorithm (dotted) displays a
similar qualitative trend, but the underestimate is reduced to $\lta
10\%$ in amplitude.  A similar trend is seen for the new algorithm in
the \lcdm\ case where the number density lies $\sims 12\%$ lower than
the J01 fit expectations.  This analysis indicates that the 
amplitude $A$ derived from fitting the space density to the Jenkins
form, to equation~(\ref{eq:JMF}), may be biased low by $\sims 10 \%$
at masses below $\sims 10^{15} \hinv \msol$.  

A further check of resolution effects is made by directly comparing 
the HV mass function to one derived from smaller volumes with
improved mass resolution.  We do this for SO(200) clusters in the
\lcdm\ model at $z \se 0$, using data from the $256^3$ particle
simulation of a $239.5 \hinv \mpc$ region from Jenkins \etal\ (1998).
The new SO(200) algorithm is used to identify clusters in the same manner
as done in the HV simulation.  The cosmological parameters
for the models are the same, except for a slight difference in the
power spectrum used to generate the initial conditions, whereas the
particle mass in the $256^3$ particle simulation is a factor 32 times
smaller than that used in the HV computations.

Figure~\ref{fig:conv_test} shows the cumulative number of clusters
found in the $256^3$ particle simulation (dotted line) along with the
number expected based on the $z \se 0$ HV population (solid).
Vertical bars on the HV results show the range in
number derived from dividing the HV volume into 1728 independent cubes
of side $250 \hinv \mpc$ and rescaling the counts in each cube to a
$(239.5 \hinv \mpc)^3$ volume.  The inset shows the correlation between
counts above $5 \times 10^{13}$ and $3 \times 10^{14} \hinv \msol$
within the subvolumes.  The small--volume simulation result is
inconsistent with the HV distribution; the count
distribution shows agreement at the $98\%$ level at $3 \times 10^{14}
\hinv \msol$ but the $5 \times 10^{13}$ count is $\sims 0.5 \%$ higher
than the maximum of the HV distribution and $\ssim 20\%$ above the
mean.  

These findings, along with the slight discrepancy in predicted versus
measured counts in the \tcdm\ octant surveys (Figure~\ref{fig:Ncum_M}) lead
to a conservative estimate of the systematic theoretical uncertainty
in the  number density of clusters above $10^{14} \hinv
\msol$ to be $20\%$.   Future studies, in particular those which
cross-calibrate results for a particular cosmology modeled by 
different simulation teams, are needed to better assess the overall
accuracy of this model of the mass function.

\setcounter{figure}{0}

\begin{figure*}
\epsfxsize=20.0cm 
\hbox{\hskip -1.0truecm  \epsfbox{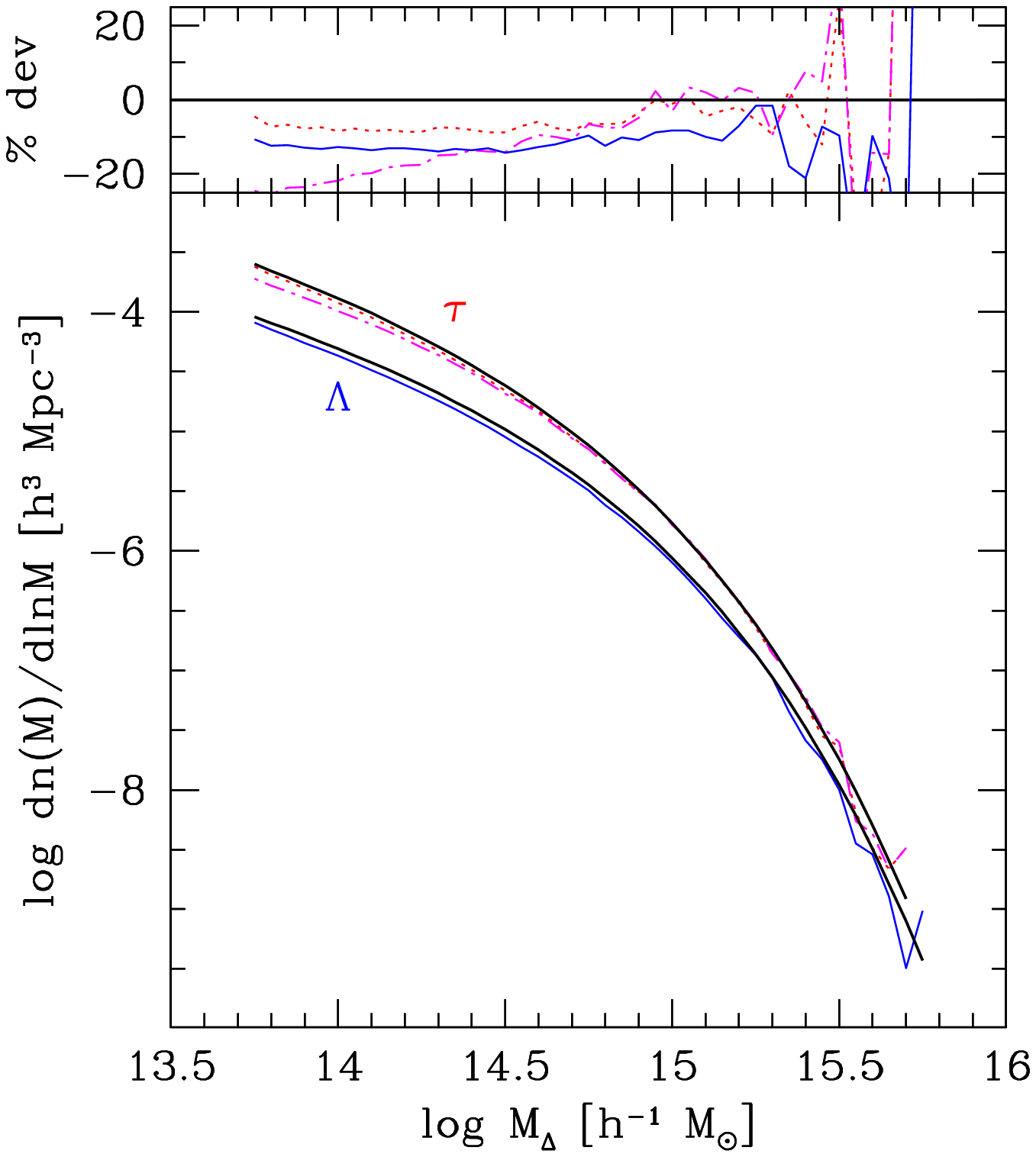} }
\vskip -1.0truecm
\caption{Differential SO mass functions at $z \se 0$ are compared to
the fits published in J01.  In the lower panel, thick lines are the
J01 expectations  
(from their Appendix B) for \tcdm\ at $\Deltac \se 180$ 
(upper curve) and \lcdm\ at mean $\Deltac \se 324$ (critical
$\Deltac \se 97.2$, lower curve).  The \lcdm\ 
simulation data are shown by the slightly jagged solid curve.  Two
simulation results are shown for \tcdm ---  the dot-dashed line
reflects the SO algorithm used by J01 and the dotted line shows
results of the algorithm used in this work.  
The upper panel displays the percent deviation in number density 
between the HV simulation data and the J01 model fits. 
\label{fig:SO180_comp}
}
\end{figure*}

\clearpage

\begin{figure*}
\epsfxsize=20.0cm 
\hbox{\hskip -1.0truecm  \epsfbox{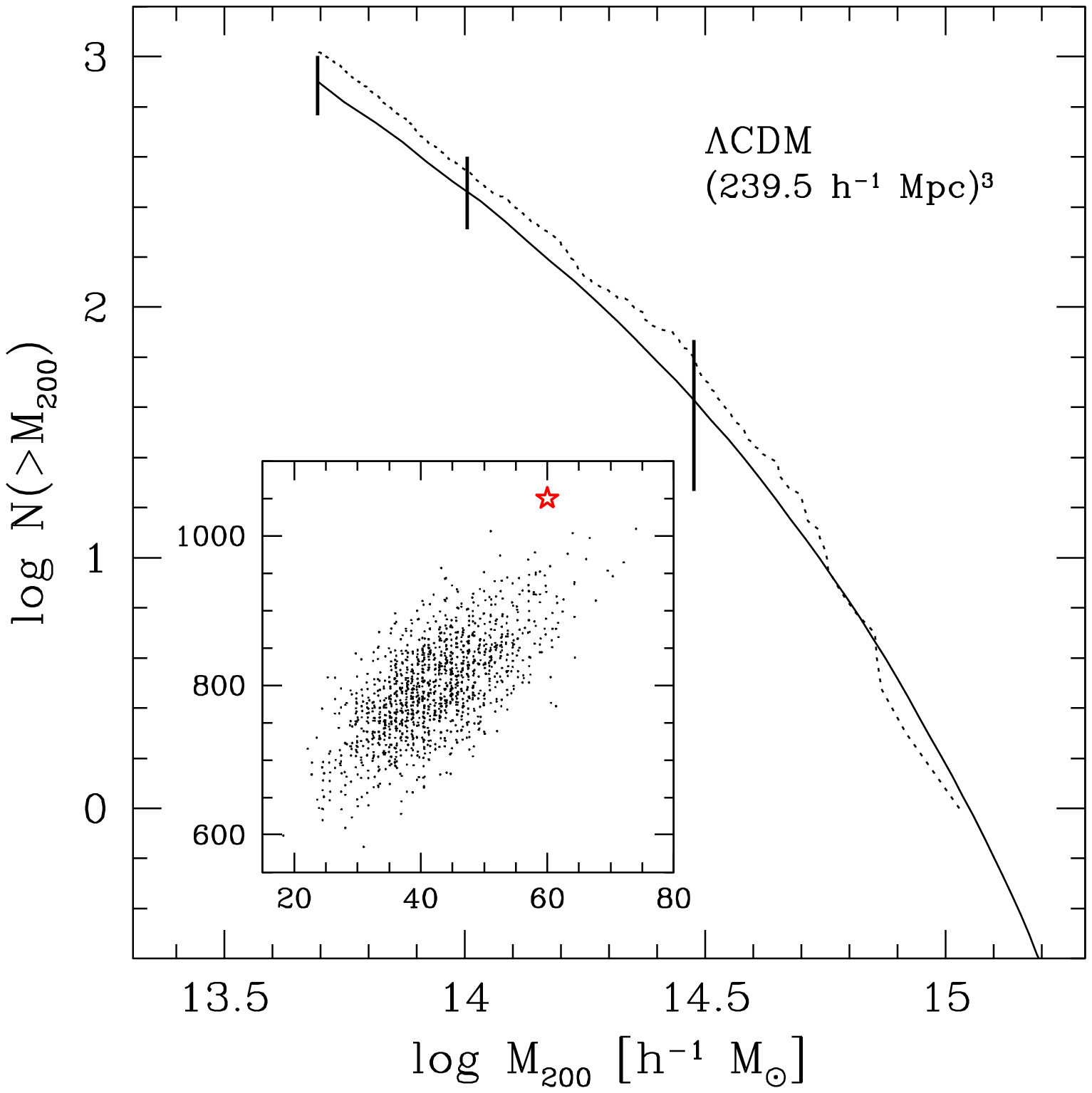} }
\vskip -1.0truecm
\caption{
The cumulative number of clusters within the volume indicated, scaled from 
the full $z \se 0$ HV simulation (solid line) and from a single
realization of a $(239.5 \hinv\mpc)^3$ volume (Jenkins \etal\ 1998;
dotted line) for the \lcdm\ cosmology.   
Vertical lines show the entire range of counts 
above masses $5 \times 10^{13}$, $10^{14}$ and $3 \times 10^{14}\hinv
\msol$ derived from subsampling 1728 cubic sub-volumes of side $250 \hinv
\mpc$ within the HV realization and scaling to $239.5 \hinv\mpc$.   
The inset plots the correlation of counts above $3 \times
10^{14}$ ($x$-axis) and $5 \times 10^{13}$ ($y$-axis) derived from 
the sub-volumes.  
The star indicates the Jenkins \etal\ (1998) values. 
\label{fig:conv_test}
}
\end{figure*}

\clearpage


\section{B. Mass scale renormalization}\label{subsec:massftnX} 

The mass scale of clusters at fixed space density is uncertain, both
theoretically and empirically, for reasons discussed in the
opening section.  The lack of a uniquely defined scale motivates
a model that would transform the JMF fit parameters
derived in \S\ref{subsec:massftn} to values appropriate for a
redefined mass scale.  As an example, we develop here a model to
estimate the SO mass function fit parameters for threshold values
$\Delta \ne 200 $.  The method is similar to that used in 
\S\ref{subsec:MassSam} to effectively vary $\sigma_8$ within the
discrete cluster samples.  

We have chosen a convention in which a spherical density threshold 
$\Delta \se 200$ defines cluster masses $M$.  A choice $\Delta \ne
200$ would lead to a new mass $M^\prime$ for each cluster related to the
original by some factor
\begin{equation} \label{eq:Mprime}
M^\prime \ = \ M \, e^\mu .
\end{equation}
The space density of a set of {\em disjoint} clusters is fixed,
implying
\begin{equation} \label{eq:nequal}
n(M^\prime) \, \dln \! M^\prime \ = \ n(M) \, \dln \! M.
\end{equation}
This condition, with equation~(\ref{eq:Mprime}), when used with the
space density, equation~(\ref{eq:JMF}), leads to a relation between
JMF parameters
\begin{eqnarray}\label{eq:JMFprime}
& A^\prime \ = \ A \, e^\mu \nonumber \\ 
B^\prime & \ = \ B - \alpha_{\rm eff}(M) \, \mu
\end{eqnarray}
to first order in $\mu$.

The factor $\mu$ can be calculated from an assumed mass profile. 
We use the form introduced by Navarro \etal\ (1996; hereafter NFW),  
\begin{equation} \label{eq:NFWmass}
M(x) \ = \ \frac{800\pi}{3} \, \rhocrit \, \rtwoh^3 \biggl[
 \frac{\ln(1+cx) - cx/(1+cx)}{\ln(1+c) - c/(1+c)} \biggr] 
\end{equation}
where $x \se r/\rtwoh$ is a scaled radius and $c$ is a concentration
parameter.  N--body simulations suggest $c \ssimeq 5$ at masses near
$10^{15} \hinv \mpc$ for the cosmologies studied here (NFW, Frenk
\etal\ 1999; Jing 2000; Bullock \etal\ 2001). 

Figure~\ref{fig:ndiff_M_X} shows the results of applying
equation~(\ref{eq:JMFprime}) to critical contrasts $\Delta \se 97.2$
and $500$ for the \lcdm\ model, assuming $c \se 5$. 
The logarithmic shifts in mass scale are 
$\mu \se 0.093$ and $-0.141$, respectively.  The
agreement between the predicted and measured values is quite good.  At
$\Delta \se 97.2$ (equivalent to the mean contrast of $324$ used by
J01), the bin-averaged mean fractional error and dispersion (for bins
with $10\%$ or less Poisson uncertainty) are only $4.1\%$ and
$3.8\%$.  At $\Delta \se 500$, the mean is $-14.5\%$ and dispersion
$4.9\%$.  

\clearpage

\setcounter{figure}{0}

\begin{figure*}
\epsfxsize=20.0cm 
\hbox{\hskip -1.0truecm  \epsfbox{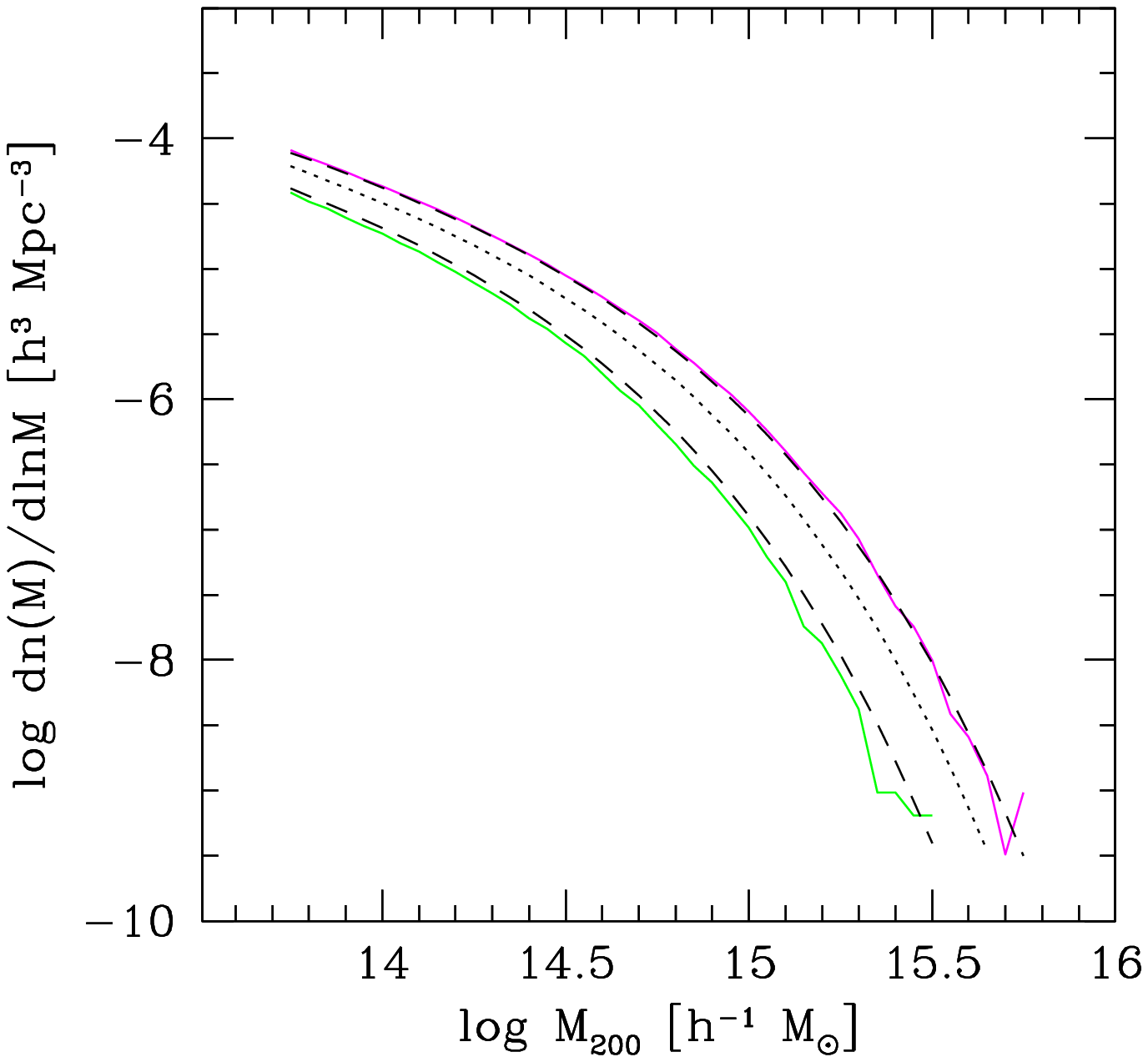} }
\vskip -1.0truecm
\caption{
The critical SO(500) and SO(97.2) mass functions (lower and upper,
respectively) at $z \se 0$ for \lcdm.  Solid lines are measured from
the HV simulation while dashed lines are predictions
based on rescaling the SO(200) JMF fit (dotted line), assuming an 
NFW profile with concentration parameter $c \se 5$.  
\label{fig:ndiff_M_X}
}
\end{figure*}

\clearpage


\section{Cluster catalogs}\label{subsec:etables} 

The SO(200) cluster catalogs derived from the sky survey and $z \se 0$
snapshot outputs of the simulations are included here as 
electronic tables.  In the print edition, Tables~6 through 17 provide 
a truncated listing of the ten most massive clusters for each survey.
Electronic versions list all clusters resolved above a
mass limit of 22 particles ($5 \times 10^{13} \hinv \msol$), counts 
of which are listed in Table~\ref{tab:counts}.  
Tables~6 through 11 are \lcdm\ simulation catalogs from the $z\se 0$
snapshot, the combined PO and XW sky surveys, and the NO, VS, MS and DW
sky surveys, respectively.  Tables~12 through 17 provide the same for
\tcdm, with the exception that there is no extended wedge associated
with the PO survey of this model (see Table~\ref{tab:surveys}).

Column entries give the mass $\Mtwoh$ (in $10^{15}\hinv\msol$),
redshift (for sky survey data) derived from Hubble flow and the radial
peculiar velocity, one-dimensional velocity dispersion $\sigmadm$
determined from a three-dimensional average (in $\kms$),  
position in comoving coordinates (in $\hinv$ Gpc), and
peculiar velocity (in $\kms$).  
The position is defined as the location of the particle having the
smallest distance to its eighth neighboring particle (see Appendix~A), 
and the peculiar velocity is defined by the mass-weighted mean within
$\rtwoh$.   

\clearpage

\begin{deluxetable}{cccccccc}
\tablewidth{0pt} 
\tablecaption{Clusters in the \lcdm\ $z=0$ snapshot
survey.\label{tab:lcdm_z0}}  
\tablehead{ 
\colhead{$\Mtwoh$} & \colhead{$\sigma$} & \colhead{$x$} &   
\colhead{$y$} & \colhead{$z$} & \colhead{$v_x$} & \colhead{$v_y$} &   
\colhead{$v_z$} \\
\colhead{$10^{15} \hinv\msol$} & \colhead{$\kms$} &   
\colhead{$\hinv \gpc$} & \colhead{$\hinv \gpc$} & \colhead{$\hinv \gpc$} &   
\colhead{$\kms$} & \colhead{$\kms$} & \colhead{$\kms$} } 
\startdata 
4.7745 & 1795 & 2.27792 & 0.44894 & 0.54195 &   -90 &   153 &    -6  \cr    
4.7475 & 1866 & 2.03410 & 0.91683 & 0.44397 &   269 &    46 &   727  \cr 
4.2840 & 1781 & 1.74103 & 2.84553 & 1.40265 &    87 &   -52 &  -154  \cr       
3.9645 & 1632 & 0.96734 & 0.65102 & 2.32061 &    63 &   312 &  -282  \cr       
3.8227 & 1891 & 1.75213 & 0.72151 & 1.80426 &   380 &   -90 &  -279  \cr       
3.4920 & 1665 & 1.50312 & 1.77336 & 2.19206 &   -26 &     1 &   -17  \cr       
3.3772 & 1531 & 1.21292 & 2.12486 & 0.27272 &  -434 &  -228 &  -821  \cr       
3.3592 & 1583 & 2.50579 & 2.18682 & 0.31021 &  -183 &   588 &   174  \cr       
3.3187 & 1519 & 2.07928 & 2.97238 & 1.91753 &  -532 &  -531 &  -156  \cr       
3.2602 & 1566 & 0.38232 & 1.70524 & 0.62298 &  -228 &   257 &    98  \cr       
\enddata
\end{deluxetable}

\medskip

\begin{deluxetable}{ccccccccc}
\tablewidth{0pt} 
\tablecaption{Clusters in the \lcdm\ PO+XW sky survey\label{tab:lcdm_POX}}  
\tablehead{ 
\colhead{$\Mtwoh$} & \colhead{redshift} & 
\colhead{$\sigma$} & \colhead{$x$} &   
\colhead{$y$} & \colhead{$z$} & \colhead{$v_x$} & \colhead{$v_y$} &   
\colhead{$v_z$} \\
\colhead{$10^{15} \hinv\msol$} & & \colhead{$\kms$} &   
\colhead{$\hinv \gpc$} & \colhead{$\hinv \gpc$} & \colhead{$\hinv \gpc$} &   
\colhead{$\kms$} & \colhead{$\kms$} & \colhead{$\kms$} } 
\startdata 
2.1465 & 1.04210 & 1740 & 2.27834 & 0.44795 & 0.54248  & 229  &  319 &  -99\cr 
1.9035 & 0.49167 & 1490 & 0.48208 & 1.20790 & 0.20341  & 287  & -806 & -725\cr 
1.8337 & 0.48731 & 1473 & 1.25433 & 0.30248 & 0.12419  & 300  & -692 &   13\cr 
1.6245 & 0.48492 & 1440 & 1.25607 & 0.30110 & 0.12310  & -520 &  45  &  327\cr 
1.6132 & 0.74772 & 1389 & 0.38256 & 1.70412 & 0.62232  & 171  &  -58 &  192\cr 
1.5052 & 0.48217 & 1339 & 1.00142 & 0.47311 & 0.64768  & -177 & 670  &  268\cr 
1.4850 & 0.63925 & 1309 & 1.21259 & 0.74437 & 0.79747  &  -4  &   -6 &  406\cr 
1.4265 & 0.33432 & 1338 & 0.15125 & 0.88516 & 0.24984  & -533 & -287 & -210\cr 
1.3927 & 0.26386 & 1273 & 0.62241 & 0.37143 & 0.16475  & 283  &   50 & -310\cr 
1.3860 & 0.95680 & 1403 & 1.28228 & 0.28511 & 1.81830  & -107 & -259 &   41\cr 
\enddata
\end{deluxetable}

\medskip

\begin{deluxetable}{ccccccccc}
\tablewidth{0pt} 
\tablecaption{Clusters in the \lcdm\ NO sky survey\label{tab:lcdm_NO}}  
\tablehead{ 
\colhead{$\Mtwoh$} & \colhead{redshift} & 
\colhead{$\sigma$} & \colhead{$x$} &   
\colhead{$y$} & \colhead{$z$} & \colhead{$v_x$} & \colhead{$v_y$} &   
\colhead{$v_z$} \\
\colhead{$10^{15} \hinv\msol$} & & \colhead{$\kms$} &   
\colhead{$\hinv \gpc$} & \colhead{$\hinv \gpc$} & \colhead{$\hinv \gpc$} &   
\colhead{$\kms$} & \colhead{$\kms$} & \colhead{$\kms$} } 
\startdata 
1.8720 & 0.47097 & 1357 & 0.78598 & 0.89484 & 0.40376 &  -13 & 365  & -193\cr
1.6762 & 0.45872 & 1332 & 0.25973 & 1.14905 & 0.34105 & -128 &  508 & -150\cr
1.5750 & 0.39013 & 1393 & 0.50888 & 0.89409 & 0.29059 & -106 &  -92 & -379\cr
1.4985 & 0.36176 & 1308 & 0.54647 & 0.73774 & 0.40355 & -467 & -203 & -432\cr
1.4625 & 0.09295 & 1230 & 0.11250 & 0.20762 & 0.13595 &  114 &   63 &   68\cr
1.4602 & 0.21513 & 1254 & 0.13535 & 0.45678 & 0.38800 & -624 &  336 &  -46\cr
1.4422 & 0.23629 & 1209 & 0.21785 & 0.32614 & 0.54597 &   41 &  299 & -195\cr
1.4310 & 0.84337 & 1304 & 1.25924 & 0.15777 & 1.59600 & -158 & -124 &   27\cr
1.3612 & 0.25340 & 1225 & 0.35217 & 0.57399 & 0.23736 &  -38 &  458 &  -21\cr
1.3567 & 0.24869 & 1277 & 0.68191 & 0.11578 & 0.14755 &   46 &  -19 &-1027\cr
\enddata
\end{deluxetable}

\clearpage

\begin{deluxetable}{ccccccccc}
\tablewidth{0pt} 
\tablecaption{Clusters in the \lcdm\ VS sky survey\label{tab:lcdm_VS}}  
\tablehead{ 
\colhead{$\Mtwoh$} & \colhead{redshift} & 
\colhead{$\sigma$} & \colhead{$x$} &   
\colhead{$y$} & \colhead{$z$} & \colhead{$v_x$} & \colhead{$v_y$} &   
\colhead{$v_z$} \\
\colhead{$10^{15} \hinv\msol$} & & \colhead{$\kms$} &   
\colhead{$\hinv \gpc$} & \colhead{$\hinv \gpc$} & \colhead{$\hinv \gpc$} &   
\colhead{$\kms$} & \colhead{$\kms$} & \colhead{$\kms$} } 
\startdata 
3.7710 & 0.36705 & 1849 & 0.77855 & 1.94855 & 2.04204 & -179 &  109 &  -38\cr
2.3985 & 0.12317 & 1427 & 1.66806 & 1.18289 & 1.50177 & -100 & -204 & -269\cr
2.1847 & 0.31494 & 1513 & 1.16682 & 1.78942 & 0.74591 & -276 &   55 & -309\cr
2.1352 & 0.22786 & 1510 & 1.83355 & 0.95563 & 1.60533 &  228  & -71 &  352\cr
2.0407 & 0.25998 & 1427 & 1.69566 & 2.12671 & 1.17188 & -529  &  84 & -410\cr
2.0385 & 0.27780 & 1536 & 1.14664 & 2.02703 & 1.04228 &  -73 &   94 &  220\cr
1.9957 & 0.21552 & 1393 & 1.20062 & 1.88904 & 1.87100 &   60 &  546 & -365\cr
1.9282 & 0.17104 & 1473 & 1.83406 & 1.48791 & 1.14040 &  367 & -434 &  -99\cr
1.9260 & 0.38687 & 1515 & 1.99907 & 1.19847 & 2.37855 &  294 & -141 &  207\cr
1.9237 & 0.21232 & 1425 & 1.62686 & 0.92821 & 1.34563 &  203 &  -10 & -516\cr
\enddata
\end{deluxetable}

\medskip
\begin{deluxetable}{ccccccccc}
\tablewidth{0pt} 
\tablecaption{Clusters in the \lcdm\ MS sky survey\label{tab:lcdm_MS}}  
\tablehead{ 
\colhead{$\Mtwoh$} & \colhead{redshift} & 
\colhead{$\sigma$} & \colhead{$x$} &   
\colhead{$y$} & \colhead{$z$} & \colhead{$v_x$} & \colhead{$v_y$} &   
\colhead{$v_z$} \\
\colhead{$10^{15} \hinv\msol$} & & \colhead{$\kms$} &   
\colhead{$\hinv \gpc$} & \colhead{$\hinv \gpc$} & \colhead{$\hinv \gpc$} &   
\colhead{$\kms$} & \colhead{$\kms$} & \colhead{$\kms$} } 
\startdata 
3.0622 & 0.12208 & 1765 & 1.61387 & 1.63047 & 1.81743 & -135 &  -43 & -404\cr
2.9452 & 0.20045 & 1583 & 1.13968 & 1.94410 & 1.59070 &   37 & -315 & -414\cr
2.7225 & 0.21132 & 1724 & 1.77946 & 1.14258 & 1.88827 &   41 & -144 &  941\cr
2.6640 & 0.26329 & 1622 & 1.50191 & 1.77218 & 2.19369 &  434 &  317 & -245\cr
2.3692 & 0.39004 & 1536 & 1.21250 & 0.74412 & 0.79839 & -210 &    0 &  684\cr
2.2320 & 0.53759 & 1482 & 1.20256 & 2.12932 & 0.26569 &  636 & -375 &  138\cr
2.1690 & 0.48039 & 1534 & 1.28226 & 0.28481 & 1.81857 &   55 & -597 &  114\cr
2.0925 & 0.54800 & 1465 & 0.38300 & 1.70434 & 0.62229 & -215 &   29 &  239\cr
2.0272 & 0.47799 & 1414 & 1.16678 & 1.69964 & 0.28919 &  308 &  159 & -474\cr
1.9147 & 0.41372 & 1393 & 0.49259 & 1.31877 & 1.03747 &   86 &  548 & -522\cr
\enddata
\end{deluxetable}

\medskip
\begin{deluxetable}{ccccccccc}
\tablewidth{0pt} 
\tablecaption{Clusters in the \lcdm\ DW sky survey\label{tab:lcdm_DW}}  
\tablehead{ 
\colhead{$\Mtwoh$} & \colhead{redshift} & 
\colhead{$\sigma$} & \colhead{$x$} &   
\colhead{$y$} & \colhead{$z$} & \colhead{$v_x$} & \colhead{$v_y$} &   
\colhead{$v_z$} \\
\colhead{$10^{15} \hinv\msol$} & & \colhead{$\kms$} &   
\colhead{$\hinv \gpc$} & \colhead{$\hinv \gpc$} & \colhead{$\hinv \gpc$} &   
\colhead{$\kms$} & \colhead{$\kms$} & \colhead{$\kms$} } 
\startdata 
1.2667 & 0.35165 & 1194 & 0.52955 & 0.51554 & 0.62439 &  236 &  159 &  160\cr
0.7425 & 0.63063 & 1020 & 0.89636 & 1.04111 & 0.84907 &   58 &  -57 &   21\cr
0.7335 & 0.54100 & 1078 & 0.83017 & 0.88976 & 0.74354 & -346 & -269 & -328\cr
0.6952 & 0.86474 & 1077 & 1.26522 & 1.12729 & 1.20595 & -417 & -122 &  217\cr
0.6795 & 0.23262 &  883 & 0.42833 & 0.36816 & 0.34631 &  184 &  -40 & -201\cr
0.6277 & 0.49817 &  968 & 0.79520 & 0.73876 & 0.75309 &   24 &  340 & -103\cr
0.6210 & 0.61920 & 1075 & 0.87578 & 0.91529 & 0.96280 & -130 & -498 &  540\cr
0.5872 & 0.16708 &  833 & 0.29741 & 0.28511 & 0.25289 &  406 & -169 & -255\cr
0.5557 & 0.24148 &  766 & 0.37255 & 0.44021 & 0.38042 &  331 & -714 & -279\cr
0.5512 & 0.52327 &  998 & 0.76163 & 0.87778 & 0.75294 & -389 & -179 & -160\cr
\enddata
\end{deluxetable}

\clearpage

\begin{deluxetable}{cccccccc}
\tablewidth{0pt} 
\tablecaption{Clusters in the \tcdm\ $z=0$ snapshot
survey.\label{tab:tcdm_z0}}  
\tablehead{ 
\colhead{$\Mtwoh$} & \colhead{$\sigma$} & \colhead{$x$} &   
\colhead{$y$} & \colhead{$z$} & \colhead{$v_x$} & \colhead{$v_y$} &   
\colhead{$v_z$} \\
\colhead{$10^{15} \hinv\msol$} & \colhead{$\kms$} &   
\colhead{$\hinv \gpc$} & \colhead{$\hinv \gpc$} & \colhead{$\hinv \gpc$} &   
\colhead{$\kms$} & \colhead{$\kms$} & \colhead{$\kms$} } 
\startdata 
4.8840 & 1912 & 1.60532 & 0.11171 & 1.05673 & -273 &  241 & -141 \cr       
4.8707 & 1946 & 1.76731 & 0.09581 & 0.98359 &  206 &  229 & -728  \cr      
4.7597 & 1907 & 0.46288 & 1.83931 & 0.73075 &  402 &  598 & -359  \cr      
4.3290 & 1787 & 1.06007 & 1.25750 & 0.15836 &  195 & -102 &  160  \cr      
4.3246 & 1816 & 0.62941 & 0.32143 & 0.59557 & -120 &  462 &  -43   \cr     
4.1270 & 1655 & 0.18549 & 1.04133 & 1.38113 &  125 & -584 & -549   \cr     
4.0448 & 1820 & 1.40794 & 0.55788 & 0.85002 & -596 &  282 & -236   \cr     
3.8473 & 1734 & 0.24495 & 0.63610 & 0.51995 &  169 &  473 &  235   \cr     
3.5209 & 1688 & 1.59185 & 0.02998 & 1.51849 &  -21 &-1044 &  -78   \cr     
3.5187 & 1569 & 0.54769 & 0.03614 & 1.24131 & -125 &  524 &  -62   \cr     
\enddata
\end{deluxetable}

\medskip

\begin{deluxetable}{ccccccccc}
\tablewidth{0pt} 
\tablecaption{Clusters in the \tcdm\ PO sky survey\label{tab:tcdm_PO}}  
\tablehead{ 
\colhead{$\Mtwoh$} & \colhead{redshift} & 
\colhead{$\sigma$} & \colhead{$x$} &   
\colhead{$y$} & \colhead{$z$} & \colhead{$v_x$} & \colhead{$v_y$} &   
\colhead{$v_z$} \\
\colhead{$10^{15} \hinv\msol$} & & \colhead{$\kms$} &   
\colhead{$\hinv \gpc$} & \colhead{$\hinv \gpc$} & \colhead{$\hinv \gpc$} &   
\colhead{$\kms$} & \colhead{$\kms$} & \colhead{$\kms$} } 
\startdata 
2.2333 & 0.15351 & 1635 & 0.02751 & 0.34796 & 0.21763 & -155 &  234 &   84\cr
1.9869 & 0.20570 & 1568 & 0.10704 & 0.52032 & 0.07637 & -219 &  -55 &  -49\cr
1.7671 & 0.39740 & 1512 & 0.62955 & 0.31885 & 0.59612 & -123 &  640 & -105\cr
1.4208 & 0.26994 & 1345 & 0.18684 & 0.32731 & 0.55685 &  934 & -132 &  219\cr
1.3942 & 0.01660 & 1259 & 0.02224 & 0.05106 & 0.01156 & -742 & -559 &   77\cr
1.2920 & 0.29180 & 1311 & 0.09007 & 0.52090 & 0.49083 & -153 &  -70 &   39\cr
1.2898 & 0.08068 & 1246 & 0.21347 & 0.08099 & 0.01491 &  122 & -443 & -133\cr
1.2521 & 0.27828 & 1341 & 0.58237 & 0.37616 & 0.02971 & -206 &  150 &   35\cr
1.1500 & 0.25740 & 1307 & 0.43210 & 0.35366 & 0.33809 & -507 & -135 &   33\cr
1.0811 & 0.19157 & 1201 & 0.42661 & 0.26000 & 0.09128 & -478 & -105 & -160\cr
\enddata
\end{deluxetable}

\medskip

\begin{deluxetable}{ccccccccc}
\tablewidth{0pt} 
\tablecaption{Clusters in the \tcdm\ NO sky survey\label{tab:tcdm_NO}}  
\tablehead{ 
\colhead{$\Mtwoh$} & \colhead{redshift} & 
\colhead{$\sigma$} & \colhead{$x$} &   
\colhead{$y$} & \colhead{$z$} & \colhead{$v_x$} & \colhead{$v_y$} &   
\colhead{$v_z$} \\
\colhead{$10^{15} \hinv\msol$} & & \colhead{$\kms$} &   
\colhead{$\hinv \gpc$} & \colhead{$\hinv \gpc$} & \colhead{$\hinv \gpc$} &   
\colhead{$\kms$} & \colhead{$\kms$} & \colhead{$\kms$} } 
\startdata 
1.8781 & 0.16574 & 1489 & 0.08750 & 0.34930 & 0.25921 &  532 &  -14 & -310\cr
1.6650 & 0.23534 & 1613 & 0.17937 & 0.04762 & 0.57891 & -102 & -429 & -661\cr
1.6228 & 0.40003 & 1540 & 0.64304 & 0.36559 & 0.55891 &   -6 &  475 &   93\cr
1.5873 & 0.23482 & 1433 & 0.19753 & 0.03359 & 0.57222 & -426 & -150 & -519\cr
1.4119 & 0.13676 & 1456 & 0.15879 & 0.33140 & 0.02784 &   70 &  401 &  403\cr
1.2388 & 0.08092 & 1148 & 0.13108 & 0.17990 & 0.05517 &  -93 &  166 & -475\cr
1.1566 & 0.14827 & 1301 & 0.23406 & 0.24841 & 0.20322 &  599 &  505 & -569\cr
1.1433 & 0.14885 & 1309 & 0.15585 & 0.14787 & 0.33972 &   36 & -297 &  142\cr
1.1255 & 0.13588 & 1194 & 0.25388 & 0.00900 & 0.26156 &  275 &  118 &  572\cr
1.0922 & 0.05791 & 1091 & 0.03287 & 0.13919 & 0.07741  &  23 &  353 &  197\cr
\enddata
\end{deluxetable}

\clearpage

\begin{deluxetable}{ccccccccc}
\tablewidth{0pt} 
\tablecaption{Clusters in the \tcdm\ VS sky survey\label{tab:tcdm_VS}}  
\tablehead{ 
\colhead{$\Mtwoh$} & \colhead{redshift} & 
\colhead{$\sigma$} & \colhead{$x$} &   
\colhead{$y$} & \colhead{$z$} & \colhead{$v_x$} & \colhead{$v_y$} &   
\colhead{$v_z$} \\
\colhead{$10^{15} \hinv\msol$} & & \colhead{$\kms$} &   
\colhead{$\hinv \gpc$} & \colhead{$\hinv \gpc$} & \colhead{$\hinv \gpc$} &   
\colhead{$\kms$} & \colhead{$\kms$} & \colhead{$\kms$} } 
\startdata 
2.5530 & 0.08805 & 1595 & 1.23821 & 1.06196 & 0.96259 &  -18 & -331 &   78\cr
2.5108 & 0.15568 & 1622 & 1.41088 & 0.90971 & 0.98357 & -261 &  -24 & -177\cr
2.2333 & 0.15351 & 1635 & 1.02751 & 1.34795 & 1.21762 & -155 &  234 &   84\cr
1.9869 & 0.20570 & 1568 & 1.10704 & 1.52032 & 1.07637 & -218 &  -55 &  -49\cr
1.9358 & 0.17072 & 1389 & 0.96303 & 1.00973 & 0.54585 &  -48 &  394 &  126\cr
1.8937 & 0.16180 & 1570 & 0.57437 & 0.89833 & 1.00483 &  410 &  211 & -184\cr
1.8781 & 0.16574 & 1489 & 0.91249 & 0.65070 & 0.74078 & -532 &   14 &  311\cr
1.7671 & 0.39740 & 1513 & 1.62955 & 1.31885 & 1.59612 & -120 &  638 & -106\cr
1.7560 & 0.37118 & 1547 & 1.46201 & 0.83718 & 1.73270 &  127 &  852 & -638\cr
1.7405 & 0.24550 & 1491 & 0.92007 & 1.16642 & 1.59436 & -310 &  137 &   81\cr
\enddata
\end{deluxetable}

\medskip
\begin{deluxetable}{ccccccccc}
\tablewidth{0pt} 
\tablecaption{Clusters in the \tcdm\ MS sky survey\label{tab:tcdm_MS}}  
\tablehead{ 
\colhead{$\Mtwoh$} & \colhead{redshift} & 
\colhead{$\sigma$} & \colhead{$x$} &   
\colhead{$y$} & \colhead{$z$} & \colhead{$v_x$} & \colhead{$v_y$} &   
\colhead{$v_z$} \\
\colhead{$10^{15} \hinv\msol$} & & \colhead{$\kms$} &   
\colhead{$\hinv \gpc$} & \colhead{$\hinv \gpc$} & \colhead{$\hinv \gpc$} &   
\colhead{$\kms$} & \colhead{$\kms$} & \colhead{$\kms$} } 
\startdata 
2.8616 & 0.08778 & 1619 & 1.12538 & 1.09734 & 0.80682 & -359 &   49 &  176\cr
2.6662 & 0.09450 & 1569 & 0.83169 & 1.09487 & 0.82472 & -444 &   32 & -178\cr
2.0446 & 0.23637 & 1552 & 0.69721 & 1.13044 & 1.50301 &  161 & -325 &  514\cr
2.0291 & 0.23655 & 1533 & 1.40187 & 1.26906 & 1.36378 & -248 & -415 &  427\cr
1.9625 & 0.36864 & 1529 & 0.62959 & 0.31918 & 0.59623 &  -93 &  539 & -159\cr
1.8448 & 0.37457 & 1541 & 1.06066 & 1.25588 & 0.15810 &   20 &  315 &   53\cr
1.7960 & 0.06498 & 1418 & 1.04302 & 1.16623 & 0.93822 &  566 &  268 &   62\cr
1.7804 & 0.35602 & 1629 & 1.35688 & 1.63401 & 1.44111 &    9 & -443 & -117\cr
1.7649 & 0.13530 & 1456 & 0.70869 & 1.17978 & 0.85280 &  571 &  211 &  141\cr
1.7027 & 0.13167 & 1369 & 0.94757 & 0.90342 & 0.65712 &  205 &   45 &  -20\cr
\enddata
\end{deluxetable}

\medskip
\begin{deluxetable}{ccccccccc}
\tablewidth{0pt} 
\tablecaption{Clusters in the \tcdm\ DW sky survey\label{tab:tcdm_DW}}  
\tablehead{ 
\colhead{$\Mtwoh$} & \colhead{redshift} & 
\colhead{$\sigma$} & \colhead{$x$} &   
\colhead{$y$} & \colhead{$z$} & \colhead{$v_x$} & \colhead{$v_y$} &   
\colhead{$v_z$} \\
\colhead{$10^{15} \hinv\msol$} & & \colhead{$\kms$} &   
\colhead{$\hinv \gpc$} & \colhead{$\hinv \gpc$} & \colhead{$\hinv \gpc$} &   
\colhead{$\kms$} & \colhead{$\kms$} & \colhead{$\kms$} } 
\startdata 
0.5683 & 0.30774 &  993 & 0.41763 & 0.43087 & 0.45194 &  292 &   69 &   54\cr
0.5461 & 0.11396 &  876 & 0.19790 & 0.17877 & 0.17220 &  -94 & -258 &  -63\cr
0.4951 & 0.46020 & 1001 & 0.65398 & 0.57027 & 0.56948 & -842 &  261 &    4\cr
0.4484 & 0.32176 &  890 & 0.47050 & 0.45774 & 0.42240 &  210 &  -75 &  -31\cr
0.4285 & 0.28315 &  837 & 0.39105 & 0.38532 & 0.44474 &   37 & -578 & -129\cr
0.4107 & 0.31961 &  986 & 0.47673 & 0.44308 & 0.42215 & -146 &  375 &   11\cr
0.3929 & 0.18746 &  818 & 0.28105 & 0.30862 & 0.25462 &  351 &  600 &  -69\cr
0.3441 & 0.17978 &  833 & 0.28576 & 0.26987 & 0.26635 &  -27 &   34 &  234\cr
0.3419 & 0.19586 &  796 & 0.29392 & 0.31370 & 0.27936 &  328 &  150 & -391\cr
0.3397 & 0.51005 &  894 & 0.60047 & 0.61408 & 0.71184 &   70 &  119 &  190\cr
\enddata
\end{deluxetable}


\end{document}